\documentclass[12pt]{article}
\pdfoutput=1
\usepackage{draft}
\usepackage{multirow}

\begin{document}

 \begin{titlepage}

\begin{center}

\hfill \\
\hfill \\
\vskip 1cm

\title{Gauge Theory and Skein Modules
}

\author{Du Pei}

\address{Centre for Quantum Mathematics, University of Southern Denmark, \\Campusvej 55, 5230 Odense, Denmark}

\email{dpei@imada.sdu.dk}

\end{center}

 \vfill

\begin{abstract} We study skein modules of 3-manifolds by embedding them into the Hilbert spaces of 4d $\cN=4$ super--Yang--Mills theories. When the 3-manifold has reduced holonomy, we present an algorithm to determine the dimension and the list of generators of the skein module with a general gauge group. The analysis uses a deformation preserving $\cN=1$ supersymmetry to express the dimension as a sum over nilpotent orbits in its Lie algebra. We find that the dimensions often differ between Langlands-dual pairs beyond the $A$-series, for which we provide a physical explanation involving chiral symmetry breaking and 't Hooft operators. We also relate our results to the structure of $\bC^*$-fixed loci in the moduli space of Higgs bundles. This approach helps to clarify the relation between the gauge-theoretic framework of Kapustin and Witten with other versions of the geometric Langlands program, explains why the dimensions of skein modules do not exhibit a TQFT-like behavior, and provides a physical interpretation of the skein-valued curve counting of Ekholm and Shende. 
\end{abstract}
 \vfill

 \end{titlepage}

 \tableofcontents

\section{Introduction}

In this paper, we investigate skein modules of 3-manifolds, using supersymmetric gauge theories and their infrared dynamics as our primary tools. The analysis involves a surprisingly rich collection of quantum field theories in two, three, and four dimensions.

We begin by introducing the cast of characters and the puzzles that we aim to solve.

\subsection{Skein modules}

Three-manifold skein modules were first introduced in the context of the skein relations for the Kauffman bracket polynomial  \cite{Przytycki1991,turaev2016quantum}. Given a connected and oriented three-manifold $M_3$, the Kauffman-bracket skein module sk$(M_3)$ can be defined as the $\bZ[A,A^{-1}]$-module generated by links in $M_3$, modulo ambient isotopies and the skein relations,
\begin{align}
&\vev{\;\vcenter{\hbox{\includegraphics[width=0.07\linewidth]{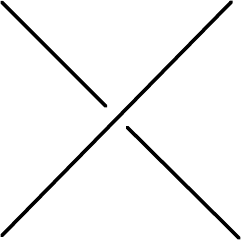}}}\,}
= A\cdot \vev{\;\vcenter{\hbox{\includegraphics[width=0.06\linewidth]{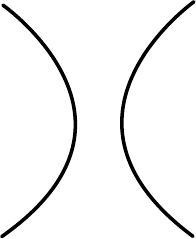}}}\;} + A^{-1} \cdot \vev{\,\raisebox{-0.4\height}{\rotatebox{90}{\scalebox{1.25}[0.8]{{\hbox{\includegraphics[width=0.06\linewidth]{skein2.pdf}}}}}}\,},\\[1.5ex]
&\vev{L \,\cup\, \tikz[baseline=-0.6ex] \draw[line width=.7pt] (0,0) circle (3ex);} = (-A^2 - A^{-2})\cdot\vev{L}.
\end{align}
This construction has a generalization for an arbitrary braided ribbon category $\cC$ \cite{turaev2016quantum}, which recovers the Kauffman-bracket skein module when $\cC$ is the Temperley--Lieb category. What concerns us in this paper is the case of $\cC=\Rep_q(G)$, where $G$ is a connected reductive algebraic group over $\bC$, and we study the corresponding skein module sk$(M_3;G)$. This is a family that directly generalizes the original skein modules, recovering sk$(M_3)$ via a change of variable when $G=\SL(2)$. Although our approach might also lead to interesting predictions about the torsion part of the module, the main focus of this paper is on the case with the value of $q\in\bC^*$ being generic. Equivalently, one can regard sk$(M_3;G)$ as a vector space over the field of fractions $\bQ(q)$.

The skein modules sk$(M_3;G)$ naturally arise in a variety of seemingly different settings in mathematics (see \cite{przytycki2006skeinmodules} for a review). Several of these contexts are closely connected to physics, including:
\begin{itemize}
    \item The skein module can be viewed as the quantization of the character variety $\cM(M_3;G)\simeq \mathrm{Hom}\left( \pi_1(M_3), G \right)/G$, and reduces to its coordinate ring when $q=1$ \cite{Turaev1991,Bullock1997,PrzytyckiSikora2000}. 
    \item In quantum topology, sk$(M_3;G)$ is the natural home for the quantum invariants of links in $M_3$, in the sense that, when $M_3$ is no longer $S^3$, the generalization for the Jones polynomial or the Kauffman bracket is no longer a polynomial but instead an element of sk$(M_3;G)$. 
    \item Skein modules also have a natural place in the geometric Langlands correspondence \cite{beilinson1991quantization}. In particular, they are naturally associated with three-manifolds in (an extension of) the Betti version of the correspondence (see e.g.~\cite{ben2016betti,ben2018integrating,jordan2022langlands}). 
\end{itemize}

While these three facts conceptually connect skein modules to quantum field theories (QFTs) in, respectively, two (via the A-model with certain boundary conditions, see e.g.~\cite{Gukov:2022gei} and references therein), three (via Chern--Simons theory \cite{WittenJones}), and four dimensions, their practical use in the study of skein modules has remained largely mysterious. One of our missions is to clarify the connection between skein modules and physics so that we can then use quantum field theories to make mathematical predictions about the structure and properties of skein modules.

\subsection{Puzzles}

We start with the interplay with 4d theories, partially because results about QFTs in higher dimensions can often subsume those in lower-dimensional ones. In fact, the last point in the previous list, when combined with the gauge-theoretic approach to the geometric Langlands program initiated in \cite{Kapustin:2006pk,gukov2008gauge}, seems to lead to the tempting identification,
\begin{equation}\label{Question}
    \sk(M_3;G) \stackrel{?}\simeq \cH(M_3;G),
\end{equation}
where $\cH(M_3;G)$ denotes the space of states (a.k.a.~the Hilbert space) of the topologically twisted 4d $\cN=4$ super--Yang--Mills theory.

The two sides of \eqref{Question} are actually comparable in the sense that both depend on the choice of the 3-manifold and the gauge group, as well as the value of a complex parameter.\footnote{On the gauge theory side, the corresponding complex parameter is essentially the coupling constant together with the theta angle. We also note that we label the Hilbert space of the gauge theory by the algebraic group $G$ to conform with the notation for skein modules. The actual gauge group is the unique (up to conjugacy) maximal compact subgroup $K$ of $G\simeq K_\bC$ viewed as a complex Lie group. The difference between the algebraic and analytic points of view is largely irrelevant for the results of this paper, and we often do not carefully distinguish between them.} However, these two vector spaces are almost polar opposites, with the skein module being always finite dimensional due to a powerful theorem of \cite{gunningham2023finiteness}, while the Hilbert space $\cH(M_3;G)$ is almost always infinite dimensional. 

Indeed, even for the simplest choice of $M=S^3$, for which $\dim\sk(M_3;G)=1$ for any $G$, the Hilbert space $\cH(M_3;G)$ contains a tower of states given by $H^*(BG)$ generated by gauge-invariant combinations of the adjoint scalar in the theory. See \cite{Gukov:2022cxv} for a more detailed analysis of the Hilbert space, where it was conjectured to be infinite dimensional for any $M_3$. 

After accepting the fact that \eqref{Question} cannot be an isomorphism between vector spaces, one might hope that the dimension on the right-hand side can be regularized---as is often needed to define the partition function of a quantum field theory---leading to an equality
\begin{equation}\label{QuestionDim}
    \dim \sk(M_3;G) \stackrel{?}= \dim'\cH(M_3;G).
\end{equation}
One natural way to regularize is to, instead of the dimension, compute the Witten index---equivalently the graded dimension---in a massive phase of the theory. This has several issues, one is that there might not be such a deformation to a massive theory that is compatible with the topological twist; another is that, when such deformations exist, there might be no canonical one.\footnote{Different massive phases of the gauge theory can have different Witten indices, which are only deformation invariant if the theory remains massive but can jump as we cross walls where massless degrees of freedom emerge.} Lastly, even when deformations exist and they all give the same answer for the Witten index, it can in general be different from the dimension of the skein module. 

Below, we compare the two quantities for $M_3=T^3$:
\begin{center}
    \begin{tabular}{c|c|c|c|c||c||c|c}
       $G$ & SL(2) & SL(3) & SL(4) & SL(5) & GL(5) & $G_2$ & $F_4$  \\\hline
       dim sk$(T^3;G) $  & 9 & 29 & 75& 131&7 &24 & 123\\\hline
       \text{Witten index}  & 10 & 30 & 84& 130& 0 & 30 & 227
    \end{tabular}\;\ldots
\end{center}
Here, the results for SL and GL groups are taken from \cite{gunningham2024skeins}, whereas the results for the exceptional groups are our predictions, which will be explained later together with the computation of the Witten indices. 

In this table, one finds that the two rows are almost never identical. For the SL-series, one might have guessed from the first two data points that the answers always differ by one, and this discrepancy for $\SL(2)$ was noted in \cite{Gukov:2022gei}  and is sometimes known as the ``10-vs-9'' paradox. But this pattern does not persist either within or beyond the $\SL$-series. The Witten index can be either larger (e.g.~SL(4)) or smaller (e.g.~SL(5)), and they can be within 15\% ``error margin'' (e.g.~SL$(N)$ for small $N$),  have a larger gap (e.g.~the exceptional series or classical groups with large rank), or be qualitatively different (e.g.~GL$(N)$).

For $M_3=\Sigma\times S^1$, with $\Sigma$ a genus-$g$ Riemann surface, the mismatch becomes worse. On the side of the skein module, one has \cite{gilmer2019skein,detcherry2021basis}
\begin{equation}\label{SL2SkeinDim}
    \dim \sk(\Sigma\times S^1;\SL(2))= 2^{2g+1}+2g-1,
\end{equation}
while the Witten index is given by\footnote{This is with a particular deformation which we will explain later. For another deformation used in \cite{Gukov:2022cxv}, the discrepancy is even larger.}
\begin{equation}\label{SL2WittenInd}
    \cI(\Sigma\times S^1)=2^{2g+1}+2.
\end{equation}
From the field-theory point of view, what is more concerning is not the fact that the difference between the two formulae grows with the genus, but rather the non-TQFT behavior of \eqref{SL2SkeinDim}, due to the presence of the term that is linear in $g$. In other words, while \eqref{SL2WittenInd} is of the form expected from the partition function of a TQFT, the formula in \eqref{SL2SkeinDim} is not, and one should not expect to obtain it from a TQFT in a simple fashion.

This touches upon a bigger issue---the non-TQFT behavior in (the Betti version of) the geometric Langlands program, which is itself puzzling if one views the correspondence as originating from the $S$-duality of the topologically twisted 4d $\cN=4$ theory, which is, by construction,  topological.

\subsection{The solution}

We will resolve these puzzles, starting with a more precise understanding of the relation between the skein module and the Hilbert space of the 4d gauge theory, which will ultimately allow us to generalize \eqref{SL2SkeinDim} to other groups. Schematically, the dimension formula for a simply connected $G$ takes the following form,
\begin{equation}\label{SkDimG}
    \dim \sk(\Sigma\times S^1;G)\sim\sum_{\lambda}|C_\lambda|^{2g+1}\cdot \Gamma\left(G_{\lambda},c_\lambda\cdot (2g-2)\right),
\end{equation}
which is a sum over conjugacy classes $\lambda$ of $\fsl_2$-triples in $\frak{g}$ (or, equivalently, nilpotent orbits $\cO_\lambda\subset\frak{g}$), with the summand being a product of two factors. The first factor counts the number of flat $C_\lambda$-connections over $\Sigma\times S^1$ with $C_\lambda:=\pi_0(G_\lambda)$ 
denoting the group of components of the centralizer subgroup $G_\lambda\subset G$ of the triple. 
The second factor $\Gamma\left(G_{\lambda},c_\lambda\cdot (2g-2)\right)$ counts the number of dominant weights of $G_{\lambda}$ at level $c_\lambda\cdot (2g-2)$---with $c_\lambda$ related to the embedding index of $G_\lambda\subset G$---modulo the action of the center symmetry $Z(G)$.

To obtain the actual dimension formula and to incorporate cases with non-simply connected $G$, there are a few modifications to \eqref{SkDimG} that will be explained later. We will also study its doubly graded refinement by electric and magnetic fluxes valued in $e\in H_1(M_3,Z(G)^\vee)$ and $m\in H^2(M_3,\pi_1(G))$ coming from the electric and magnetic 1-form symmetries of the 4d gauge theory. For now, we will give some examples for $G=\SL(N)$ or $\GL(N)$, for which no modification of \eqref{SkDimG} is needed.

\subsubsection{Examples}

When the genus $g=1$, we have $\Gamma(G_{\lambda},0)=1$. For $G=\SL(N)$ or $\GL(N)$, the nilpotent orbits are labeled by partitions of $N$. The formula \eqref{SkDimG} becomes
\begin{equation}\label{SkDimGT3}
    \dim \sk(\Sigma\times S^1;G)=\sum_{\lambda\vdash N}|C_\lambda|^{3}.
\end{equation}
 While the group $C_\lambda$ is trivial for GL($N$), for $G=\SL(N)$ it is given by $C_\lambda=\bZ_{\gcd(\lambda)}$, a cyclic group of order equal to the greatest common divisor of all parts of the partition, and thus \eqref{SkDimG} recovers the results in \cite{gunningham2024skeins}. 

On the other hand, for general $g$, the formula reproduces \eqref{SL2SkeinDim} when $G=\SL(2)$, but yields new predictions for other Lie groups. For example, when $G=\SL(3)$ or PSL(3), the formula gives \begin{equation}\label{SL3}
    \dim \sk (\Sigma\times S^1;\SL(3))=3^{2g+1}+3(2g+1)(g-1)+1+\delta_{g,1}.
\end{equation}
The appearance of the Kronecker delta $\delta_{g,1}$ turns out to be a general feature of contributions from ``intermediate'' $\fsl_2$-triples with a non-semisimple centralizer $G_\lambda$.

\subsubsection{Strategy}

In deriving \eqref{SkDimG}, we utilize the following three main ideas.

\paragraph{Embedding into gauge theory.}  We propose an embedding of sk$(M_3;G)$ into the dual of $\cH(M_3;G)$, 
\begin{equation}\label{Embed}
     \sk(M_3;G) \subset \cH(M_3;G)^\vee,
\end{equation}
which we believe clarifies the relation between them. Alternatively, this realizes the dual of the skein module as a quotient, 
\begin{equation}
    \sk(M_3;G)^\vee \simeq \cH(M_3;G)/\!\sim .
\end{equation}
This embedding/quotient utilizes a setup similar to those in \cite{Witten:2011zz,Gaiotto:2011nm,Gukov:2016gkn,Gukov:2017kmk} used to study quantum invariants of knots and 3-manifolds and their categorifications. This proposal allows us to readily recover several known results on sk$(M_3)$ in cases where $\cH(M_3)$ has a simple description, such as when $M_3\simeq S^3/\Gamma$ is spherical. 

\paragraph{An $\cN=1$ deformation.} When an $M_3$ has reduced holonomy, a powerful computational tool for the dimension of $\sk(M_3;G)$ is the $\cN=1$ deformation studied in \cite{Witten:1994ev,Vafa:1994tf}. Combined with the proposed embedding \eqref{Embed}, this gives the dimension of $\sk(T^3)$ for any $G$ via counting vacua in the deformed theory modulo certain equivalence relations, which explains the 10-vs-9 paradox. Furthermore, it leads to a surprising finding that the dimensions are often different for Langlands dual pairs, 
\begin{equation}
    \dim\sk(T^3;G)\neq \dim\sk(T^3;^\mathrm{L}\!G)
\end{equation}
once we go beyond the $A$-series. When $M_3=\Sigma\times S^1$, the deformation will give rise to  ``cosmic strings''---surface defects with massless degrees of freedom on their worldsheet, and one technical challenge to overcome is to understand how they contribute to sk$(M_3)$. This leads to the next point.

\paragraph{Bulk--string coupling.} It turns out that we will need to understand the action of the bulk Wilson lines on the Hilbert space of the cosmic string, which we argue can be modeled by the coupled system of a Chern--Simons bulk and a WZW/free-fermion boundary theory after compactifying on the ``meridian circle.'' This allows us to obtain the dimension formula \eqref{SkDimG} and to give an algorithm to produce the set of generators.

\subsubsection*{Organization of the paper}

The paper is organized around the sequential introduction and exploration of the three ideas outlined above. In Section~\ref{sec:embedding}, we discuss the relation between sk$(M_3)$ and the Hilbert space of the 4d $\cN=4$ super--Yang--Mills theory.
In Section~\ref{sec:deform}, we use the $\cN=1$ deformation to study skein modules on $T^3$. In Section~\ref{sec:cosmic}, we incorporate cosmic strings to complete the computation for sk($\Sigma\times S^1$). We then examine the extent to which our results are compatible with Langlands duality in Section~\ref{sec:checks}, where we also study the connection with the geometry of the moduli space of Higgs bundles. In Section~\ref{sec:future}, we briefly explore some other applications of our approach---emphasizing new perspectives on existing results---and outline some interesting future directions.  

\section{Embedding}\label{sec:embedding}

We first outline the construction for embedding the 3-manifold skein module into the Hilbert space of the 4d $\cN=4$ theory. 

\subsection{Summary of the construction}

To study the $G$-skein module sk$(M_3;G,q)$ for the connected reductive algebraic group $G$ via gauge theory, it is convenient to view $G$ instead as a complex Lie group. Its maximal compact Lie subgroup is $K$, which will be the gauge group of our UV theory. A well-known result on the structure of $G$ and $K$ states that $K=(U(1)^r\times \widetilde H)/F$ is a finite quotient of a product of an abelian part with a simply-connected semisimple compact Lie group $\widetilde H$. The ``non-abelian part'' of $K$ is given by the commutator subgroup $H:=[K,K]$ and is a quotient of $\widetilde H$ by a subgroup of its center $Z(\widetilde H)$.

We then consider the topologically twisted 4d $\cN=4$ super--Yang--Mills theory on $M_3\times \bR_+$ with gauge group $K$. The boundary condition at the origin $x=0$ of $\bR_+$ is the one used in \cite{Witten:2011zz} for discussions of the analytically continued Chern--Simons theory and the categorification of the Jone polynomial, which is the ``tip of the cigar'' boundary condition when lifted to higher dimensions (see also~\cite{Gukov:2016gkn,Gukov:2017kmk} where a similar setup is used in the study of invariants of 3-manifolds), and it is expected that insertions of Wilson lines along $M_3$ at $x=0$ are equivalent up to ambient isotopy and skein relations \cite{Witten:2011zz,Gaiotto:2011nm}. 

\begin{figure}[htb!]
     \centering
     \includegraphics[width=0.75\linewidth]{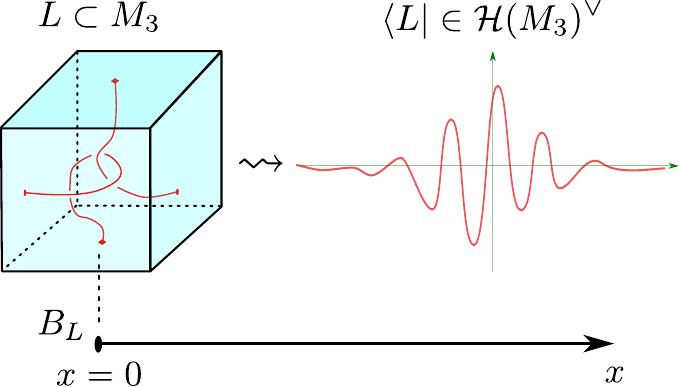}
     \caption{A boundary condition of the topologically twisted 4d theory on $M^3\times \bR^+$ with insertions of Wilson lines along $L\subset M_3$ is illustrated on the upper left. Upon compactification on $M_3$, this gives a boundary condition $B_L$ for the quantum mechanics on $\bR^+$ (lower part of the figure), which gives a state $\bra{L}$ in the dual of its Hilbert space $\cH(M_3)^\vee$ (upper right).}
     \label{fig:construction}
 \end{figure}

Viewing $\bR_+$ as the time direction and $x=0$ as either the beginning or the end of time gives a state $|L\rangle$ in the Hilbert space $\cH(M_3)$ of the 4d theory on $M_3$ or $\langle L|$ in its dual, where $L$ encodes information about the insertion of Wilson lines. We will take the latter perspective,\footnote{One reason is that we typically prefer to talk about normalizable states, whereas $|L\rangle$ is expected to be not normalizable. On the other hand, the dual $\cH(M_3)^\vee$ is bigger and can actually contain $\bra{L}$.} viewing the collection $L$ of line operators as a functional on $\cH(M_3)$ and define 
\begin{equation}\label{embedding}
\cS(M_3;G,q)=\text{span}\{\bra{L}\}\subset  \cH(M_3)^\vee   
\end{equation}
 as the space of functionals spanned by all such insertions. Here, $q:=e^{2\pi i \Psi}$ with $\Psi$ being the ``canonical parameter'' formed out of the coupling constant $\tau$ of the gauge theory and a twist parameter $t$ (see Appendix~\ref{app:Details} for more details). This construction is illustrated in Figure~\ref{fig:construction}.

In one direction, it is obvious that 
\begin{equation}
    \text{sk}(M_3;G,q)\rightarrow \cS(M_3;G,q)
\end{equation}
is surjective by construction. How to show that the map is also injective? Injectivity equivalently means that any pair $L$ and $L'$ in sk$(M_3;G,q)$ that are not equivalent can always be distinguished by a state $|\psi\rangle$ with $\vev{L|\psi}\neq \vev{L'|\psi}$. As sk$(M_3)$ is the quantization of the character variety and the wave function of $L$ and $L'$ can be probed by $\vev{L|\alpha}$ and $\vev{L'|\alpha}$ with $\alpha$ being a flat connection. Intuitively, if they agree for all $\alpha$, then they have the same quantum wave function and should also agree as elements in the skein module.

Based on this reasoning---as well as the ``naturalness'' of the construction \eqref{embedding}---we conjecture that 
\begin{equation}
    \sk(M_3;G,q)\simeq \cS(M_3;G,q)
\end{equation}
are isomorphic as vector spaces. This enables us to access the skein modules of three-manifolds using QFT techniques, which, in turn, allows us to test this conjecture by comparing its predictions with known results.

\subsection{'t Hooft operators and magnetic sectors}

One might be concerned that the construction of the subspace/quotient $\cS(M_3)$ is not compatible with the $S$-duality of the 4d $\cN=4$ theory, which exchange Wilson and 't Hooft line operators \cite{goddard1977gauge,montonen1977magnetic}. Therefore, the identification of $\cS(M_3)$ with the skein module of $M_3$ does not automatically guarantee the duality between sk$(M_3,G)$ and sk$(M_3,^\mathrm{L}\!G)$. 

Therefore, it might be tempting to define a duality-invariant version $\tilde \cS$ by allowing also 't Hooft operators in the construction \eqref{embedding}, which live on a D5-like boundary condition. Due to Witten effect, they will become dyonic under $T$-transformations, and, to have the full SL$(2,\bZ)$ symmetry, one should include all dyonic lines on all $(p,q)$-type boundary conditions. We will denote the resulting vector space as $\tilde \cS$. Then, automatically, one has 
\begin{equation}
\tilde \cS(M_3;G,q)\simeq \tilde \cS(M_3;^\mathrm{L}\!G,^\mathrm{L}\!q),
\end{equation}
where $^\mathrm{L}\!q=e^{-2\pi i\frac{1}{n_{\frak g}\Psi}}$, with $n_{\frak g}$ being the lacing number of $\frak g$.\footnote{When $G$ is not simple, one can make the coupling constant for each abelian or simple summand of $\frak g$ independent. Therefore, $q$ is in fact a collection of complex numbers, with $n_\frak{g}$ in the formula also understood as a collection of integers. As our focus is on the case with $q$ generic, we will not keep track of this collection and will simply denote it as $q$ or suppress it altogether.}
From the point of view of duality-invariance, it is certainly advisable to work with $\tilde \cS$ instead, but its connection with the skein module is less straightforward. Indeed, the dual boundary conditions are not very ``Chern--Simons--like'' and it is not clear how to incorporate the line operators on these boundary conditions in a simple fashion into skein theory. Despite of this, we will refer to $\tilde \cS$ as the  ``enriched skein module.''

One might hope that going from $\cS$ to $\tilde \cS$ would not actually enlarge $\cS$ in a meaningful way. One justification would be that Wilson and 't Hooft operators behave as canonical dual pairs (e.g.~like position and momentum), and one could generate the entire representation, on which both of them act, only using one set of the operators. 

However, there is one important subtlety, as the above is only true if one starts with a ``generic state.'' On the other hand, if the state one starts with is not generic enough, one would not be able to generate the entire representation only using Wilson lines, but instead get only a subspace. This clearly happens when $\pi_1(K)=\pi_1(G)$ has torsion and there are different topological types of $G$-bundles labeled by the magnetic flux $m\in H^2(M_3, \pi_1(G))$. Indeed, in such a situation, an 't Hooft line can change the topological type of the gauge bundle which can never be achieved by a Wilson line insertion. 

Partly motivated by this, we consider a slight generalization of $\cS$ when $G$ is not simply connected by also taking into account the existence of different ``magnetic sectors,''
\begin{equation}
    \cS(M_3;G)=\bigoplus_{m}\cS^m(M_3;G),
\end{equation}
where each $\cS^m(M_3;G)$ is generated by Wilson lines alone in the presence of the given magnetic-flux background.\footnote{ In this notation, what the construction \eqref{embedding} gives is the sector $\cS^{0}$ with $m=0$. It might be a better idea to refer to the version with $m$ summed over as $\cS^{\text{total}}$. But as the meaning of ``$\cS$'' is often clear from the context, we do not carefully distinguish them by adopting different notations. } Adding back the magnetic sectors ``by hand'' allows us to only work with Wilson lines, as well as providing a conceptual explanation why the Langlands duality for skein modules can (sometimes) hold even though the skein modules do not seem to incorporate the 't Hooft lines.

It turns out that there can be other ``degeneracies'' leading to states in $\tilde \cS$ that cannot be reach by Wilson lines, even in the $m=0$ sector. In some cases (e.g.~for SL($N$) vs PSL($N$)), they appear on both sides and ignoring them does not ruin the duality. However, we will also encounter cases (e.g.~dual $B$--$C$ pairs) where the duality for $\cS$ fails, but can be restored by passing to the enriched version $\tilde \cS$.

\subsection{Electric sectors}

The Hilbert space $\cH$ (or its dual) also admits a grading by electric fluxes $e\in H_1(M_3,Z(G)^\vee)$ coming from the center $Z(G)$ 1-form symmetry of the 4d gauge theory. Combined with the magnetic 1-form symmetry, it refines the subspace/quotient $\cS$ by a double grading of both $e$ and $m$. (See also \cite{Witten:2000nv,Kapustin:2006pk,Gukov:2020btk,Gukov:2025dol} for related discussions.) For a state in $\cS$ represented by a configuration of Wilson lines, their charges under $Z(G)$ determine a $Z(G)^\vee$-valued 1-cycle which in turn determines $e$. 

A ``dual'' way of interpreting the electric flux is the following. The 1-form symmetry in 4d becomes an $H^1(M_3,Z(G))$ 0-form symmetry acting on the Hilbert space $\cH(M_3)$ of the quantum mechanics after compactification. This gives an isotypical decomposition of the Hilbert space into sectors labeled by characters,
\begin{equation}
    \cH(M_3)=\bigoplus_{e} \cH^e(M_3),
\end{equation}
with $e\in H^1(M_3,Z(G))^\vee$. The two perspectives are related by the universal coefficient theorem, which identifies this group of characters with $ H_1(M_3,\bZ)\otimes Z(G)^\vee\simeq H_1(M_3,Z(G)^\vee)$. Equivalently, one can use the perfect pairing between $H^1(M_3,Z(G))$ and $H_1(M_3,Z(G)^\vee)$ that describes the action of elements of the group on charges.

To make $e$ and $m$ more symmetrical, one can use Poincar\'e duality to regard $e$ as a 2-cocycle living in $H^2(M_3,Z(G)^\vee)$. The $S$-duality then exchanges $e$ and $m$ via the natural identification
\begin{equation}
    \pi_1(G)\simeq Z\big(^\mathrm{L}\!G\big)^\vee,\quad \text{and}\quad Z(G)^\vee\simeq \pi_1\big(^\mathrm{L}\!G\big),
\end{equation}
between Langlands/GNO dual groups.

In fact, it is often convenient to work with the Hilbert space of the ``relative'' theory, which has non-local electric and magnetic charges but is consistent as a boundary condition of a 5d non-invertible TQFT. When $\frak g$ is semisimple, the Hilbert space of the relative theory has sectors labeled by the pair $(e,m)$ where $e$ and $m$ both live in $H^2(M_3,Z(\frak g))$ with $Z(\frak g)$ denoting the center of the simply-connected group with Lie algebra $\frak g$.\footnote{Here, we have used an canonical isomorphism between $Z(\frak g)$ and is dual given by the pairing 
\begin{equation}\label{Zpairing}
    Z(\frak g)\times Z(\frak g)\to \bC/\bZ.
\end{equation}
 If we identify $Z(\frak g)\simeq \mathrm{coker}(A_{\frak g})$ as the co-kernel of the Cartan matrix $A_{\frak g}$, the pairing \eqref{Zpairing} is then given by the inverse of the Cartan matrix $A^{-1}_{\frak g}$. When $\frak g$ is not semisimple and $Z({\frak g})$ is not discrete, it turns out that one only needs the component group $\pi_0Z({\frak g})$ as only the torsion part of $Z({\frak g})^\vee$ is expected to give a non-trivial Hilbert space in the topologically twisted theory. This is dual to the statement that the meaningful labels for magnetic fluxes live in the torsion subgroup of $\pi_1(G)$. } This also leads to a doubly graded ``universal version'' of $\cS$,
\begin{equation}
    \cS(M_3;\frak g)=\bigoplus_{e,m}\cS^{e,m}(M_3;\frak g).
\end{equation}
Just like one can build the Hilbert space of any absolute version of the 4d theory by assembling different sectors of the relative theory, one can build the space $\cS(M_3;G)$ for any global form $G$ from sectors  $\cS^{e,m}(M_3;\frak g)$ of the universal version. Also, one constraint is that the space $\cS^{e,m}$ can only be non-trivial when $e$ and $m$ are mutually local (i.e.~they com from an ``Lagrangian'' subgroup of $Z({\frak g})\times Z({\frak g})$ with respect to the pairing \eqref{Zpairing}).

This formulation will be convenient later (e.g.~in Section~\ref{sec:duality}) when we test the Langlands duality for skein modules, which can be naturally interpreted as an electric-magnetic duality for the doubly graded $\cS^{e,m}$.

\subsection{Examples}

Without much work, this embedding of sk$(M_3)$ into gauge theory already allows us to recover some familiar yet non-trivial results about skein modules. 

\paragraph{Connected sum.} For a test of the construction, consider the connected sum $M_3\#M_3'$ of two 3-manifolds $M_3$ and $M_3'$. Using topological invariance, one can stretch the two summands far apart, connected via a long tube $S^2\times [-l,l]$ with $l\gg1$. One can then view $\cH(M_3\#M_3')$ as the Hilbert space of the $S^2$-compactification of the 4d theory on the interval $[-l,l]$ with two boundary conditions. At low energy, the $S^2$-compactification is described by a 2d $(4,4)$ sigma model with target $(\frak k^4)/W$ parametrized by four of of the six adjoint scalars that remain scalars on $S^2$. This can be constructed as a symmetric orbifold of a free theory, and the UV Wilson lines are heavy and become defect operators. Therefore, Wilson lines stretched between the two boundaries will cost energy and such states will not be in the IR Hilbert space. Topological invariance then implies that they pair trivially with any incoming states. As a consequence, the Wilson lines can only be inserted at the two boundaries, and we have a factorization,
\begin{equation}
    \cS(M_3\#M_3')=\cS(M_3)\otimes \cS(M_3'),
\end{equation}
recovering a well-known property of skein modules.

\bigskip

Another case where our construction can be readily applied to compute the skein module is when the character variety $\cM(M_3;G)$ is a collection of points. In these cases, one can argue that the gauge theory on $M_3$ at low energy decomposes into vacua labeled by the flat connections. In each vacuum, the expectation values of all Wilson lines are determined by the flat connection, and their span $\cS(M_3;G)$ is the same as the span of the flat connections or, equivalently, $\bC[\cM(M_3;G)]$.

\paragraph{Lens spaces.} To give a concrete example, consider a lens space $M_3=L(p,r)$ and $G=\SL(2)$. The moduli space consists of $\lfloor\frac{p}{2}\rfloor+1$ points. The space $\cS$ can be thought of as being generated by Wilson lines labeled by weight $\lambda\in \bZ_{\ge0}$ wrapping a generator of $\pi_1(M_3)\simeq \bZ_p$, as a multi-wrapped Wilson line can always be decomposed. The sectors of the 4d theory on $L(p,r)$ are labeled by SL$(2)$ flat connections diag$\{\alpha_j,\alpha_j^{-1}\}$ with  $\alpha_j:=e^{2\pi i j/p}$. The $\lambda$-th Wilson line in the $j$-th sector has expectation value
\begin{equation}
\vev{W_\lambda}_j=\vev{W_\lambda|\alpha_j}=\frac{\sin\left(\frac{2\pi j \lambda}{p}\right)}{\sin\left(\frac{2\pi j }{p}\right)}.
\end{equation}
This gives equivalence relations $\bra{W_\lambda}\sim \bra{W}_{\lambda+p}$ and  $\bra{W_\lambda}\sim-\bra{W_{p-\lambda}}$, and therefore carving out a subspace of dimension $\lfloor\frac{p}{2}\rfloor+1$ inside $\cH(L(p,r))^\vee$. The space $\cS$ can be dually thought of as being a quotient of $\cH(L(p,r))$, which collapses $\cH(L(p,r))$ by identifying infinite towers of states above each vacuum as they cannot be distinguished from each other using Wilson lines.\footnote{At least part of the tower can be seen perturbatively and is identified with $H^*(\mathrm{B}G)\oplus H^*(\mathrm{B}T)^{\lfloor\frac{p}{2}\rfloor-1}$, which can be captured by the \textit{derived} skein module of the lens space (we thank David Jordan and Penghui Li for explaining this fact to the author). It is an interesting question how to incorporate derived skein modules in our construction, which we will leave for future work. Another interesting question is whether there are more states beyond these towers. When $p>1$, there are hints of additional states from a higher-dimensional perspective by considering the 6d $(2,0)$ theory on $L(p,1)\times \Sigma \times \bR$, which gives rise to a quantum mechanics whose index is an infinite power series for a general Riemann surface $\Sigma$ \cite{Gukov:2015sna}. When $\Sigma=T^2$, there is a massive deformation that yields a finite-dimensional quantum mechanics, which is the $S^1$-compactification of a 2d theory with massive vacua in bijection with the flat connections on $L(p,1)$ \cite[Sec.~3.1]{Pei:2015jsa}.  }

\medskip

While these infinite towers do not cause much problem in examples with isolated $\cM(M_3)$, they appear to be more troublesome to deal with when $\cM(M_3)$ has higher-dimensional components, where one can no longer count the number of flat connections but instead has to quantize the classical configurations to find the quantum vacua. The extreme case of this, in the sense of being the polar opposite of having an isolated $\cM$, is when $M_3=\Sigma\times S^1$.

Understanding skein modules for such manifolds is one of the main goals of this paper, and the solution neatly showcases the power of the QFT approach. One essential ingredient is a deformation of the 4d $\cN=4$ super--Yang--Mills theory to $\cN=1$, which we will discuss next. Readers interested in learning more about the details of the construction are directed to Appendix~\ref{app:Details}.

\section{Deformation}\label{sec:deform}

The idea of using a deformation preserving $\cN=1$ supersymmetry to study the topologically twisted theory originates from \cite{Witten:1994ev} while what is more relevant for the present work is the application of this technique to the Vafa--Witten theory \cite{Vafa:1994tf}, as $\cH(M_3)$ in our setup can be also identified with that of the Vafa--Witten theory. One difference is that our focus is more on the Hilbert space as opposed to the partition function.

Recall that the $\cN=4$ theory, when viewed as an $\cN=1$ theory, has three adjoint chiral multiplets $T,U$ and $V$ with the superpotential,
\begin{equation}
    W=\Tr T[U,V],
\end{equation}
and we consider the $\cN=1$ deformation given by adding another term,\footnote{This is sometimes known as the $\cN=1^*$ deformation. In fact, we need a slightly different deformation when considering the theory on a more general Kähler manifold, which we will review later.}
\begin{equation}
    W'=-\frac{m}{2}\cdot \Tr \left(T^2+U^2+V^2\right).
\end{equation}
Turning on such an interaction will cause the theory to decompose into a collection of sectors labeled by critical points of $W+W'$, where the fields are constant and satisfy
\begin{align}
    [U,V]=&mT,\\
    [V,T]=&mU,\\
    [T,U]=&mV.
\end{align}
Modulo gauge transformations, the solutions are labeled by conjugacy classes $\lambda$ of $\fsl_2$-triples in $\frak{g}$, which are in 1-to-1 correspondence with nilpotent orbits of $\frak{g}$.

\subsection{Phases and nilpotent orbits}

In principle, one can avoid talking about nilpotent orbits and just study the vacua via the $\fsl_2$-triple. However, it turns out to be quite useful and illuminating to make the connection, bringing the field-theory analysis closer to the standard toolkit of (geometric) representation theorists. Even for physicists, this provides a cleaner way to organize the IR phases of the theory, especially when $G$ is not a classical Lie group. See \cite{collingwood1993nilpotent,Jantzen2004} for nice introductions of the subject of nilpotent orbits and also \cite{gukov2010rigid,Gaiotto:2008ak} for closely related appearances of them in the $\cN=4$ super--Yang--Mills theory.

For a general complex reductive $G$, we label the nilpotent orbits $\cO_\lambda$ also by $\lambda$, which in turn labels a phase of the deformed theory. The gauge group $K$ will be broken by the vevs of the adjoint scalars to the centralizer subgroup $K_\lambda$ of the $\fsl_2$-triple. 

In general, this group, as well as its complexification $G_\lambda$, has more than one component. We have the short exact sequence,
\begin{equation}
    G_\lambda^0\to   G_\lambda \to C_\lambda,
\end{equation}
with $G_\lambda^0$ the identity component of $G_\lambda$ and  $C_\lambda:=G_\lambda/G_\lambda^0$ the group of components.

A well-known result is that, when $G$ is semisimple and simply-connected, $C_\lambda$ is isomorphic to the fundamental group of the nilpotent orbit $\pi_1(\cO_\lambda)$. For more general $G$, it is a quotient of $\pi_1(\cO_\lambda)$ and can be found in tables (e.g.~\cite[Ch.~8.4]{collingwood1993nilpotent}). On the opposite end, when $G$ is of adjoint type, $C_\lambda\simeq A(\cO_\lambda)$ which is known as the ``component group.'' When $\frak{g}$ is either $\fsl(N)$ or $\frak{sp}(N)$, $C_\lambda$ is always abelian, which is an important simplification, as its number of flat connections on any $M_3$ is simple to count and given by $|H^1(M_3,C_\lambda)|$.\footnote{When $C_\lambda$ is non-abelian, it is constructed out of the permutation groups $S_3$ (for $G_2$ and $E_\ell$), $S_4$ (for $F_4$), $S_5$ (for $E_8$), the dihedral group D$_8$ of order 8, and the quaternion group Q$_8$ (both only arise for $\frak{so}(N)$). Therefore, the computation for $|\cM(M_3,C_\lambda)|$ is still a manageable task. We list the results for $M_3=\Sigma\times S^1$ with $C_\lambda$ being the above finite groups in Appendix~\ref{app:DW}.} 

Inside $K_\lambda^0$, where can be a non-abelian commutator subgroup $H_\lambda:=[K_\lambda^0,K_\lambda^0]$. The 4d $\cN=1$ theory with the non-abelian gauge group $H_{\lambda}$ will have confinement in the infrared, and the classical vacuum for the $H_{\lambda}$ gauge theory will split into a collection of massive quantum vacua. (See \cite{Tachikawa:2018sae} for an excellent review on the dynamics of 4d $\cN=1$ gauge theories.) Aside from this degeneracy, at long distances, one only sees the gauge group $K_\lambda^{\text{IR}}=K_\lambda/H_{\lambda}$. It is given by the semi-direct product $C_\lambda\ltimes K_\lambda^0/H_\lambda$, with the two factors describing the finite and continuous parts of the IR gauge group. The structure of $G_\lambda^0$ can be described uniformly for classical groups \cite{Carter1985} and can be found in \cite{elashvili1975centralizers,ELKINGTON1972137} for exceptional groups. See also \cite{Carter1985,LawtherTesterman2011} where much detail on nilpotent orbits and their centralizers is presented.

The phase diagram of the theory is parametrized by the $\fsl_2$-triple and, modulo gauge transformations, becomes the Hasse diagram for the partial order of $\cO_\lambda$ given by orbit closure. The largest orbit in the partial order is known as the principal/regular orbit, for which $K$ is broken down to its center $Z(K)$, and the infrared dynamics is described---assuming $K$ is semisimple---by a finite-group gauge theory. On the opposite end is the zero orbit that corresponds to the trivial $\fsl_2$-triple. In this phase, $K_\lambda=K$ in the UV but the non-abelian part $H_{\lambda}=H:=[K,K]$ becomes confined in the IR, leaving a free abelian theory with gauge group $K_\lambda^{\text{IR}}=K^{\text{ab}}:=K/H$, the abelianization of $K$. Another two orbits (hence phases) with special names are the sub-regular and the minimal orbits. They also play some interesting roles in our analysis as we will see later.

When $G=\SL(N)$, each $\lambda$ corresponds to a partition of $N$ and $C_\lambda=\bZ_{\gcd(\lambda)}$ with $\gcd(\lambda)$ being the greatest common divisor of all the numbers in the partition. Below we list the phases together with their relevant data for $N=2,3,$ and 4.\footnote{We mostly use additive notation for abelian groups in this paper. Therefore, ``0'' denotes the trivial group with one element in our tables.}
\begin{center}
    \begin{tabular}{c||c|c||c|c|c||c|c|c|c|c}
                 $N$ & \multicolumn{2}{c||}{$2$}& \multicolumn{3}{c||}{$3$} & \multicolumn{5}{c}{$4$} \\\hline
                 $\lambda$ & [2] & $[1^2]$ & [3] & [2,1]& $[1^3]$ & $[4]$ & $[3,1]$ & $[2^2]$ & $[2,1^2]$ & $[1^4]$ \\\hline
                 $C_\lambda$ & $\bZ_2$ & 0 & $\bZ_3$ & 0 & 0 & $\bZ_4$ & 0 & $\bZ_2$ & 0 & 0\\\hline
                 $K_\lambda^0$ & 0 & SU(2) & 0 & U(1)& SU(3) &0 & U(1) & SU(2) & SU(2)$\times$U(1) & SU(4) \\\hline
                $K_\lambda^{\text{IR}}$ & $\bZ_2$ & 0 & $\bZ_3$ & U(1) & 0 &$\bZ_4$ & U(1) & $\bZ_2$ & U(1) & 0 \\\hline
                type & E & M & E & N & M & E & N & HO & N & M
                \\\hline\hline
                index & 8 & 2 & 27 & 0 & 3 & 64 & 0 & 16& 0 & 4\\\hline
                skein & 8 & 1 & 27 & 1 & 1 & 64 & 1 & 8& 1 & 1
    \end{tabular}
\end{center}
Here, ``type'' is referring to a classification of phases based on the fate of the center of the semisimple part $H$ of the gauge group $K$ in the IR. To make the classification independent of the choice of the global form of $K$ (e.g.~SU vs PSU), we consider the variant with $H$ being simply connected. There are in general four possibilities.

\paragraph{Electric phase (E).} This is when (a part of) $Z(H)$ survives at low energy, becoming either $C_\lambda$ by itself or a non-trivial subgroup of $C_\lambda$ and part of the discrete gauge symmetry in the IR. When $K$ is semisimple with a non-trivial center, the phase corresponding to the principal orbit is always an electric phase, where the gauge symmetry is completely broken down to $Z(K)$ by the vevs of the scalars and one has a $Z(K)$ gauge theory at long distances. Electric phases cannot exist if either $Z(H)$ is trivial or contained in the identity component of $Z(K)$ when $K$ is not semisimple. One example of such is $K=\U(N)$, for which the center of $H=\SU(N)$ is inside $Z(K)=\U(1)$ and becomes part of a U(1) gauge theory in the infrared in every phase.

\paragraph{Magnetic phase (M).} This is when (a part of) $Z(H)$ lives in $H_\lambda$ and disappears in the IR due to confinement. We additionally require that the IR phase is gapped.\footnote{In fact, to make it dual to the definition of the electric phase, one should instead require that the extension $Z(H)\to Z(K)\to K/H$ is trivial when restricting to (a part of) the subgroup of $Z(H)$ that lives in $H_\lambda$, which is a slightly weaker condition. However, its difference with the gapped condition is only relevant in ``boring'' cases such as when $K=\U(1)\times H$ has an abelian factor, or in highly contrived cases not relevant for us. Either the gapped condition or this technically more accurate condition is meant to ensure that we can turn on 't Hooft fluxes for $H$ at no cost of energy. When the extension is non-trivial, an 't Hooft flux requires an accompanying first Chern class for the abelian part, and such a sector will have a non-zero ground-state energy.} The zero orbit always corresponds to such a phase when $K$ is semisimple with a non-trivial center, as the entire $K$ survives the deformation and becomes confined in the IR. One non-example is again $K=\U(N)$, for which the center of $H=\SU(N)$ can be confined, but the IR phase is never gapped. 

\paragraph{Neutral phase (N).} This is when $Z(H)$ doesn't appear in either $C_\lambda$ or $Z(H_\lambda)$. In other words, it is part of the identity component of the IR gauge group $K^\text{IR,0}_\lambda$. When this happens, the Hilbert space of the IR theory doesn't have non-trivial electric- or magnetic-flux sectors---hence the term ``neutral.'' This can be the only type of phases when $Z(H)=0$ (e.g.~$K=G_2,F_4,E_8$) or when $Z(H)$ is contained in the identity component of $Z(K)$ (e.g.~$K=\U(N)$). A more interesting neutral phase (i.e.~when $K$ is not of the kind mentioned above) is the $\lambda=[21]$ phase with $K=\SU(3)$, where the unbroken gauge symmetry is U(1) and the center $\bZ_3$ of $K$ is fully contained in it.   

\paragraph{Hybrid phase (H).} This is the opposite of the neutral phase: in a hybrid phase, one can turn on either electric or magnetic background. There are actually two subtypes: HO and HE. The former stands for ``ordinary hybrid'' phase. In such a phase, a subgroup of $Z(H)$ is in $Z(H_\lambda)$ and disappears in the IR due to confinement, while a quotient is in $C_\lambda$ and becomes part of the discrete gauge symmetry in the IR. Therefore, strictly speaking, it is also both an electric and a magnetic phase (although we will only count them as HO phases in order to make the labels as mutually exclusive as possible). One example for an HO phase is the one labeled by the partition $\lambda=[22]$ for $K=\SU(4)$. A $\bZ_2$ subgroup of the $\bZ_4$ center will become the center of the unbroken $\SU(2)$ that confines, while the $\bZ_2$ quotient will be the gauge group in the gapped IR phase. On the other hand, an exceptional hybrid (HE) phase is one that looks naively like an E/M/N/HO phase with (a part of) $Z(H)$ inside $K^\text{IR,0}_\lambda$ but has non-trivial flux sectors because $K^\text{IR}_\lambda$ is non-abelian and built out of $C_\lambda$ and $K^\text{IR,0}_\lambda$ via a non-trivial extension. Because of this, the electric symmetry could still act non-trivially on $\pi_0\cM(M_3,K^\text{IR}_\lambda)$ and one might still be able to turn on a magnetic flux. We will study this phenomenon in more detail later, but we remark now that being HE is in principle not mutually exclusive with the other labels. In other words, one can have phases of subtypes E-HE, M-HE, N-HE, and HO-HE. However, in the examples we will study, the HE phases encountered are often of the type N-HE, for which we will omit the N part in the notation.

\medskip

The last two rows in the table list the contribution of each phase to the Witten index $\cI$ and the dimension of the $T^3$ skein module. While it is clear that the index can be computed as a sum over IR phases, for the dimension of the skein module, this is not quite clear (and turns out to be affected by subtle phenomena such as ``overlaps'' between phases). We now examine more carefully how the deformation interacts with our proposed embedding of the skein module.

\subsection{Skein modules from deformation}\label{sec:SkeinDeform}

An $\cN=1$ theory cannot be topologically twisted on a general 4-manifold, but can be twisted if the 4-manifold is Kähler. In our setup, where the 4-manifold is taken to be $M_3\times \bR$ with a closed $M_3$, a topological twist of the deformed theory is only possible when $M_3=\Sigma\times S^1$, which is what we will mostly focus on in the remainder of the paper. 

In the Vafa--Witten theory, the partition function is expected to be invariant under the deformation, since the deformation (modulo $Q$-exact terms) has a positive ``ghost number.'' Also, all the ``intermediate vacua'' with $K_\lambda^{\text{IR}}$ having a continuous part will not contribute to the partition function due to the presence of fermionic zero modes. 

Our focus is more on the Hilbert space as opposed to the partition function, though the $M_3\times S^1$ partition function on---equivalently the Witten index of the 1d theory---serves as an important reference point. As a consequence, vacua with continuous $K_\lambda^{\text{IR}}$ \emph{do} contribute to $\cS$, and taking them into account correctly turns out to be crucial to reproducing known results on skein modules.

\subsubsection*{A common supercharge}

In order to apply this deformation to study skein modules, one needs to make sure that it is compatible with our construction. Naively, two different topological twists and supercharges are used in the process, with $\cH(M_3)$ being the cohomology of a Kapustin--Witten supercharge $\cQ_t$, while the deformation is the one used in the Vafa--Witten theory. Although these two twists are different on general $M_4$, their R-symmetry backgrounds become identical when $M_4=M_3\times \bR$. Also, the supercharge $\cQ_{t}$ actually becomes a supercharge for the Vafa--Witten theory when $t\to\infty$ (see Appendix~\ref{app:Details}).

From now on, we will work with this ``shared'' supercharge---which we simply denote by $Q$---so that the deformation interacts nicely with the embedding of skein modules. One price we pay for this choice is that, as $\Psi=\tau$ for this value of $t$ has a positive imaginary part, $q=e^{2\pi i \tau}$ now lives in the interior of the unit disk, making it difficult to access roots of unity and hence torsion phenomena in skein modules.\footnote{It would be interesting to understand whether there are similar ``controllable'' deformations for other values of $t$, so that $q$ being a root of unity is no longer a strong-coupling limit. Alternatively, one can work in the $S$-dual frame with the Nahm-pole boundary condition decorated with 't Hooft loops, which is now weakly coupled.\label{foot:strong}}

The strategy we will employ to study skein modules is to consider pairings between states $\bra{L}\in \cS$ and vacuum states in various IR phases. In addition to the standard assumptions about the IR physics of gauge theories, we will require a few further assumptions to maintain control over $\cS$ along the RG flow.

\subsubsection*{Assumptions on RG flows}

Each $\lambda$ labels a phase of the theory with an IR Hilbert space $\cH_\lambda(M_3)$. In nice cases, one expects that $\bigoplus_\lambda\cH_\lambda(M_3)\subset \cH(M_3)$ is a subspace, which is more likely to hold for the topologically twisted theory compared to the full theory. One can ask whether this leads to a homomorphism 
\begin{equation}\label{IRMap}
\cS\to \bigoplus_\lambda\cH_\lambda(M_3)^\vee.
\end{equation}  Physically, this map should be essentially about the expectation values of UV Wilson lines over different IR vacua. To make sure that it is well defined, one needs to check that this vev only depends on skein classes of the Wilson loops. This is expected to follow from the fact that two states $\bra{L}$ and $\bra{L'}$ only differ by a $Q$-exact vector if $L$ and $L'$ are in the same skein class. Then, modulo non-compactness issues (e.g.~boundary terms associated with total derivatives), their pairings with any $Q$-closed state are identical.

One can in principle directly verify this invariance in each of the IR phases. When the IR theory is gapped, either due to Higgsing or confinement, it is clear that the vev of Wilson loops only depends on the skein class. In more complicated cases (e.g.~HE phases), this is less obvious and requires a more careful analysis. In the next section, we will incorporate ``cosmic string'' defects, which may also support massless degrees of freedom on their worldsheet even when the bulk is gapped. While it would be interesting to understand whether these massless modes could lead to subtleties, we will not pursue this here. Instead we assume this map to be well defined for all $\lambda$, even when there are massless photons and/or cosmic strings in the IR.

We will in fact make a stronger assumption that the map \eqref{IRMap} is still injective---in other words, the skein module still embeds into the IR Hilbert space with no new equivalence relations. The justification for this assumption is that the pairing, $\vev{L|\psi}$, if it is non-zero to begin with, is expected to remain unchanged by the deformation based on a similar ghost-number argument. This is not yet a complete explanation with some details that remain to be checked.\footnote{For example, one should argue that a sufficiently large collection of test states $\ket{\psi}$ survive the deformation and the RG flow so that they can continue to discern different skein classes. Notice that this should be a non-trivial consequence of the ghost-number conservation and other properties of the twisted theory, as in general, one can ``lose'' states in the IR (e.g.~due to confinement). As a starting point for this analysis, it would be beneficial to better understand the behavior of the boundary condition and the Wilson lines in the twisted theory under deformation.} We hope to return to this technical point in the future, but our strategy in this paper is to justify this assumption by demonstrating that it leads to sensible results. 

Another assumption we will make to facilitate the computation is that the pairing $\vev{L|v,\lambda}$ between $\bra{L}$ and a vacuum state $\ket{v,\lambda}$ in the phase $\lambda$ is proportional to the vacuum expectation value of the Wilson loop(s) in that vacuum, $\vev{L}_{v,\lambda}$. One justification is that, in the IR theory which is gapped except for a decoupled free abelian sector, the Neumann-like boundary condition at $x=0$ is expected to behave as a ``mirror,''\footnote{This is in line with the higher-dimensional perspective viewing it as the tip of a cigar.} and the state $\bra{\emptyset}$ pairing with any vacua gives a constant (which we can normalize to 1). Also, the vacuum states are now eigenstates of any Wilson loops, with the eigenvalue being the expectation value. Combining the two points naturally gives
\begin{equation}
\vev{L|v,\lambda}=\bra{\emptyset}L\ket{v,\lambda}=\vev{L}_{v,\lambda}.
\end{equation}
Equivalently, we can express the state $\bra L\in \cH^\vee$ as
\begin{equation}
    \bra L=\sum_{{v,\lambda}}\vev{L}_{v,\lambda}\cdot(\bra{v,\lambda}+\cdots).
\end{equation}
Here ``$\cdots$'' stands for the non-vacuum states above $\bra{v,\lambda}$, which will play interesting roles for $M_3=\Sigma\times S^1$ later. They are also the reason why $\bra L$ is in general not normalizable and lives in the dual of the Hilbert space, even after the deformation.

With these assumptions, it seems that we have a concrete procedure to determine the skein module:
\begin{enumerate}
    \item Use the mass deformation to $\cN=1$ to obtain a collection of vacua $\ket{v,\lambda}$.
    \item Find the vacuum expectation values of Wilson loops for different vacua.
    \item Identify Wilson loops with the same expectation values for any choice of $\ket{v,\lambda}$.
\end{enumerate}

There turns out to be an important subtlety when the genus $g$ of $\Sigma$ is not equal to 1, as the deformation parameter is actually a section of a line bundle, which can have poles or zeroes. When it must have a pole, it is not clear how to turn on the deformation continuously, as the deformation parameter is always large in some area of $M_3$. This can potentially lead to jumps of $\cS$. When the deformation parameter is forced to have a zero, there is no problem with continuity, but the opposite problem arises---there are regions in $M_3$ where the deformation is small, and the RG argument can fail. We won't offer a solution to the problem caused by the pole, which only arises for $g=0$ (i.e.~$M_3=S^2\times S^1$), but will deal with the latter problem in the next section. In the remainder of the present section, we will focus on the $g=1$ case, where this subtlety doesn't arise and the above procedure already gives new predictions for skein modules.

\subsection{\texorpdfstring{Examples for $\sk(T^3)$}{Examples for sk(T^3)}}

We are now well-equipped to perform the computation for $M_3=T^3$. To be pedagogical, we first analyze the cases of $G=\SL(2)$, $\SL(3)$ and their PSL versions in great detail before dealing with the general case.

\subsubsection{$\SL(2)$}
When the gauge group of the UV theory is $K=\SU(2)$, there is an electric phase labeled by the non-trivial embedding (corresponding to the principal orbit) $\lambda=[2]$ described by a $\bZ_2$ gauge theory in the infrared. It has eight different vacua on $T^3$ given by the $2^3=8$ choices of holonomies. In this infrared phase, $\cH_{\lambda=[2]}(T^3)\simeq \bC^8$. Any UV Wilson line labeled by an irreducible representation of SU(2) becomes a multiple of either the trivial line $1$ or the non-trivial $\bZ_2$ Wilson line $\varphi$. Any vacuum can be distinguished from the rest via vevs of $\bZ_2$ Wilson loops, but one needs all eight configurations of Wilson loops obtained by inserting them along some of the three cycles of $T^3$. In the $\bZ_2$ phase, the fusion rule for Wilson lines is $\varphi^2=1$. It is clear that any other configurations are equivalent to one of the eight. Therefore, the map $\cS\to \cH_{\lambda=[2]}(T^3)^\vee\simeq \bC^8$ is surjective.

For the trivial $\fsl(2)$-triple (zero orbit) $\lambda=[1^2]$, the theory is in a magnetic phase. It has an unbroken $\SU(2)$ gauge group, which is confined in the IR. There are two additional quantum vacua with gaugino condensate $\langle\lambda\lambda\rangle=\pm\Lambda^{1/2}$. In the deep IR where all massive degrees of freedom is integrated out, one has $\cH_{\lambda=[1^2]}(T^3)\simeq \bC^2$ being a two-dimensional space spanned by the two vacua, $\ket{+\Lambda^{\frac12}}$ and $\ket{-\Lambda^{\frac12}}$. However, they are mutually indistinguishable by any insertion of Wilson loops along the boundary condition at $x=0$. One way to think about this fact is that the two vacua are swapped by $\theta\mapsto\theta+2\pi$, which leaves $q$ invariant.\footnote{If one works with the $\SO(3)$ theory, this $2\pi$ rotation of the $\theta$ angle will change the theory by a discrete theta angle (e.g.~from $\SO(3)_+$ to $\SO(3)_-$). However, this won't change the spectrum of Wilson lines and their skein relations in our setup.} As a consequence, $\cS\to \cH_{\lambda=[1^2]}(T^3)^\vee\simeq \bC^2$ is the 1-dimensional diagonal subspace, generated by the image of $\bra{\emptyset}$ which pairs identically with the two vacua, giving the functional  $a\ket{+\Lambda^{\frac12}}+b\ket{-\Lambda^{\frac12}}\mapsto a+b$.

Before concluding that the skein module is 8+1=9 dimensional, it remains to show that any one of the confining vacua is distinguishable from the other 8 states in the $\bZ_2$ phase. This can be argued by noticing that, as these two vacua are confining, insertion of any non-trivial Wilson line should give zero correlation function. To detect it, it is not enough to just use the eight configurations of Wilson lines, as a linear combination of the eight states in the $\bZ_2$ vacuum of the form 
\begin{equation}
    \ket{\rightarrow}=\frac{1}{2\sqrt{2}}(\ket{+}+\ket{-})^{\otimes3},
\end{equation}
which is a superposition of all the 8 states, cannot be distinguished from one of the confining vacua using only these configurations. Indeed, the vev for any of the 7 non-trivial configurations will be zero on both this state and a confining vacuum. To distinguish them, one needs to introduce another configuration, which can be taken to be an adjoint Wilson loop along an arbitrary cycle of $T^3$. This would be the same as the trivial configuration in the $\bZ_2$ vacua, evaluating to $3$---the dimension of the adjoint representation---on $\ket{\rightarrow}$, but gives zero in the confining phase. Alternatively, one can use two parallel Wilson lines in the fundamental representation of $\SU(2)$ winding along the same cycle. In the tensor product, it contains a trivial representation and an adjoint, and hence distinguishes the two by giving 4 in the $\bZ_2$ phase and 1 in the confining phase.

\begin{figure}[htb!]
    \centering
    \includegraphics[width=0.75\linewidth]{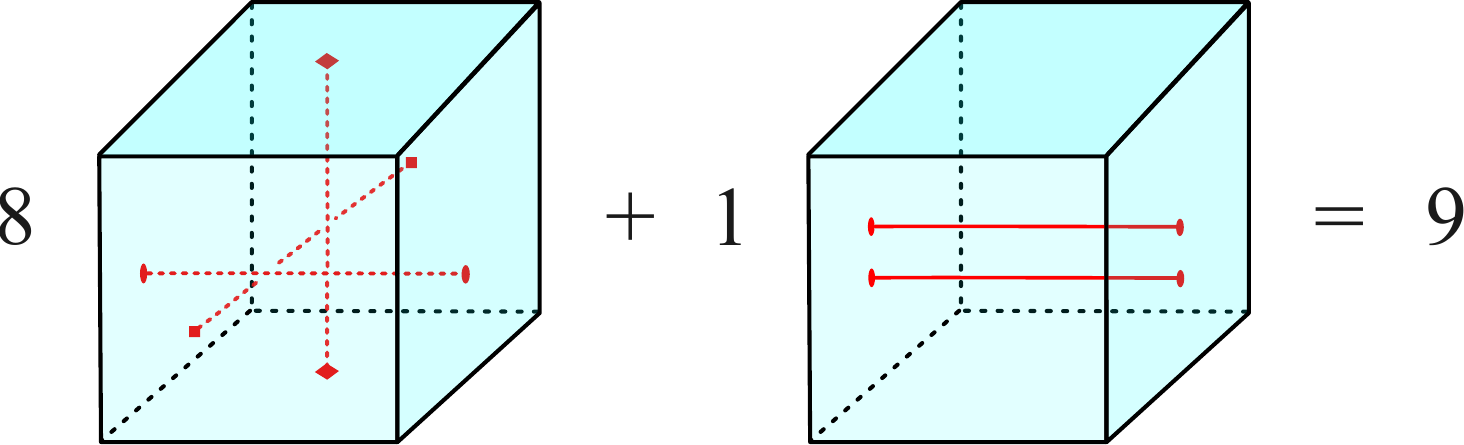}
    \caption{The nine generators of sk$(T^3)$. The configuration on the left represent $2^3=8$ different configurations with possible insertions of Wilson loops along the three directions, ranging from the empty configuration to the one with all three insertions. They can be associated with $\bZ_2$ phase. The need for the additional configuration on the right can be explain by the existence of the confinement phase.}
    \label{fig:T3}
\end{figure}

Therefore, $\cS(T^3)\subset \cH(T^3)^\vee$ is a 9-dimensional space. Phrasing it in the dual picture, any possible incoming states in $\cH(T^3)$ will become a linear combination of the 10 states in $\cH_{\lambda=[2]}\oplus\cH_{\lambda=[1^2]}$ after mass deformation and RG flow, but they only define 9 independent functionals on $\cS(T^3)$ as the two confining vacua cannot be distinguished by Wilson lines. In other words, we have
\begin{equation}
    \cS(T^3)^\vee\simeq \cH_{\lambda=[2]}\oplus\cH_{\lambda=[1^2]}/\left(\ket{+\Lambda^{\frac12}}-\ket{-\Lambda^{\frac12}}\right)=\bC^8\oplus \bC\simeq \bC^9.
\end{equation}
This explains the result of Carrega and Gilmer \cite{carrega2017nine,gilmer2018kauffman} and resolves the ``9-vs-10'' paradox. The 9 generators are illustrated in Figure~\ref{fig:T3}.

In a sense, the 9-th generator exists because of the confinement phase, as it corresponds to the following state in $\cH(T^3)^\vee$ in the IR,
\begin{equation}
    \bra{L_9}=4\sum_{x,y,z=\pm}\bra{xyz} + \bra{+\Lambda^{\frac12}}+\bra{-\Lambda^{\frac12}}=4\bra{\emptyset}-3\bra{+\Lambda^{\frac12}}-3\bra{-\Lambda^{\frac12}},
\end{equation}
where $\{x,y,z\}$ are the $\bZ_2$ holonomies along the three cycles. Without the confinement phase, $\bra{L_9}$ would simply be a multiple of the trivial configuration $\bra {\emptyset}$.

To refine $\cS$ with electric fluxes, one simply puts each generators into the correct grading according to the electric charge of the Wilson lines. Then the 9 decomposes into 2+7 with two in the trivial flux sector (no line plus 2 parallel lines) and one for each of the seven non-trivial fluxes in $H_1(T^3,\bZ_2)\simeq H^2(T^3,\bZ_2)$.

\subsubsection{$\SL(3)$}

As $\SL(2)$ is very special, before moving on to the most general case, we will analyze another case with $K=\SU(3)$, which has an ``intermediate vaccum'' after deformation. In general, such vacua won't contribute to the partition function of the Vafa--Witten theory or the Witten index due to the presence of fermionic zero modes, but they can contribute to the skein module. Therefore, it is necessary for us to better understand and correctly incorporate them.

\paragraph{The $\bZ_3$/electric phase.} The other two types of vacua for the $\SU(3)$ theory are pretty similar to the $\SU(2)$ case. The $\bZ_3$ vacuum associated with the irreducible embedding (principal nilpotent orbit), $\lambda=[3]$, will lead to $3^{3}=27$ states on $T^3$ that can be distinguished by 27 configurations of $\bZ_3$ Wilson lines. Notice that in each direction, one needs both of the non-trivial $\bZ_3$ Wilson lines, otherwise $\ket{\omega^2}$, where $\omega$ is the generator of $\bZ_3$, cannot be distinguished from a linear combination of $\ket{1}$ and $\ket{\omega}$.

\paragraph{The confinement/magnetic phase.} For the trivial $\fsl(2)$-triple (zero orbit), $\lambda=[1^3]$, one has three confining vacua, which can't be distinguished from each other via Wilson lines. To distinguish them from the 27 states in the $\bZ_3$ phase, one can use three parallel Wilson lines in the fundamental representation (equivalently, one can simply use the adjoint or a pair of fundamental and anti-fundamental). As this tensor product is decomposed as $\bf3\times\bf3\times\bf3=1+8+8+10$, the vev of this configuration is 27 in the $\bZ_3$ phase but 1 in the confinement phase. Therefore, the three different quantum vacua collectively contribute a one-dimensional space to $\sk(T^3,\SL(3))$, but 3 to the Witten index.

\paragraph{The Maxwell/neutral phase.} The intermediate vacuum corresponds to the embedding of $\fsl(2)$  as a ``$2\times 2$ block,''
\[
\begin{pmatrix}
* \;&\; *\; &\; 0 \\
*\; & \;*\; & \;0 \\
0\; & \;0\; & \;0
\end{pmatrix}\subset  \fsl(3).
\]
 This breaks the gauge group to a U$(1)$ subgroup parametrized as diag$\{e^{i\theta},e^{i\theta},e^{-2i\theta}\}$. In this phase, there is a massless vector multiplet, leading to both bosonic and fermionic zero modes. The latter makes the contribution of such vacua to the partition function vanish, while the former are parametrized by holonomies U$(1)^3$ along the $T^3$. A ground state wave function is the constant function on U$(1)^3$, up to the action of fermionic zero modes. When a Wilson line is inserted along a non-trivial cycle of $T^3$, the correlation function is expected to be the multiplicity of the trivial representation of U$(1)$. This allows one to distinguish this vacuum from the rest. For example, the fundamental of $\SU(3)$ does not have an invariant subspace of this U$(1)$ subgroup, but the adjoint does. Notice that all the first 27 configurations cannot distinguish this vacuum $\ket{{\rm U}(1)}$ with the superposition $\ket{\psi}_{\bZ_3}=(\ket{1}+\ket{\omega}+\ket{\omega^2})^{\otimes3}$ or a confining vacuum $\ket{c}$. The 28th generator, taken to be a parallel pair of fundamental and anti-fundamental (which can be viewed as a fundamental but with the opposite orientation), equivalent to a sum of a trivial and adjoint line, evaluates to $1$ on $\ket{c}$, $9$ on $\ket{\psi}_{\bZ_3}$, and $5$ on $\ket{{\rm U}(1)}$, still cannot distinguish $\ket{{\rm U}(1)}$ from a linear combination of $\ket{c}$ and $\ket{\psi}_{\bZ_3}$. Therefore, an additional generator is really needed (which is also expected simply on dimensional grounds). The $29$th generator can be taken to be three parallel Wilson lines in the fundamental representation. This gives 1 on $\ket{c}$, 12 on $\ket{{\rm U}(1)}$ and 27 on $\ket{\psi}_{\bZ_3}$, which enables to distinguish between all three of them when combined with the 28th generator. In the above, we used the values of the adjoint \textbf{8} and the decuplet \textbf{10} on the three vacua, which is summarized below:
\begin{center}
    \begin{tabular}{c|c|c|c}
         & $\ket{c}$& $\ket{{\rm U}(1)}$ &  $\ket{\psi}_{\bZ_3}$\\\hline
       \textbf{8}  & 0 & 4 & 8\\\hline
      \textbf{10}   & 0 & 3 & 10\\
    \end{tabular}
\end{center}
They can themselves serve as the 28th and 29th generators, but we chose to use only the fundamental in the description above which contains them in the tensor product.

In the end, we have the dimension for the skein module being 29 while the Witten index is 30. The difference is only one, but this is somewhat misleading---among the three difference phases, only one gives the same contribution to both $\dim\sk(T^3)$ and the index.

\subsubsection{$\mathrm{PSL}(2)$ and $\mathrm{PSL}(3)$}
When $G$ is non-simply connected, there can be non-trivial magnetic sectors, and our proposal is that the ``total'' skein module is the direct sum of $\cS^m$ over all sectors, which we now apply to study the non-simply-connected versions of the previous two examples, starting with $K=\SO(3)$.

As $\pi_1(G)=\bZ_2$, there are 8 magnetic sectors labeled by $H^2(T^3,\bZ_2)=\bZ_2^3$. They form two SL$(3,\bZ)$-orbits with one consisting of $m=0$ only while the other containing all the non-trivial $m$'s. Therefore, we only need to find the contributions of each phase to $\cS^m$ with $m=0$ and another non-trivial $m$.

\paragraph{Electric phase (principal orbit).} In this phase, the $\SO(3)$ gauge group will be completely broken by the vevs of the scalars with $C_\lambda=0$. The trivial connection can still contribute to $m=0$, but no configuration can contribute to a sector with non-trivial $m$. One can also interpret this statement in the same phase of the $\SU(2)$ theory. There, one has a $\bZ_2$ gauge theory, but there is not a $\bZ_2$-connection that can accommodate a magnetic flux. This always happens to theories with $K$ of adjoint type in an electric phase. (Recall that the phase classification is uniform across groups with the same Lie algebra and previously stated using the global form with simply-connected $H$.)

\paragraph{Magnetic phase (zero orbit).} In this phase, the $\SO(3)$ gauge group is confined, and, similar to the $\SU(2)$ case, the two quantum vacua are not distinguishable by Wilson lines. Therefore, it contributes a one-dimensional space to $\cS^m$ for any $m$. Another way of understanding this is via the low-energy theory which consists of two quantum vacua corresponding to the condensation of either a dyon or a charge-2 monopole. In the $m=0$ sector, the two are indistinguishable by Wilson lines which are confined, and are exchanged under a shift of the theta angle by $2\pi$. The monopole condensation leaves unbroken a $\bZ_2$ magnetic 1-form symmetry, which can be viewed as the center symmetry of a magnetic $\bZ_2$ gauge theory with background being exactly $m$. It then contributes a one-dimensional space to $\cS^m$ for all values of $m$.

\medskip

Therefore, we again have 
\begin{equation}
    \dim \sk (T^3,\SO(3))=1+8=9.
\end{equation}
However, notice that the Langlands duality between $\SU(2)$ and $\SO(3)$ only holds after summing over the orbits:
\begin{center}
    \centering
    \begin{tabular}{c|c|c|c}
       $\lambda$ & type & SU(2) & SO(3)   \\\hline
        [2] & E& 1+{\color{blue}7}  &  1  \\\hline
        [$1^2$] &M&  1  &  1+{\color{red}7}
    \end{tabular}
\end{center}
Here the contributions of the non-trivial electric- and magnetic-flux sectors are colored blue and red respectively. Alternatively, one can map an orbit to the one labeled by the dual partition and switch the electric flux with a magnetic flux. This turns out to be how the Langlands duality for $T^3$ skein module in the $A$-series works in general.

We similarly list the results for $K=\mathrm{PSU}(3)$ by comparing it to the SU(3) case:
\begin{center}
    \centering
    \begin{tabular}{c|c|c|c}
       $\lambda$ & type &SU(3) & PSU(3)   \\\hline
        [3] & E& 1+{\color{blue}26}  &  1  \\\hline
        [21] & N& 1  &  1 \\\hline
        [$1^3$] & M& 1 & 1+{\color{red}26}
    \end{tabular}
\end{center}
The only additional analysis needed is about the abelian phase labeled by $\lambda=[21]$ in the presence of a non-trivial magnetic flux. There, the U(1) gauge group is required to have a non-trivial flux with the first Chern class congruent to $m \pmod 3$. This makes the ground-state energy in such a sector non-zero. It is therefore not visible at low energy and cannot contribute to the skein module. 

Also, in such tables with contributions of phases to different global forms of $K$ listed, it is fairly easy to tell which phase is electric (contributing to the non-trivial electric-flux sectors for the simply-connected group but not to these with non-trivial magnetic flux for the adjoint-type group), magnetic (the exact opposite), neutral (only contributing to the trivial sector regardless of the global form), or hybrid (contributing to both non-trivial electric and magnetic sectors).

\subsection{General $G$}

For a general reductive $G$, the procedure is similar. When $\pi_1(G)$ is non-trivial, one needs to again remember the magnetic flux sectors (i.e.~different topological type of bundles). If $\pi_1(G)$ has a free part, it allows non-trivial first Chern class for the gauge bundle to be turned on. However, to have a flat connection (and hence visible at low energy and contribute to the skein module), $c_1$ has to be torsion. Since we will be dealing with $M_3$ of the form $\Sigma\times S^1$---whose cohomology groups over $\bZ$ are free---in this section and the next, this forces the Chern class to be trivial. As a consequence, only the torsion part of $\pi_1(G)$ is relevant, and the magnetic fluxes will be classified by $m\in H^2(M_3,\Tor\,\pi_1(G))$.

Turning on the mass deformation, one again has a collection of IR phases, classified by the conjugacy class $\lambda$ of the $\mathfrak{sl}_2$-triple in $\mathfrak{g}$, and, for each $\lambda$, one needs to study the vacuum structure of the 4d $\cN=1$ theory with gauge group $K_\lambda$ whose complexification $G_{\lambda}$ is the centralizer of the image of $\lambda$ under the adjoint action of $G$. In general, $K_\lambda$ has a connected piece $K_\lambda^0$ that contains the identity and different components indexed by the group of components $C_\lambda=G_\lambda/G_\lambda^0$. Alternatively, one can divide $K_\lambda$ into the following two pieces, the ``confined part'' $H_\lambda:=[K_\lambda^0,K_\lambda^0]$, and the ``free part'' $K_\lambda^\text{IR}=K_\lambda/H_\lambda$. The two ways of dissecting $G_\lambda$ are quite similar, with the difference being that the abelianization of $K^0_\lambda$ is placed in different parts. We organize the discussion by first looking at the different parts of $K_\lambda$ separately and then putting them together across the different phases.

\subsubsection*{The discrete/free part}

For the zero-flux sector $m=0$, one can first count flat $C_\lambda$-connections over $M_3$ to find $|\cM(M_3,C_\lambda)|$. In addition, we need to incorporate all non-trivial magnetic backgrounds and count the cardinality of $\cM^m(M_3,C_\lambda)$.

However, the true IR gauge group is $K^\text{IR}_\lambda$, and there can be more deformation classes of flat $K^\text{IR}_\lambda$-connections than $C _\lambda$-connections when the extension \begin{equation}
    K^\text{IR,0}_{\lambda}\to K^\text{IR}_\lambda\to C _\lambda
\end{equation}
is non-trivial and causes $K^\text{IR}_\lambda$ to be non-abelian. If this is not an issue (e.g.~when $G$ is of type $AGF$), one can simply count the $|\cM^m(T^3,C_\lambda)|$ vacua on $T^3$ for the $C_\lambda$ gauge theory. On the other hand, the HE phases are exactly these where $\cM^m(M_3,K^\text{IR}_\lambda)$ has more components compared to $\cM^m(M_3,C_\lambda)$ making it necessary to remember the continuous part in $K^\text{IR}_\lambda$. This can lead to extra complications when $G$ is of $BC$- or $D$-type (and potentially also for $E_7$ and $E_8$), where O$(2)$ subgroups of $K^\text{IR}_\lambda$ can have more components in their moduli space of flat connections compared with the component group $\pi_0\mathrm{O}(2)\simeq \bZ_2$ (e.g.~29 vs 8 on $T^3$). For $G=E_6$, there is an even more complicated scenario with $K_\lambda=\U(1)^2\rtimes S_3$.

Turning on a magnetic flux $m$ often kills the entire moduli space $\cM^m(M_3,K^\text{IR}_\lambda)$, as it is difficult to incorporate a magnetic flux in a finite-group/abelian gauge theory. For many $K$ and $\lambda$, this moduli space can only be non-empty when $C_\lambda$ ``doesn't know'' about $m$, meaning that $m\in \pi_1(G)$ becomes trivial under the map to $\pi_1(\cO_\lambda)$. In other words, $m$ being non-trivial in $\pi_1(\cO_\lambda)$ is very often sufficient to render $\cM^m$ empty, but this could fail when $\pi_1(\cO_\lambda)$ is a ``non-abelian extension'' of $C_\lambda$ (i.e.~the 2-cocycle classifying the central extension is not symmetric, which can happen in $B$- and $D$-series) or if $\lambda$ is an HE phase (e.g.~$K^\text{IR}_\lambda$ contains O$(2)$).

This can be more clearly understood from the perspective of the relative theory,  which, effectively, can be thought of as a theory with a gauge group $\widetilde K$ which is a $\Tor\,\pi_1(G)$-covering of $K$ and has free $\pi_1$, but allowing insertions of 't Hooft fluxes corresponding to different magnetic backgrounds.\footnote{Notice that, although the electric and magnetic fluxes are treated on equal footing in this way, there is an asymmetry in our terminology (the root cause being the preference of Wilson lines over 't Hooft lines in our construction). For example, when talking about a given magnetic sector $m$, we are taking the sum over all (compatible) electric sectors, $\cH^m=\bigoplus_e\cH^{e,m}$, while a given electric sector $e$ is often viewed as a further refinement within a given magnetic sector. We will however treat them more symmetrically in Section~\ref{sec:duality}.} The component group is now $\widetilde C_\lambda =\widetilde G_\lambda/\widetilde G_\lambda^0\simeq \pi_1(\cO_\lambda)$, and the moduli space of flat $\pi_1(\cO_\lambda)$-connection with the 't Hooft flux $m$ is $\cM^m(M_3,\pi_1(\cO_\lambda))$. Now, the effect of turning on the flux $m$ is described by its image in $H^2(M_3,\pi_1(\cO_\lambda))$ under $\pi_1(G)\to Z(\widetilde G) \to \widetilde G_\lambda \to \pi_1(\cO_\lambda)$, which is central in $\pi_1(\cO_\lambda)$. Assuming that the flux on $M_3=T^3$ is through the (12)-plane, we have $[f_1,f_2]:=f_1f_2f_1^{-1}f_2^{-1}=m$ with $f_{1,2}$ being the holonomies along the 1 and 2 directions. This can only have a solution in $\pi_1(\cO_\lambda)$ if it is non-abelian and, furthermore, the 2-cocycle in $H^2(C_\lambda,\pi_1(G))$ has an antisymmetric part, which determines the commutator. Alternatively, there can be a solution in $K^\text{IR}_\lambda$ if it is non-abelian. When both are abelian and the flux is non-trivial, there cannot be any flat $\pi_1(\cO_\lambda)$- or $K^\text{IR}_\lambda$-connections as such a central flux cannot be screened and has nowhere to go on a closed manifold (i.e.~the meridian of the flux tube in $T^3$ bounds a one-holed torus in the complement, requiring the flux to be trivial).

When the image of $m$ is indeed trivial, $\cM^m(M_3,C_\lambda)$ is then independent of $m$, and one can just count flat $C_\lambda$-connections.\footnote{One can also work with $\widetilde C_\lambda$-connections, which can contribute to multiple electric flux sectors, and the extra flat connections (compared with $C_\lambda$) are in the sectors not needed to reproduce the Hilbert space of the $K$ gauge theory. For example, if $K$ is of adjoint type, then its Hilbert space is the sum of $(0,m)$ sectors of the relative theory with no electric flux but all magnetic fluxes. If one is only interested in such sectors, working with $\widetilde C_\lambda$- and $C_\lambda$-connections would give the same answer.} This deals with ``half of'' $G_\lambda$---except in the more subtle cases mentioned above which we will deal with later in examples---and we are left with the task of analyzing the $H_\lambda$ part. 

\subsubsection*{The continuous/confined part}

For each such $\lambda$, one expects confinement to make the non-abelian part $H_{\lambda}$ of $K_\lambda^0$ invisible in the IR where there are discrete quantum vacua characterized by gaugino condensates. The number of vacua equals the dual Coxeter number  $h^\vee(G_{\lambda})$---interpreted as the product of dual Coxeter numbers for the simple summands of Lie($G_{\lambda}$)---but different quantum vacua are related by shifts of theta angles and are mutually indistinguishable by Wilson lines. Therefore, one only needs to take into account the contribution of a single vacuum. The low-energy theory at each vacuum is an abelian gauge theory, the continuous part of which contributes one dimension to the skein module. 

If one turns on a magnetic flux $m$, it does not affect the above analysis as long as the flux will be carried entirely by the non-abelian part $H_{\lambda}$ and hence invisible in the IR theory. Therefore, we have the opposite situation compared to the $C_\lambda$ part, as now every compatible $m$ in the image of 
\begin{equation}
    H^2(M_3,\pi_1(H_\lambda))\to H^2(M_3,\pi_1(G))
\end{equation}
can receive a contribution from this phase, but electric flux are confined with only these becoming trivial for $H_{\lambda}$ under the map 
\begin{equation}
    e\in H_1(M_3,Z(G)^\vee)\to H_1(M_3,(Z(H_\lambda)\cap Z(G))^\vee )
\end{equation}
can receive a contribution. Here we have used the fact that the intersection $Z(H_\lambda)\cap Z(G)\subset G_\lambda^0$ is a subgroup of $Z(G)$, giving the above map for the homology groups with coefficients in the duals of them.

\subsubsection*{Summing over the phases}

There is a remaining subtlety before one can straightforwardly combine contributions from all phases: one still needs to argue that all of them are distinguishable by Wilson lines labeled by representations of $K$. It is not too hard to convince oneself that, after fixing the phase $\lambda$, all connections in $\cM^m(T^3,C_\lambda)$---or all components in $\cM^m(T^3,K_\lambda^\text{IR})$ if there are more in the latter---can be distinguished by Wilson lines, based on the fact that they give a complete basis for class functions for $K$, and hence are expected to distinguish connected components of (conjugacy classes of) holonomy representations $\pi_1(M_3)\to K_\lambda^\text{IR}$.\footnote{In general, class functions for $K$ might not be able to separate conjugacy classes of a subgroup. However, in our setup, this is not what we are aiming for when there is a continuous part of $K_\lambda^\text{IR}$. Instead, we want to ``collapse'' higher-dimensional components into a single vaccum and ask about the expectation value of a Wilson line. We will assume that, after this, finite-dimensional representations of $K$ always distinguish components, but it would be interesting to test and understand this point better.} However, it is not immediately clear that each flat connection can always be distinguished from other ones in \textit{different} phases. We will see that this, in general, does fail, and summing up the number of flat connections in all phases could result in an overcount. 

Because of this subtlety and the one that involves the difference between flat $K^\text{IR}_\lambda$- and $C_\lambda$-connections, we write the dimension of the skein module of $T^3$ as
\begin{equation}
    \dim \cS^m (T^3;G)=\sum_{\lambda}|\cM^m(T^3,C_\lambda)|'
\end{equation}
where the prime is a reminder that the cardinality of $\cM^m(T^3,C_\lambda)$ sometimes needs to be modified to take into account of the two issues. 

The cardinality of $\cM(T^3,C_\lambda)$ for the various component group $C_\lambda$ that can arise for simple $G$ is gathered below. (See Appendix~\ref{app:DW} for computation.)
\begin{center}
    \centering
    \begin{tabular}{c|c|c|c|c|c}
        $C_\lambda$ & $S_3$ & $S_4$ & $S_5$ & D$_8$ & Q$_8$ \\\hline
        $|\cM(T^3,C_\lambda)|$ & 21 & 84 & 206 & 92 & 92\\
    \end{tabular}
\end{center}

A basis for the dual of $\cS^m (T^3;G)$ is generated by $\ket{\alpha;\lambda,m}$ with $\alpha\in \pi_0\cM^m(T^3,K^{\text{IR}}_\lambda)$ labels the  components of the moduli space of flat $K^{\text{IR}}_\lambda$-connections on $T^3$, except that one needs to identify linear combinations (within a given flux sector) that are not distinguishable by Wilson lines. With this answer in hand, to find a basis for $\cS^m$, one just needs to find a collection of Wilson loops with the same cardinality as $\dim \cS^m$, with the condition that they give linearly independent functionals on the states $\ket{\alpha;\lambda,m}$ with fixed $m$.

\subsection{More examples}

We now analyze additional cases in varying depth, covering all types of simple Lie algebras. We will start with the $A$-series where none of the two subtleties arise.

\subsubsection{$\GL(N)$, $\SL(N)$ and $\mathrm{PSL}(N)$}

In the $A$-series, the $\fsl_2$-triples or nilpotent orbits $\lambda$ are labeled by partitions of $N$, $\lambda=[1^{r_1}2^{r_2}\cdots]$. Therefore, they all have the same phase classification, but the unbroken gauge group in each phase can be different for $\GL$, $\SL$ and $\mathrm{PSL}$.

\paragraph{The case of $G=\GL(N)$.}  We have 
\begin{equation}
    K_\lambda=\prod_i\U(r_i)
\end{equation}
is connected with $C_\lambda$ being trivial for each $\lambda$. With confinement in the infrared, $K^\text{IR}_\lambda$ is a product of U$(1)$'s, one for each distinct number in the partition. The center of $H=\SU(N)$ is always contained in $K^\text{IR}_\lambda$, and hence all phases are neutral according to our four-fold classification. Therefore, one has
\begin{equation}\label{SKGL}
    \dim \sk (T^3;\GL(N))=\#|\lambda\vdash N|,
\end{equation}
given by the number of partitions of $N$. On the other hand, the Witten index is zero, since $K_\lambda$ is never semisimple. Also, as $\pi_1(G)$ is free, there are no non-trivial flux sectors that can contribute to the skein module.

\paragraph{The case of $G=\SL(N)$.} Here, $\lambda$ is again labeled by partitions of $N$, but the unbroken gauge group in each phase is
\begin{equation}
        K_\lambda=\mathrm{S}\left(\prod_i\U(r_i)\right).
\end{equation}
Here, ``S'' means taking the kernel of the determinant map $\prod_i\U(r_i)\to \U(1)$, which is the natural one compatible with its embedding in $\GL(N)$. In particular, for each $\U(r_i)$ factor, the map is $i$ times the usual one as $\U(r_i)$ acts on $r_i$ blocks, each of size $i\times i$, and the pre-image of an element of U(1) has $i$ components. Different components can be connected after combining all $i$'s, but when a number $d>1$ divides all of the $i$'s that appear in the partition, $K_\lambda$ can have non-trivial components. In fact, it is easy to verify that $C_\lambda=\bZ_{\gcd(\lambda)}$. It is a direct factor of $K^\text{IR}_\lambda=\mathrm{S}(\prod_i\U(1))$ with its generator being represented by the constant matrix $e^{2\pi i/\gcd(\lambda)} \cdot \mathbf{1}\subset K^\text{IR}_\lambda$.\footnote{Note that this factorization is a special property for the $A$-series. Also, the analogous statement with $K^\text{IR}_\lambda$ replaced with $K_\lambda$ is not ture. The simplest counter-example is $N=4$ and $\lambda=[22]$ for which $K_\lambda=\SU^\pm(2)$ is the group of unitary two-by-two matrices with $\pm1$ determinant. The component group $\bZ_2$ is not a direct factor but a quotient of the $\bZ_4$ center generated by $i\cdot\mathbf{1}_{2\times 2}$.\label{foot:SUpm}} Therefore, for each $\lambda$, one can simply count flat $\bZ_{\gcd(\lambda)}$-connections, leading to
\begin{equation}\label{SKSL}
    \dim \sk (T^3;\SL(N))=\sum_{\lambda\vdash N}\gcd(\lambda)^3.
\end{equation}
This formula and \eqref{SKGL} reproduce the results for SL and GL in \cite{gunningham2024skeins}. 

In obtaining \eqref{SKSL}, one also needs to argue that different phases genuinely contribute with no overlaps. In other words, one can distinguish different phases by Wilson lines. Conceptually, this is because the symmetry-breaking/confinement patterns of $K$ are different for different $\lambda$. And this can be detected by choosing Wilson lines in different representations of $K$ whose expectation values can only be non-trivial when they contain the trivial representation of $K_\lambda$. Using this method, one can probe the sizes of the blocks and hence recover the partition.

In contrast, the Witten index only receives contributions from partitions in which only a single number appears, $\lambda=[d^{r_d}]$ with $r_d=N/d$. In this phase, one has an $\cN=1$ gauge theory with the gauge group $K_\lambda$ being the group of $r_d\times r_d$ unitary matrices with determinant a $d$-th root of unity. In other words, it is given by the extension 
\begin{equation}
    K_\lambda^0=\SU(r_d) \longrightarrow K_\lambda\longrightarrow C_\lambda=\bZ_{d},
\end{equation}
which becomes the central extension $\bZ_{r_d}\to\bZ_N\to \bZ_d$ when restricting to the center. After confinement, the $r_d$ quantum vacua become decoupled from the $C_\lambda$ gauge field, and this phase contribute $r_d\cdot d^3$ to the index.\footnote{There is a potential sign that one can in principle fix by more carefully analyzing the fermionic modes. In our setup, it seems natural that all the phases contribute with the same sign, which also matches with other regularization schemes of the Witten index of the $\cN=4$ theory such as the one in \cite{Gukov:2022cxv}. Notice that this is different from the result in \cite{Witten:2000nv}, where a sign $(-1)^\text{rank}$ appears. In other words, we are assuming that the fermionic parity of the vacuum for the $K_\lambda$ theory on $T^3$ is in fact shifted, canceling this additional factor. } Therefore, the Witten index of the full theory is
\begin{equation}
    \cI(T^3,\SL(N))=\sum_{d|N} r_d\cdot d^3=N\cdot\sum_{d|N}  d^2.
\end{equation}

One can also refine \eqref{SKSL} by electric fluxes labeled by $H_1(T^3,\bZ_N^\vee)\simeq \bZ_N^3$. A phase $\lambda$ will contribute once to every electric sectors in the image of 
\begin{equation}
    H_1(T^3,\bZ_{\gcd(\lambda)}^\vee)\to H_1(T^3,\bZ_N^\vee).
\end{equation}
Although the Pontryagin dual of a finite abelian group is isomorphic to itself, we keep the ``$\vee$'' to indicate that the coefficients in the above are actually the duals of $C_\lambda$ and the center of $K$.

\paragraph{The $\mathrm{PSL}$ case and duality.} Now the remaining gauge group in the phase $\lambda$ is 
\begin{equation}
        K_\lambda=\mathrm{S}\left(\prod_i\U(r_i)\right)/\bZ_N
\end{equation}
with $\bZ_N$ being the center of SU$(N)$. This is always connected, and, therefore, only contributes a one-dimensional space to the $m=0$ sector with trivial topology. On the other hand, the phase $\lambda$ can contribute only to $m$ in the image of
\begin{equation}
    H^2(T^3,\pi_1(H_\lambda)) \to H^2(T^3,\pi_1(K))\simeq \bZ_N^3.
\end{equation}
The connected commutator subgroup $H_\lambda$ is a quotient of $\prod_i\SU(r_i)$ by a subgroup of the center $\bZ_N$, and it is easy to see that this subgroup is generated by $e^{2\pi i/\gcd(r_i)}$. However, $\gcd(r_i) $---the greatest common divisor all of $r_i$'s---equals to the gcd of the dual partition, leading to $\pi_1(H_\lambda)\simeq \bZ_{\gcd(\lambda^T)}$ where $\lambda^T$ is obtained by the transpose of the Young tableau for $\lambda$. Summing over all $m$, we have that the total contribution of the phase is $\gcd(\lambda^T)^3$. 

Therefore, we see a ``phase-wise'' electric-magnetic duality---two phases related by transposition of the Young tableaux give the same contribution to the SL and PSL skein modules but with electric and magnetic flux exchanged. The four-fold classification of phases is determined by $\gcd(\lambda)$ and $\gcd(\lambda^T)$ as
\begin{center}
    \centering
    \begin{tabular}{c|c|c}
         & $\gcd(\lambda)=1$ & $\gcd(\lambda)>1$\\\hline
        $\gcd(\lambda^T)=1$ & neutral & {\color{blue}electric}\\\hline
        $\gcd(\lambda^T)>1$ & {\color{red}magnetic}  & {\color{purple}hybrid}
    \end{tabular}
\end{center}
And the duality matches different types of phases by transposing this table. Also notice that for the $A$-series, all hybrid phases are ordinary, as $C_\lambda$ is a quotient of the center of $K$, and cannot act non-trivially on $K^\text{IR,0}_\lambda$. 

On the other hand, the matching between Witten indices no longer works in a ``phase-wise'' fashion, but requires some reorganization. We have again only equal partitions $[d^{r_d}]$ contributing to the index of the PSU theory. For this phase, there are $r_d$ quantum vacua, labeled by discrete theta angle $n\in\{1,2,\ldots,r_d\}$, and the $n$-th vaccum can contribute to the magnetic sectors if $\gcd(n,r_d)>1$, when there is a remaining magnetic gauge symmetry $\bZ_{\gcd(n,r_d)}\subset \bZ_N$ not screened by the gaugino condensate. Across all phases, the number of times for the $\bZ_k$ subgroup to appear is $N/k$.\footnote{One way to obtain this count is to notice that the $j$-th time for the $\bZ_k$ subgroup to appear is for the $\frac{jk}{d}$-th vacuum in the phase labeled by $d=\gcd(jk,N)$.} Therefore, we have
\begin{equation}
    \cI(T^3,\mathrm{PSL}(N))=\sum_{d|N}\sum_{n=1}^{N/d}\gcd(n,N/d)^3=\sum_{k|N}\frac{N}{k}\cdot k^3=\cI(T^3,\mathrm{SL}(N)).
\end{equation}
Notice that for the SU$(N)$ computation, the sum is over phases, as each vacuum in the same phase has the same electric gauge symmetry (hence also the same electric 1-form symmetry). On the PSU$(N)$ side, we are reorganizing the vacua not according to the phases where they come from, but their remaining magnetic symmetry. This is not just about taking the transpose of the Young diagrams and quite different from the duality of the skein module.

In general, the computations for $\cI$ and $\cS$ and their behavior are quite different. While the $S$-duality of $\cN=4$ theories combined with the invariance of the Vafa--Witten partition function under the $\cN=1$ deformation always gives
\begin{equation}
    \cI(T^3,G)=\cI(T^3,^\mathrm{L}\!G),
\end{equation}
it is not as simple for the dimension of $\cS$ computed from the deformation, due to the various subtleties involved. Our focus will be on the latter in the rest of the paper, and we only include the index calculation when it serves to illustrate certain phenomena for $\cS$. To explain the duality for $\cS$ from the $S$-duality of the $\cN=4$ theory, one should perform a computation for $\tilde \cS$ and argue that the subspaces added by incorporating 't Hooft and dyonic operators are isomorphic on the two side.

One can test the duality for $\tilde \cS$ with the simplifying assumption that, once all dyonic operators are added, one can distinguish between the quantum vacua. The computation is largely a combination of the skein one and the index one. When $N=p$ is prime, all the intermediate phases still contribute equally to both the SU  and PSU sides. For the partition $[1^p]$, one still add $p-1$ states on both side in the $e=m=0$ sector, as the magnetic $\bZ_p$ symmetry on the PSU$(p$) side is only supported on one of the quantum vacuum. Therefore, adding or removing the 't Hooft operators would add or remove an isomorphic subspace on the two sides. This observation, combined with the $S$-duality of $\tilde \cS$, leads to the Langlands duality for $\cS$ for $A_{p-1}$. When $N$ is not a prime, the analysis will be slightly more complicated, and the verification of duality would requires reorganizing the quantum vacua similar to the Witten index computation. We will not perform this calculation here in full generality, but only in the $A_3$ example---disguised as $D_3$---in a later subsection.

We will next explore simple groups of $B$- and $C$-types, where we will encounter several complications not present for the $A$-series, such as the presence of the exceptional hybrid phases. In particular, they make it difficult to match the skein modules of the Langlands dual groups.

\subsubsection{$B$- and $C$-series}

As $B$- and $C$-type groups are Langlands/GNO dual, we will study them together. For both, the two subtleties---the appearance of O$(2)$ theories in the IR and some flat connections being ``not independent''---arise, and one of our focus is to illustrate their effects in examples. 

\paragraph{A first look at $G=\mathrm{Sp}(2N)$.} The nilpotent orbits---or phases of the deformed theory---are again labeled by partitions $\lambda\vdash2N$ but now with the condition that odd numbers appear even number of times. In other words, $r_i\equiv 0 \pmod 2$ for any odd $i$. The unbroken gauge group in the phase $\lambda$ is given by\footnote{We have used the convention that the compact part of Sp$(2N,\bC)$ is denoted as Sp$(N)$, with Sp$(1)\simeq \SU(2)$.} 
\begin{equation}
    K_\lambda=\prod_{i \text{ odd}}\mathrm{Sp}(r_i/2)\times \prod_{i \text{ even}} \mathrm{O}(r_i).
\end{equation}
This gives $C_\lambda=\bZ_2^{b(\lambda)}$ where $b(\lambda)$ is the number of distinct even $i$'s that appear in the partition. The center $\bZ_2$, when projected to the O$(r_i)$ factor, is a rotation when $r_i$ is even, and a reflection when $r_i$ is odd. This is the starting point for the phase classification:
\begin{itemize}
    \item A phase is electric unless all even parts in $\lambda$ have even multiplicity. Then the center $\bZ_2$ acts as a reflection on some O$(r_i)$ factors and becomes part of $C_\lambda$.  This includes $[4]$ and $[21^2]$ for $N=2$.
    \item The phase is magnetic if all even parts have even multiplicities not equal to two. Notices that zero is also counted as even. Indeed, if only odd numbers appear, the center of Sp$(N)$ will be the diagonal of the center of Sp$(r_\text{odd})$ groups and becomes confined in the IR. For $N=4$, they include $\lambda=[3^21^2],\, [2^4],\,$ and $[1^8]$. 
    \item The phase is of type HE if some even part has multiplicity two. Then the infrared dynamics is given by a product of O(2) and $\bZ_2$ gauge theories. Notice that being HE and E are not mutually exclusive in the Sp case. For $N=4$, there are three HE phases, $[4^2]$, $[2^21^4]$ and $[42^2]$, with the last also being electric (hence E-HE).  
\end{itemize}
On the other hand, neutral phases cannot be present in the Sp cases.

If one looks at $C_\lambda$ for each phase and naively sums up the number of flat connections, one gets
\begin{equation}\label{SKSp}
    \dim \cS (T^3;\mathrm{Sp}(2N,\bC))\stackrel{?}=\sum_{\lambda}2^{3b(\lambda)}.
\end{equation}
For example, when $N=2$, the allowed partitions are $[4],[2^2],[2,1^2]$ and $[1^4]$, and this formula becomes
\begin{equation}
    2^3+2^3+2^3+1=25.
\end{equation}

However, for each even number $i$ that appears exactly twice in the partition, $K_\lambda^\text{IR}$ contains a copy of O$(2)$, and $\lambda$ becomes an exceptional hybrid (HE) phase. One then needs to count the connected component of the moduli space of flat O$(2)$-connections, which we turn to now. 

\paragraph{Counting O$(2)$-connections on $T^3$.} One main difference between O$(2)$ and its component group $\bZ_2$ is that a reflection $S$ in O$(2)$ commutes with a $\pi$-rotation $R$, as well as another reflection $R'=SR$. Together with the identity $I$, they form a $\bZ_2\times \bZ_2$ subgroup. However, counting flat O$(2)$-connections also differs from counting $\bZ_2\times \bZ_2$ ones, as $R$ and $R'$ are conjugate. We organize the flat connections in terms of the number of reflections involved along the three 1-cycles of $T^3$:
\begin{itemize}
    \item No reflection: There is a single component parametrized by SO$(2)$ holonomies modulo conjugation.
    \item One reflection: The other two holonomies will have to be $I$ or $R$, resulting in $2^2\cdot3=12$ components.
    \item Two reflections: Both reflections can be $S$, or one is $S$ while the other is $S'$, giving $2\cdot 3=6$ choices; for each option, the remaining holonomy will have to be $I$ or $R$, giving again $12$ components.
    \item Three reflections: Once a chosen reflection is fixed to be $S$, the other two can be either $S$ or $S'$, giving $2^2=4$ components.
\end{itemize}
Summing them up, there are 29 components in total.

Depending on how the O$(2)$ embeds into the UV gauge group $K$, it can have either electric or magnetic symmetry contributing to different flux sectors.\footnote{In a sense, there are two different versions of the O$(2)$ theory. The electric version can be viewed as the normalizer of the Cartan of SU$(2)$, whose $\bZ_2$ center is identified with the center of SU(2). It can realize the electric $\bZ_2$ 1-form symmetry of the mother theory in the IR phase. On the other hand, the magnetic version can be viewed as the normalizer of the Cartan of SO(3), whose $w_2$ is identified with that of SO(3). It can instead realize the magnetic $\bZ_2$ 1-form symmetry. In addition, there can also be a ``neutral'' version that does not carry either electric or magnetic symmetries of the UV theory, although we have not yet encounter one.} The electric symmetry is associated with the action of $H^1(M_3,\bZ_2)$ on the moduli space (and the induced action on the space of locally constant functions). In the present case of $M_3=T^3$, the action of $H^1(M_3,\bZ_2)\simeq \bZ_2^3$ can permute different components, but in a way that preserves the number of reflections. It is easy to see that for components with 1, 2, or 3 reflections, they are organized into orbits of size 4, with a $\bZ_2$ acting trivially while a $\bZ_2\times \bZ_2$ acting transitively. Therefore, the space of functions split as $1+3$ as characters of $\bZ_2^3$ (i.e.~a trivial character and three non-trivial ones). Summing them up, one has the total decomposition $29=8+3\times 7$ with an $8$-dimensional $\bZ_2^3$-invariant subspace and a three-dimensional space for each of the seven non-trivial characters. 

If one gauges the electric $\bZ_2$ 1-form symmetry, the resulting theory still has an O$(2)$ gauge group---although, in a sense, with smaller radius---but it will be ``a magnetic O$(2)$'' subgroup of $K/\bZ_2$ and contributes to different magnetic flux sectors. For this purpose, one grades the components by $H^2(M_3,\bZ_2)$ using the second Stiefel--Whitney classes of the O$(2)$ bundle. The class $w_2$ will be non-trivial along a plane if the two holonomies are distinct in the non-trivial part of $\bZ_2\times \bZ_2\subset \mathrm{O}(2)$. One then has exactly the same decomposition $29=8+3\times 7$. A quick way to see the 8 is the following. When $w_2=0$, all the reflections have to be the same and the rest has to be the identity (unless there is no reflection, but then they are all in the same connected component). So one can simply count $2^3=8$. Using SL$(3,\bZ)$ invariance, we only need to explain the $3$ for a given $w_2$. Choosing it to be $(1,0,0)$, we can enumerate the three components to be $(1,S,S')$, $(1,S,R)$ and $(1,R,S)$.

We are now better equipped to understand the $T^3$ skein module for  $C_2\simeq\mathrm{Sp}(4,\bC)$.

\paragraph{$\Sp(2)$ vs $\SO(5)$.} For the partition [22], the centralizer O$(2)$ is the one acting as O$(2)\otimes\mathbf{1}_{2\times 2}$ on the two 2-dimensional symplectic blocks of $\bC^4$. All of $S$, $S'$ and $R$ are realized as diagonal four-by-four matrices in Sp$(4,\bC$),
\begin{equation}
    S =
\begin{pmatrix}
+1 & 0 & 0 & 0 \\
0 & +1 & 0 & 0 \\
0 & 0 & -1 & 0 \\
0 & 0 & 0 & -1
\end{pmatrix},
\qquad
S' =
\begin{pmatrix}
-1 & 0 & 0 & 0 \\
0 & -1 & 0 & 0 \\
0 & 0 & +1 & 0 \\
0 & 0 & 0 & +1
\end{pmatrix},
\qquad
R =
\begin{pmatrix}
-1 & 0 & 0 & 0 \\
0 & -1 & 0 & 0 \\
0 & 0 & -1 & 0 \\
0 & 0 & 0 & -1
\end{pmatrix}.
\end{equation}
One can then sum over all $\lambda$ to find the total dimension of sk$(T^3,C_2)$. This would also be the answer for Spin$(5)$ due the exceptional isomorphism $B_2\simeq C_2$. The Langlands duality would predict that the answer is the same for SO$(5)$.\footnote{The $S$-dual of the Sp(2) theory is the SO$(5)_+$ theory with a trivial discrete theta angle. Unlike the SO(3) case, it will not become the SO$(5)_-$ theory when $\theta\mapsto\theta+2\pi$. See \cite{Aharony:2013hda} for more details on such pairs of theories. In this work, we only discuss the SO$(N)_+$ theories and will omit the subscript, but the existence of the SO$(N)_-$ theory would indicate that there is a different kind of skein module for SO$(N)$, which would be interesting to explore.} However, while the Witten index matches, there is a mismatch for the dimensions of skein modules.
\begin{center}
    \centering
    \begin{tabular}{c|c|c|c|c||c|c|c|c}
          \multicolumn{5}{c||}{Sp(2)=Spin(5)}&  \multicolumn{4}{c}{SO(5)} \\\hline
     type &  $\lambda_{\Sp}$& $K_\lambda$ & skein & index  &  $\lambda_\SO$ &  $K_\lambda$ & skein & index \\\hline
     E&   [4] &  $\bZ_2$ & 1+{\color{blue}7} & 1+{\color{blue}7} & [5] & 0 & 1 &   1  \\\hline
      HE &  [$2^2$] & O(2) & 8+{\color{blue}21}  & 0 & [$31^2$] & O(2)  &  8+{\color{red}21} & 0 \\\hline
      E &  [$21^2$]  & Sp(1)$\times\bZ_2$ & 1+{\color{blue}7} & 2$\cdot$(1+{\color{blue}7}) & [$2^21$] & Sp(1) & 1 & 2 \\\hline
      M&  [$1^4$] & Sp(2) & 1 & 3 & [$1^5$]& SO(5) &  1+{\color{red}7} &  $3\cdot$(1+{\color{red}7}) \\\hline
        \multicolumn{3}{c|}{total:} & 11+{\color{blue}35} & 6+{\color{blue}21} & \multicolumn{2}{c|}{total:} & 11+{\color{red}28} & 6+{\color{red}21} 
    \end{tabular}
\end{center}
Notice that for the equality of the index, the difference in the number of quantum vacua in different phases make up for the fact that there are two electric phases while only one magnetic phase. However, for the skein module, we are out of luck as the quantum vacua are indistinguishable by Wilson lines.

There are several possible interpretations for the apparent mismatch:
\begin{itemize}
    \item There are overlaps on the Sp(2) side, meaning that the five states in an $e\neq 0$ sector that we counted actually satisfy a linear relation. Someone who believes in this possibility might be encouraged by the observation that the flat $\bZ_2$-connections in the $\lambda=[4]$ phase all embed into the continuous component of the O(2) moduli space and, therefore, maybe they can ``tunnel'' into each other and should only be counted as one quantum state. However, there are further problems when comparing Sp and SO for higher ranks, and it seems that one needs an increasingly long list of ad-hoc rules to maintain the Langlands duality between them.
    \item The skein modules actually match, but our method of computing it via the $\cN=1$ deformation has certain subtleties. This is a logical possibility and it seems that the mismatch is always associated with HE phases where O(2) or a higher-rank analogue appears. It would be interesting to understand more precisely the physics of the deformation such as the detailed behavior of the states $\bra{L}$ in the Hilbert space. 
    \item The duality between skein modules for Langlands dual groups actually fails for $B$--$C$ pairs. This interpretation is preferred by the author, as physics only predicts a duality for $\tilde \cS$, while the one for $\cS$ is only expected to hold when ignoring 't Hooft and dyonic operators removes isomorphic subspaces on the two sides. If this is the correct interpretation, one should be able to remedy the duality by adding back the ``missing'' line operators. Indeed, assuming that the quantum vacua can now be distinguished with dyonic operators, we have an increase of dimensions of $1+{\color{blue}7}$ vs 1 for the $[21^2]$ electric phase on the two sides, but $2$ vs $2\times(1+{\color{red}7})$ for the $[1^4]$ magnetic phase. Therefore, one has $\dim\tilde \cS=56=14+42$ on both sides.\footnote{Notice that a difference between the quantum vacua of SO(5) and SO(3) plays a crucial role: for the SO(5) theory, each of the three quantum vacua has the magnetic $\bZ_2$ symmetry, while for SO(3) only one of them has the magnetic symmetry. Therefore, for the SU(2) vs SO(3) pair, removing 't Hooft operators from $\tilde\cS$ does not create a mismatch for $\cS$, but it does for the Sp(2)--SO(5) pair. There are a few closely related perspectives on this difference: 1) whether a theory has oblique confinement or not, 2) the existence of ``fractional'' instantons in the theory, 3) whether the ``spectral flow'' in the sense of \cite{Witten:2000nv} involves fractional coefficients, 4) the relation between the (Pontryagin) square of the $w_2(G)$ and the $p_1(G)$ of the gauge bundle, and 5) the Postnokov $k$-invariant of B$G$.}
\end{itemize}
Although the author prefers the third interpretation, we will not take an official stance,\footnote{Including more general line operators can lead to various complications in the analysis, and treating them systematically is beyond the scope of this work. So far our discussion of them mostly involves separating different quantum vacua in the same phase, and it would be interesting to understand whether they also play other roles, such as making certain indistinguishable electric phases distinguishable.} but instead proceed to analyze some higher-rank cases, pinpoint the source for mismatches, and give our best attempt to restore duality first without invoking 't Hooft operators. It turns out that many phases match perfectly, and the problem seems to be always associated with hybrid phases, which perhaps shows that the first and the second interpretations also have some merits.

We now discuss some general aspects of the $B$-series (which we didn't have to use for the computation in the SO(5) case above due to the exceptional isomorphism), starting with the non-simply-connected SO groups.

\paragraph{Phase structures for $B_N$.} To label nilpotent orbits in $B_N$, the partition $\lambda$ is required to have even multiplicity for all (non-zero) even numbers, $r_i\equiv 0 \pmod 2$ for all even $i$. (To emphasize the difference between the three versions of partitions, we sometimes use the notation $\lambda_\SL$, $\lambda_\Sp$ and $\lambda_\SO$.) The unbroken gauge group in the phase labeled by $\lambda$ is
\begin{equation}
    K_\lambda=\prod_{i \text{ even}}\mathrm{Sp}(r_i/2)\times \mathrm{S}\left(\prod_{i \text{ odd}} \mathrm{O}(r_i)\right).
\end{equation}
The component group is $C_\lambda=\bZ_2^{a(\lambda)-1}$ where $a(\lambda)$ counts the distinct odd $i$'s appearing in the partition. From this, one might naively expect
\begin{equation}
    \dim \cS^0 (T^3;\mathrm{SO}(2N+1))\stackrel{?}=\sum_{\lambda}8^{a(\lambda)-1}
\end{equation}
in the sector with a trivial magnetic flux. In this sector, the difference between O$(2)$ and $\bZ_2$ is irrelevant, but one can still have the ``overlap problem,'' which we will see later in examples.

For the Spin group, $K_\lambda$ is a spin double cover of the one above. This would not increase $C_\lambda$ unless all $r_i=1$ for odd $i$, for which the partition is referred as ``rather odd.'' Therefore, the phase classification is as follows:
\begin{itemize}
    \item When $\lambda$ is rather odd, the phase will be electric. The $\bZ_2$ center of Spin$(2N+1)$ is part of the discrete gauge symmetry in the IR. For the SO(5) case, such phases are $[5]$ and $[2^21]$.
    \item When one of the odd $i$ has multiplicity larger than 2, we have a magnetic phase. The center of Spin$(2N+1)$ is inside the double cover of  $\mathrm{S}\left(\prod \mathrm{O}(r_i)\right)$ with the product over $r_i>2$ with odd $i$ and becomes confined in the IR. This allows turning on magnetic fluxes in the SO$(2N+1)$ theory. For SO(5), the only such phase is $\lambda=[1^5]$. 
    
    \item When $r_i=2$ for one or more odd $i$, it is an HE phase. The IR theory will be a product of O(2) and (if any) $\bZ_2$  gauge theories. For the SO(5) case, $\lambda=[31^2]$ gives such a phase. Notice that an HE phase can also be magnetic (hence M-HE) in the $B$-series, dual to the E-HE phases in $C$-series. One example is $\lambda=[3^21^3]$ for SO(9). For such a phase, the center doesn't completely disappear in the IR, but one can still turn on a flux via the O(2) theory.

\end{itemize}
Furthermore, it is not possible to have an HO phase, as $\bZ_2$ has no non-trivial proper subgroups, or a neutral phase, as it would require only having a single pair of odd parts with $r_i=2$ (which will become possible for groups of $D$-type).

Now we are prepared to test the duality between $B_3$ and $C_3$.

\paragraph{$B_3$ and $C_3$.} For the $B$--$C$ dual pair, a general concern is that the number of phases or nilpotent orbits are not the same on the two sides. This issue seems to be always avoided by taking into account of overlaps. The phases and their contributions for $B_3$ and $C_3$ are listed below, and the contributions from the phase in green are not counted exactly because of the overlap phenomenon---the discrete gauge group $\bZ_2\times\bZ_2$ for the $\lambda=[42]$ phase already includes the gauge group in the $\lambda=[6]$ phase, and any vaccum given by a flat connection in the latter phase cannot be distinguished from certain vacuum in the former. 
\begingroup
\setlength{\tabcolsep}{0.1em}
\begin{center}
    \centering
    \begin{tabular}{c|c|c|c|c|c||c|c|c|c|c|c}
          \multicolumn{4}{c|}{Sp(3)}& \multicolumn{2}{c||}{Sp(3)$/\bZ_2$} &\multicolumn{4}{c|}{Spin(7)} & \multicolumn{2}{c}{SO(7)}  \\\hline
     $\lambda_{\Sp}$& type &   $K_\lambda$ & skein & $K_\lambda$ & skein  &  $\lambda_\SO$ & type& $K_\lambda$ & skein & $K_\lambda$ & skein \\\hline
       \cellcolor{green!25}[6] & \cellcolor{green!25}E&  \cellcolor{green!25}$\bZ_2$ & \cellcolor{green!25}1+{\color{blue}7} & \cellcolor{green!25}0&\cellcolor{green!25}1  & [$1^7$] & M&  Spin(7) &   1 & SO(7) & 1+{\color{red}7}\\\hline
        [42] & E& $\bZ_2\times\bZ_2$ & 8+{\color{blue}56} & $\bZ_2$& 8 & [$31^4$] & M & Pin(4) & 8 & O(4) &  8+{\color{red}56}  \\\hline
       [$3^2$] &M  & Sp(1)& 1 &Sp(1)$/\bZ_2$& 1+{\color{red}7} & [$32^2$] & E & Sp(1)$\times\bZ_2$ & 1+{\color{blue}7} & Sp(1) & 1\\\hline
       [$2^3$]  & E & O(3) & 1+{\color{blue}7} &SO(3) & 1 & [$3^21$] & HE &O(2) & \cellcolor{orange!25}8+{\color{blue}21} & O(2) & \cellcolor{orange!25}8+{\color{red}21} \\\hline
        [$2^21^2$] & HE&Sp(1)$\times$O(2) & 8+{\color{blue}21}  & Sp(1)$\times$O(2) & 8+{\color{red}21} & [$51^2$] &HE &O(2) & 8+{\color{blue}21} & O(2) & 8+{\color{red}21}\\\hline
       [$1^6$] & M& Sp(3) & 1 &Sp(3)$/\bZ_2$&  1+{\color{red}7} & [$7$]& E & $\bZ_2$ &  1+{\color{blue}7} & 0  &1 \\\hline\hline
      [$41^2$] &E&   Sp(1)$\times\bZ_2$ & 1+{\color{blue}7} & Sp(1)&1 & [$2^21^3$]& M& Sp(1)$^2$ &  1 & Sp(1)$\times$SO(3)& 1+{\color{red}7}\\\hline
        [$21^4$] &E& Sp(2)$\times\bZ_2$ & 1+{\color{blue}7} & Sp(2)&1  \\\hline\hline
        \multicolumn{2}{c|}{total} & \multicolumn{2}{c|}{$126=21+7\cdot${\color{blue}14}} & \multicolumn{2}{c||}{$56=21+7\cdot${\color{red}5}} & \multicolumn{2}{c|}{total} & \multicolumn{2}{c|}{$84=28+7\cdot${\color{blue}8}} & \multicolumn{2}{c}{$140=28+7\cdot${\color{red}16}}
    \end{tabular}
\end{center}
\endgroup
Notice that this overlap exactly compensates for the additional phase present on the $C_3$ side. We have organized the orbits on the two sides by the Spaltenstein duality \cite{spaltenstein2006classes} that relates the ``special orbits'' on the two sides (see e.g.~\cite{collingwood1993nilpotent} for more details). The non-special orbits are below the doubled horizontal line. 

Notice that there is again a mismatch, either for the Sp(3)--SO(7) pair or for the Sp(3)$/\bZ_2$--Spin(7) pair. One possible way to remedy this mismatch is to identify a certain overlap phenomenon on the Spin/SO(7) side as well. A candidate for such an overlap is highlighted in orange. What is special about this $\lambda=[3^21]$ phase is that the $\pi$-rotation of its O(2) coincides with a reflection of the O(2) gauge group of the $\lambda=[51^1]$ phase. Furthermore, for Spin(7), it overlaps with the $\lambda=[7]$ phase, in a manner similar to what we observed in the $C_2$ case. If we take this observation as a blessing, allowing us to reduce the contributions in orange to $1$ for Spin(7) and $1+{\color{red}7}$ for SO(7), then the duality could be restored. 

I hope the reader would agree with the author that haphazardly introducing new rules of this kind is not particularly aesthetically pleasing. It might be therefore preferable to accept that the skein modules are not dual by themselves, and that one should consider the enriched version $\tilde\cS$ incorporating 't Hooft and dyonic operators to have the duality restored. We will not demonstrate the full duality for $\tilde \cS$ here, but note that one would obtain an additional copy of the O(2) contribution from the HE phase on the Sp(3) side by ``splitting off'' a quantum vacuum, thereby matching the extra copy appearing on the SO(7) side. If the remaining contributions match in a fashion similar to the Witten index (to which the HE phases do not contribute), this yields a duality for the dimensions of $\tilde \cS$.\footnote{In fact, there is at least one modification needed when compared with the matching of the index. Either the electric phase $\lambda_{\Sp}=[6]$ becomes detectable after 't Hooft and dyonic operators are included, or there should be a similarly undetectable quantum vacuum (or a combination of vacua) on the SO side, which potentially lives in the $\lambda_\SO=[2^21^3]$ phase. 
}

\paragraph{$B_4$ and $C_4$.} The last comparison that we will make is between Sp(4) and SO(9). The results are listed below:
\begin{center}
    \centering
    \begin{tabular}{c|c|c|c||c|c|c|c}
          \multicolumn{4}{c||}{Sp(4)}&  \multicolumn{4}{c}{SO(9)}  \\\hline
     $\lambda_{\Sp}$& type &   $K_\lambda$ & skein   &  $\lambda_\SO$ & type&  $K_\lambda$ & skein \\\hline
       \cellcolor{green!25}[8] & \cellcolor{green!25}E&  \cellcolor{green!25}$\bZ_2$ & \cellcolor{green!25}1+{\color{blue}7} & [$1^9$] & M & SO(9) & 1+{\color{red}7}\\\hline
        [62] & E& $\bZ_2\times\bZ_2$ & 8+{\color{blue}56}  & [$31^6$] & M & O(6) &  8+{\color{red}56}  \\\hline
       [$4^2$] &HE  & O(2)& 8+{\color{blue}21} & [$32^21^2$] & HE & O(2)$\times$Sp(1) & 8+{\color{red}21}\\\hline
       [$42^2$] &E-HE  & $\bZ_2$$\times$O(2)& 29$\cdot$(1+{\color{blue}7}) & [$3^21^3$] & M-HE & S(O(2)$\times$O(3)) & 29$\cdot$(1+{\color{red}7})\\\hline
       [$3^22$] &E  & $\bZ_2$$\times$Sp(1)& 1+{\color{blue}7} & [$3^3$] & M & SO(3) & 1+{\color{red}7}\\\hline
              [$421^2$] &E  & $\bZ_2^2$$\times$Sp(1)& 8+{\color{blue}56} & [$51^4$] & M & O(4) & 8+{\color{red}56}\\\hline
                     [$3^21^2$] &M  & Sp(1)$^2$& 1 & [$52^2$] & E & Sp(1) & 1\\\hline
       [$2^4$]  & M & O(4) & 8& [$531$] & E  & $\bZ_2\times\bZ_2$ & \cellcolor{orange!25}22+{\color{red}42} \\\hline
        [$2^21^4$] & HE&O(2)$\times$Sp(2) & 8+{\color{blue}21}  &  [$71^2$] &HE & O(2) & 8+{\color{red}21}\\\hline
       [$1^8$] & M& Sp(4) & 1& \cellcolor{green!25}[$9$]& \cellcolor{green!25}E  & \cellcolor{green!25}0  &\cellcolor{green!25}1 \\\hline\hline
      [$21^6$] &E&   $\bZ_2\times$Sp(3) & 1+{\color{blue}7} & [$2^21^5$]& M & Sp(1)$\times$SO(5)& 1+{\color{red}7}\\\hline
       [$2^31^2$] &E&   O(3)$\times$Sp(1) & 1+{\color{blue}7}  & [$2^41$]& E & Sp(2)& 1\\\hline
         [$41^4$] &E&   $\bZ_2\times$Sp(2) & \cellcolor{orange!25}1+{\color{blue}7}  & [$4^21$]& E & Sp(1)& 1\\\hline
        \cellcolor{green!25}[$61^2$] &\cellcolor{green!25}E& \cellcolor{green!25}$\bZ_2\times$Sp(1) & \cellcolor{green!25}1+{\color{blue}7}  \\\hline\hline
        \multicolumn{2}{c|}{total} & \multicolumn{2}{c||}{$460=75+7\cdot${\color{blue}55}}  & \multicolumn{2}{c|}{total} & \multicolumn{2}{c}{$509=89+7\cdot${\color{red}60}}
    \end{tabular}
\end{center}
Most of the entries follows from a straightforward computation, giving an almost perfect agreement between the two sides, and we only explain the colored entries:
\begin{itemize}
    \item The three phases $\lambda_\Sp=[8],\,[61^2]$ and $\lambda_\SO=[9]$ colored in green are indistinguishable from subsectors of, respectively, $\lambda_\Sp=[62], \,[421^2]$, and $\lambda_\SO=[531]$. After removing them, the number of phases are again identical on the two sides.\footnote{In principle, to match the numbers on the two sides for $T^3$, we don't need a phase-wise correspondence. However, if the duality for $T^3$ involves certain highly non-trivial conspiracies between phases, it would be hard to imagine that it can generalize to $M_3=\Sigma\times S^1$ and beyond.}
    \item The rather odd partition $\lambda_\SO=[531]$ can actually contribute to sectors with a non-trivial magnetic background, due to a subtlety we mentioned before involving non-abelian central extensions. In SO(9), the three non-trivial elements of the $\bZ_2\times \bZ_2$ gauge group can be represented as diagonal matrices with $\pm1$ entries. They can be chosen to have respectively $4,\,6$ and 8 negative entries. They commute in SO(9), but after lifting to Spin(9), they are represented by product of gamma matrices as $a=\gamma_{1234}$, $b=\gamma_{456789}$ and $c=ab=\gamma_{12356789}$. They satisfy $a^2=c^2=1$, $b^2=-1$, and any pair anti-commutes. They now generate a non-abelian extension of $\bZ_2\times \bZ_2$ that is isomorphic to the order-8 dihedral group D$_8$. To have $w_2(\SO(9))=0$, the three holonomies can only contain one of the three non-trivial elements of $\bZ_2\times\bZ_2$, giving $22=1+3\cdot7$ choices. For each non-trivial $w_2$, there are 6 choices. To see this, one can use the SL$(3,\bZ)$ symmetry to set $w_2=(1,0,0)$. Then the $6=3\times 2$ choices of holonomies are of the form $(e,x,y)$ with $x$ and $y$ being distinct elements of $\{a,b,c\}$. 
    \item On the $C_4$ side, after substituting the indistinguishable phases with some of the non-special ones, there is an additional electric phase $\lambda=[41^4]$ that doesn't have a dual magnetic phase. Notice that for this phase, the flat $\bZ_2$-connections are embedded in the O(2) moduli space of the $[2^21^4]$ phase. This is analogous to some of the previous cases, and if we perform a similar modification and eliminate the contributions to the $e\neq 0$ sectors, then it can be pair with one of the remaining non-special electric phase on the SO(9) side.
\end{itemize}
After this, the only mismatch concerns the rather odd partition $\lambda_\SO=[531]$, which again seems to be related to an overlap with an HE phase, now the one given by $\lambda=[711]$. Indeed, a generator of $\bZ_2\times \bZ_2$ with eight $-1$ and one $+1$ on the diagonal is a reflection in the O(2) gauge group for the $[711]$ phase. If it is killed by this overlap, the gauge group for the $[531]$ phase will be reduced to $\bZ_2$, which gives an $8$-dimensional space for (the dual of) $\cS$, matching exactly the $\lambda=[2^4]$ phase on the Sp(4) side. After these modifications, both sides will sum up to be $453=75+7\cdot 54$.

\medskip
This example concludes our discussion of the $B$- and $C$-type groups. We will now turn to the $D$-series, where we will have less trouble verifying the Langlands duality. This (hopefully) suggests that the difficulties encountered for the $B$--$C$ pairs reflect a genuine mismatch, rather than a lack of skill or effort on the author's part.

\subsubsection{$D$-series}

For the $D$-series, the simply-connected global form is Spin$(2N)$, whose center is $\bZ_4$ for odd $N$ and $\bZ_2\times \bZ_2$ for even $N$. The classification of phases is similar to the $B$-series, except that for ``very even'' partitions (only even numbers appear in the partition which can only happen for even $N$) there are two nilpotent orbits hence two phases. 

For $K=\SO(2N)$, the unbroken gauge group in a phase $\lambda$ labeled by the partition $[1^{r_1}2^{r_2}\ldots]$ is
\begin{equation}
    K_\lambda=\prod_{i \text{ even}}\mathrm{Sp}(r_i/2)\times \mathrm{S}\left(\prod_{i \text{ odd}} \mathrm{O}(r_i)\right).
\end{equation}
For Spin$(2N$), one can simply take its spin double cover. So far, this is almost identical to the SO($2N+1$) case, but now we also have two additional global forms, the semi-Spin group Ss$(2N)$ and the adjoint-type group PSO$(2N)$. For PSO(2N), the gauge group in each phase can be obtain from the above by modding out the diagonal $\pm\mathbf{1}$ action, and one can then take a double cover to obtain the gauge group associated with the semi-Spin group. The phase classification is also analogous to the $B$-series, except that N- and HO-type of phases are now possible:
\begin{itemize}
    \item When $\lambda$ is rather odd (each odd part appearing exactly once), the phase will be electric, and the full center of Spin$(2N)$ becomes part of the discrete gauge symmetry in the IR. For the SO(8) case, such phases are $[71]$, $[53]$ and $[32^21]$.
    \item When all of the odd $i$ has multiplicity larger than 2, we get a magnetic phase. The full center of Spin$(2N)$ is inside the double cover of  $\mathrm{S}\left(\prod \mathrm{O}(r_i)\right)$ with the product over $r_i>2$ with odd $i$ and becomes confined in the IR. 

    \item For ordinary hybrid phases, they can arise when some odd parts appear only once while some others appear with multiplicity larger than 2. Alternatively, they also arise for very even partitions. For these phases, a $\bZ_2$ subgroup of the center is in $H_\lambda$ while a $\bZ_2$ quotient is in $C_\lambda$. Examples include $\lambda=[2^2],\,[4^2],\,[51^3],\,[31^5]$, etc.

    \item When there is exactly one odd $i$ which also has $r_i=2$, one gets a neutral phase. This is only possible when $N$ is odd. The IR theory has a SO($r_i)\simeq \U(1)$ gauge field, and the $\bZ_4$ center of the Spin$(2N)$ theory is fully contained in it.
    
    \item When one odd $i$ has $r_i=2$ and there are other odd parts, one gets an HE phase. The smallest example is $\lambda=[3^21^2]$, which is a rather special one with $K_\lambda=\mathrm{S(O(2)\times O(2))}$ which we have to deal with in the $D_4$ case. 

\end{itemize}

We will work through some low-rank cases until $D_4$, starting with $D_2\simeq A_1\oplus A_1$, which already demonstrates the ``very even'' phenomenon unique to the $D$-series.

\paragraph{Groups associated with $D_2$.} There are several different global forms with Spin$(4)\simeq\SU(2)\times\SU(2)$ and PSO$(4)\simeq\SO(3)\times \SO(3)$ at the two ends of the spectrum, and SO$(4)$ and the semi-spin group Ss$(4)\simeq\SO(3)\times \SU(2)$ in the middle.\footnote{There is a different quotient of Spin group, the ``semi-co-spin'' group Sc. It is abstractly isomorphic to Ss and in fact complex conjugate (i.e.~the $\bZ_2$ outer-automorphism of $D_n$) to it. Therefore, we omit it from the discussion as it would be straightforward to recover it by basically swapping the very even orbits.}
The phases of the mass-deformed $D_N$ theories are labeled by partitions of $2N$ with even parts occurring even number of times, but there are two orbits for ``very even'' partitions where no odd parts appear. For $N=2$, these are listed below:  
\begingroup
\setlength{\tabcolsep}{0.4em}
\begin{center}
    \centering
    \begin{tabular}{c|c|c|c|c|c|c|c|c|c}
       $\lambda$  &type& \multicolumn{2}{c|}{Spin(4)} & \multicolumn{2}{c|}{SO(4)}  &  \multicolumn{2}{c|}{Ss(4)} & \multicolumn{2}{c}{PSO(4)}  \\\hline
        [31] & E& $\bZ_2\times\bZ_2$ & 1+{\color{blue}63} & $\bZ_2$ &   1+{\color{blue}7} &  $\bZ_2$ & 1+{\color{blue}7}  & 0 & 1 \\\hline
        [$2^2$]$_{\mathrm{I}}$ & HO& $\bZ_2\times \SU(2)$ & 1+{\color{blue}7} & SU(2) & 1 &  SU(2) & 1 & SO(3) &  1+{\color{red}7}\\\hline  [$2^2$]$_{\mathrm{II}}$ &HO&  $\bZ_2\times \SU(2)$ & 1+{\color{blue}7} & SU(2) & 1 &  $\bZ_2\times$SO(3) & (1+{\color{red}7})(1+{\color{blue}7}) & SO(3) &  1+{\color{red}7}  \\\hline
        [$1^4$] & M& Spin(4) & 1 & SO(4) & 1+{\color{red}7} & Ss(4) & 1+{\color{red}7} & PSO(4) & 1+{\color{red}63}
    \end{tabular}
\end{center}
\endgroup
For each phase, we have listed their $K_\lambda$ and contribution to the skein module for all the global forms. Again, non-trivial electric and magnetic sectors are colored blue and red. We see that there is a duality between Spin$(4)$ and PSO$(4)$, while SO(4) and Ss(4) are self-dual.

\paragraph{The case of $D_3$.} Due to the exceptional isomorphism $D_3\simeq A_3$, we have already partially covered this case. Here, the focus is on how electric-magnetic duality works for the intermediate SO(6). For $D_\text{odd}$, there are no very even partitions and no semi-Spin groups, while the center for the simply-connected Spin group is $\bZ_4$. We list the phases below, using both SL  and SO labels:  
\begin{center}
\begin{tabular}{c|c|c|c|c|c|c|c|c}
   $\lambda_\SL$ & $\lambda_\SO$
   & type & \multicolumn{2}{c|}{Spin(6)}
   & \multicolumn{2}{c|}{SO(6)}
   & \multicolumn{2}{c}{PSO(6)}
   \\\hline
   [4] & [51] &E& $\mathbb{Z}_4$ & 1+{\color{blue}63}
   & $\mathbb{Z}_2$ & 1+{\color{blue}7}
   & 0 & 1
   \\\hline
   [31] & [$3^2$] &N& $\mathrm{U}(1)$ & 1
   & U(1) & 1
   & U(1) & 1
   \\\hline
   [$2^2$] & [$31^3$]
   &HO& $\mathrm{Pin}(3)$ & 1+{\color{blue}7}
   & O(3) & (1+{\color{red}7})(1+{\color{blue}7})
   & SO(3) & 1+{\color{red}7}
   \\\hline
   [$21^2$] & [$2^21^2$]&N
   & SU(2)$\times\mathrm{U}(1)$ & 1
   & SU(2)$\times\mathrm{U}(1)$ & 1
   & U(2) & 1
   \\\hline
   [$1^4$] & [$1^6$]&M
   & Spin(6) & 1
   & SO(6) & 1+{\color{red}7}
   & PSO(6) & 1+{\color{red}63}\\\hline\hline   \multicolumn{3}{c|}{total} & \multicolumn{2}{c|}{5+{\color{blue}70}}& \multicolumn{2}{c|}{5+{\color{blue}14}+{\color{red}14}+{\color{purple}49}}  & \multicolumn{2}{c}{5+{\color{red}70}}
\end{tabular}
\end{center}
Most of the entries need no further explanation, and we will only remark on two of them. For $\lambda_\SO=[31^3]$, $K_\lambda=\mathrm{Pin}(3)$ in Spin(6) can be identified in the SL convention as SU$^\pm(2)$ (unitary matrices with determinant $\pm1$, cf.~Footnote~\ref{foot:SUpm}) with the center being $\bZ_4$. For $\lambda_\SO=[2^21^2]$, $H_\lambda\subset\mathrm{PSO(6)}$ is the simply-connected SU$(2)$ (as opposed to SO(3) which will give contributions to non-trivial magnetic sectors) because $K_\lambda\simeq \U(2)$ is the quotient of SU$(2)\times \U(1)\subset\SO(6)$ by the $\bZ_2$ generated by the ``diagonal $-1$.'' Alternatively, it is the quotient of SU$(2)\times \U(1)\subset\Spin(6)$ by a $\bZ_4$ generated by $(-1,i)$.

This is an interesting example to examine again the difference between $\cS$ and $\tilde \cS$. Once the 't Hooft and dyonic operators are introduced---again assuming their only role is to distinguish quantum vacua---the contributions from the last three phases are modified, listed in the table below:
\begin{center}
\begin{tabular}{c|c|c|c|c|c|c|c|c}
\multicolumn{3}{c|}{phase} & \multicolumn{2}{c|}{Spin(6)} & \multicolumn{2}{c|}{SO(6)} & \multicolumn{2}{c}{PSO(6)}\\\hline
   $\lambda_\SL$ & $\lambda_\SO$
   & type & \raisebox{-0.15em}{$\dim \tilde\cS$}&\raisebox{-0.15em}{$\dim \tilde\cS/\cS$}
   & \raisebox{-0.15em}{$\dim \tilde\cS$}& \raisebox{-0.15em}{$\dim \tilde\cS/\cS$}
   & \raisebox{-0.15em}{$\dim \tilde\cS$}&\raisebox{-0.15em}{$\dim \tilde\cS/\cS$}
   \\\hline
   [4] & [51] &E&  1+{\color{blue}63}&0
   &  1+{\color{blue}7}&0 & 1& 0
   \\\hline
   [31] & [$3^2$] &N& 1&0&1& 0&1
   & 0 
   \\\hline
   [$2^2$] & [$31^3$]
   &HO& 2$\cdot$(1+{\color{blue}7})
   & 1+{\color{blue}7}&(2+{\color{red}7})(1+{\color{blue}7})& 1+{\color{blue}7}
   &  2+{\color{red}7}& 1
   \\\hline
   [$21^2$] & [$2^21^2$]&N
   & 2& 1
   & 2& 1
   & 2& 1
   \\\hline
   [$1^4$] & [$1^6$]&M
   & 4&3
   & 4+2$\cdot${\color{red}7}& 3+{\color{red}7}& 4+{\color{red}7}+{\color{red}63}
   & 3+{\color{red}7}\\\hline\hline
   \multicolumn{3}{c|}{total} & 10+{\color{blue}77}& 5+{\color{blue}7} &10+{\color{blue}21}+{\color{red}21}+{\color{purple}49}  & 5+{\color{blue}7}+{\color{red}7} & 10+{\color{red}77}& 5+{\color{red}7}
\end{tabular}
\end{center}
The HO phase $\lambda_\SO=[31^3]$ will ``split off'' an additional quantum vacuum previously unrecognizable by the Wilson lines. It does not have a magnetic gauge symmetry, but its contribution is multiplied by the number of the ``electric'' $C_\lambda$ flat connections. For the neutral phase $\lambda_\SO=[2^21^1]$, there will be another neutral vacuum now being detectable. Lastly, for the magnetic phase, there will be three additional quantum vacua, one of which will carry an $\bZ_2$ magnetic gauge symmetry for both PSO and SO. Summing over their contributions, we see a duality between the Spin--PSO pair, as well as the self-duality of the SO(6) case. 

\medskip

The previous examples might give the false impression that the Langlands duality also holds automatically for $D_N$ in a simple manner. We will demonstrate with the $N=4$ case that, while the duality is still true, the realization is more subtle.

\paragraph{$D_4$ and triality.} The global forms are Spin$(8)$, SO$(8)$ and PSO$(8)$, with the two semi-Spin groups being isomorphic to SO$(8)$ via the triality. As we are labeling phases by SO partitions, the triality is not manifest, and we still include one of the semi-Spin groups for comparison: 
\begingroup
\setlength{\tabcolsep}{0.1em}
\begin{center}
    \centering
    \resizebox{\textwidth}{!}{
    \begin{tabular}{c|c|c|c|c|c|c|c|c|c}
       $\lambda$  & type & \multicolumn{2}{c|}{Spin(8)} & \multicolumn{2}{c|}{SO(8)}  &  \multicolumn{2}{c|}{Ss(8)} & \multicolumn{2}{c}{PSO(8)}  \\\hline
        [71] &E& $\bZ_2\times\bZ_2$ & 1+{\color{blue}63} & $\bZ_2$ &   1+{\color{blue}7} &  $\bZ_2$ & 1+{\color{blue}7}  & 0 & 1 \\     
        
        [53] &E& $\bZ_2\times\bZ_2$ & -- & $\bZ_2$ &  -- &  $\bZ_2$ & --  & 0 & -- \\\hline
        
[$51^3$] &HO& $\bZ_2$$\times$$ \SU(2)$ & 1+{\color{blue}7} & $\bZ_2\times$SO(3) & (1+{\color{red}7})(1+{\color{blue}7}) &  SU(2) & 1 & SO(3) &  1+{\color{red}7}\\
        
        [$4^2$]$_{\mathrm{I}}$ &HO& $\bZ_2$$\times$$ \SU(2)$ & 1+{\color{blue}7} & SU(2) & 1 &  SU(2) & 1 & SO(3) &  1+{\color{red}7}\\
        
        [$4^2$]$_{\mathrm{II}}$ & HO& $\bZ_2$$\times$$ \SU(2)$ & 1+{\color{blue}7} & SU(2) & 1 &  $\bZ_2\times$SO(3) & (1+{\color{red}7})(1+{\color{blue}7}) & SO(3) &  1+{\color{red}7}  \\\hline

                [$3^21^2$] & HE& S(O$(2)^2$) & 8\!+\!{\color{blue}105} & S(O$(2)^2$) & 1\!+\!7(1\!+\!{\color{red}3})(1\!+\!{\color{blue}3}) &  S(O$(2)^2$) & 1\!+\!7(1\!+\!{\color{red}3})(1\!+\!{\color{blue}3}) & S(O$(2)^2$) &  8\!+\!{\color{red}105}  \\\hline

        {\color{orange} [$32^21$]} & E& $\bZ_2^2$$\times$$ \SU(2)$ & 1+{\color{blue}63} & $\bZ_2\times$SU(2) & 1+{\color{blue}7}  &  $\bZ_2\times$SU(2) & 1+{\color{blue}7} & SU(2) &  1  \\\hline

[$31^5$] &HO& $\bZ_2$$\times$$\Sp(2)$ & 1+{\color{blue}7} & $\bZ_2$$\times$SO(5) & (1+{\color{red}7})(1+{\color{blue}7}) &  Sp(2) & 1 & SO(5) &  1+{\color{red}7}\\
        
        [$2^4$]$_{\mathrm{I}}$ &HO& $\bZ_2$$\times$$\Sp(2)$ & 1+{\color{blue}7} & Sp(2) & 1 &  Sp(2) & 1 & SO(5) &  1+{\color{red}7}\\
        
        [$2^4$]$_{\mathrm{II}}$ &HO&  $\bZ_2$$\times$$ \Sp(2)$ & 1+{\color{blue}7} & Sp(2) & 1 &  $\bZ_2\times$SO(5) & (1+{\color{red}7})(1+{\color{blue}7}) & SO(5) &  1+{\color{red}7}  \\\hline

        [$2^21^4$] &M& SU(2)$^3$ & 1 & SU(2)$\times$SO(4) & 1+{\color{red}7} &SU(2)$\times$SO(4) & 1+{\color{red}7} & SU(2)$^3/\bZ_2^2$ & 1+{\color{red}63}\\\hline
        
        [$1^8$] &M& Spin(8) & 1 & SO(8) & 1+{\color{red}7} & Ss(8) & 1+{\color{red}7} & PSO(8) & 1+{\color{red}63}
    \end{tabular}
    }
\end{center}
\endgroup
We give multiple remarks about the table above:
\begin{itemize}
    \item The phase $[53]$ always ``overlaps'' with $[71]$ and is indistinguishable from it by Wilson lines, as the holonomies are both in the center of $G$. Therefore, one should only count one of them.
    \item This spoils the Lustzig--Spaltenstein duality between the special orbits in the sense that, although dual orbits still have the same contribution, the overlap patterns are in general different. For example, on the opposite end, there is no problem distinguishing the two phases labeled by $[2^21^4]$ and $[1^8]$ using confinement patterns.
    \item The non-special orbit $[32^21]$ (in orange) plays a crucial role in restoring the Langlands duality for skein modules to make up for the ``lost part'' due to the overlap between the $[53]$ and $[71]$ phases. It ensures that there are the same number of electric phases as magnetic phases and they can be paired together.
    \item For $\lambda=[2^21^4]$, the group $K_\lambda=\SU(2)^3\subset \Spin(8)$ has the action of the center $\bZ_2\times \bZ_2$ that flips even copies of SU$(2)$'s. The quotient by any of the three $\bZ_2$ subgroups is identified with SU(2)$\times$SO(4), although with three different SO$(4)$'s. The quotient $\SU(2)^3/\bZ_2^2$ inside PSO(8) can be thought of in a triality-invariant way as the $\bZ_2$ central extension of SO$(3)^3$ in a way symmetric under permutation. This $\lambda$ corresponds to the minimal nilpotent orbit, whose weighted Dynkin diagram has only a non-zero label for the central node. 
    \item There are HO phases that are not triality invariant but instead form size-3 orbits. There are two such groups and they both consist of two very even partitions and an additional one. To restore triality, one can include the Sc(8) theory by swapping the very even $\lambda$'s in the Ss(8) theory. Then the SO-Ss-Sc theories form a ``triality triangle.''
    \item For the HE phase labeled by $\lambda=[3^21^2]$, we have $K_\lambda$ being S(O(2)$\times$O(2)), which is perhaps the most complicated one we have encountered so far. One can obtain its moduli space of flat connections by ``gluing together'' two O(2) moduli spaces by requiring that the reflections for the two copies are always paired up. For each chosen flat connection for O(2)$_1$ with a non-zero number of reflections, one always has four choices for the holonomies of the second copy, as one can conjugate a reflection to a standard one while the other two holonomies can each be flipped. Then we see that the previous $29=8+21=1+7\times (1+3)$ now becomes $113=8+105=1+7\times (1+3)\times(1+3)$. One can then argue that the refinement is simply about ``coloring the $3$'s.'' The fact that S(O(2)$\times$O(2)) can be electric, magnetic or mixed is important in ensuring the Langlands duality. If one throws away the continuous part and just work with the component group $C_\lambda=\bZ_2$, it will be difficult for the duality to work.
\end{itemize}

Although these examples suggest that the $D$-series is better behaved under Langlands duality, we will not try to go beyond $D_4$ in the present work but instead turn our attention to the exceptional groups for the remainder of this section. 

\subsubsection{Exceptional cases}

For $G_2$ and $F_4$, there is no non-trivial check of the duality, and our main job will be to correctly putting together different phases and try to avoid overcounts.

\paragraph{The case of $G_2$.} For $G_2$ there are five nilpotent orbits,
\begin{center}
    \begin{tabular}{c||c|c|c|c|c}
        $\lambda$ & $G_2$ & $G_2(a_1)$ & $\widetilde A_1$ & $A_1$ & 0 \\\hline\hline
        $C_\lambda$ & 0 & $S_3$ & 0 & 0 & 0\\\hline
        $G_\lambda^0$ & 0 & 0 & $A_1$ & $A_1$ & $G_2$ \\\hline
        $h^\vee(G_\lambda^0)$ & 1 & 1 & 2 & 2 & 4 \\\hline
        skein & -- & 21 & 1 & 1 & 1\\\hline
        index & 1 & 21 & 2 & 2 & 4 \\
    \end{tabular}
\end{center}
Here we have used the Bala--Carter labels for the nilpotent orbits, which are more convenient in the exceptional cases. The Lie algebra in the label denotes the derived/non-abelian part of the minimal Levi subgroup (``Bala--Carter Levi'') of $G$ intersecting the orbit. Its rank complements that of $G_\lambda$. In four out of the five phases, the discrete part of the gauge group $C_\lambda$ is trivial, and each only contributes a one-dimensional space to the skein module. For the phase labeled by $G_2(a_1)$, one needs to count the number of $S_3$-connections, which turns out to be 21 (see Appendix~\ref{app:DW}). One additional subtlety is that the phase $\lambda=G_2$ with no gauge symmetry cannot be distinguished with the vacuum with trivial $S_3$-connection in the $\lambda=G_2(a_1)$ phase. Taking this into account leads to
\begin{equation}
    \dim \sk(T^3,G_2)= 24
\end{equation}
while the index is 30. 

Notice that $G_\lambda^0$ in the $G_2$ case are all semisimple, which is also true in the case of $F_4$ but in sharp contrast with the simple groups of other types.

\paragraph{The case of $F_4.$} There are in total 16 phases of the deformed theory listed below:
\begin{center}
    \centering
    \begingroup
    \renewcommand{\arraystretch}{1.15}
    \begin{tabular}{c|c|c|c||c|c|c|c}
        $\lambda$ & $C_\lambda$ & $G_\lambda^0$ & $h^\vee$ & $\lambda$ & $C_\lambda$ & $G_\lambda^0$ & $h^\vee$ \\\hline
        0 & 0 & $F_4$ & 9 & $\widetilde A_2+A_1$ & 0 & $A_1$ & 2\\
        $A_1$ & 0 & $C_3$ & 4 & $C_3(a_1)$ & $\bZ_2$ & $A_1$ & 2\\
        $\widetilde A_1$ & $\bZ_2$ & $A_3$ & 4 & $F_4(a_3)$ & $S_4$ & 0 & 1 \\
        $A_1+\widetilde A_1$ & 0 & 2$A_1$ & 4 & $B_3$ & 0 & $A_1$ & 2 \\
        $A_2$ & $\bZ_2$ & $A_2$ & 3 & $C_3$ & 0 & $A_1$ & 2\\
        $\widetilde A_2$ & 0 & $G_2$ & 4 & $F_4(a_2)$ & $\bZ_2$ & 0 & 1\\
        $A_2+\widetilde A_1$ & 0 & $A_1$ & 2 & $F_4(a_1)$ & $\bZ_2$ & 0 & 1 \\
        $B_2$ & $\bZ_2$ & 2$A_1$ & 4 & $F_4$ & 0 & 0 & 1
    \end{tabular}
    \endgroup
\end{center}
The potentially indistinguishable phases are $F_4,F_4(a_1),F_4(a_2)$ from certain vacua in the $F_4(a_3)$ phase, and $C_3$ from the vacuum with trivial holonomy in the $C_3(a_1)$ phase. Assuming that these are the only overlaps, it is straightforward to count the dimension of $\cS$, which leads to the prediction that 
\begin{equation}
    \dim \sk(T^3,F_4)= 7+4\cdot 8 +84 = 123,
\end{equation}
where we used the fact that $|\cM(T^3,S_4)|=84$. One can compute the Witten index by summing over the phases, weighted by $h^\vee$. However, There is a small subtlety that prevents one from directly taking the product of $h^\vee$ and $|\cM(T^3,C_\lambda)|$, as $C_\lambda$ can act on the $h^\vee$ massive vacua. This has been analyzed in \cite{Bourget:2015lua} for all simple $G$.\footnote{Notice that what are being counted in \cite{Bourget:2015lua} are massive vacua on $\bR^4$ modulo gauge symmetry. Therefore, the results differ from ours, which count states on $T^3$.} For $G_2$ this is not an issue while in the $F_4$ case this only arises for the $B_2$ orbit in the $F_4$ case, where the $\bZ_2$ exchanges the two copies (hence the $\ket{+-}$ vaccum with the $\ket{-+}$ one). When $C_\lambda$ acts non-trivially on the quantum vacua, one should, for each flat $C_\lambda$-connection, form an $h^\vee$-covering of $M_3$ and count the number of components of the covering space. In the present case, one gets 4 for the trivial $\bZ_2$ connection, but 3 for the each of the 7 non-trivial ones.

Assuming again that there are no relative signs, one finds $\cI(T^3,F_4)=227$. 

\paragraph{$E_6$ and $E_6/\bZ_3$.} We now analyze the case of $G=E_6$ and test the Langlands duality with $E_6/\bZ_3$. We organize phases using the four-fold classification:
\begin{itemize}

\item Electric phases. These are phases where the center $\bZ_3$ survives as part of the discrete gauge symmetry in the IR. Their $C_\lambda$ contains a $\bZ_3$ factor coming from the center of $E_6$, which disappears if we pass to the $E_6/\bZ_3$ theory.

\item Magnetic phases. These are phases where one can turn on a magnetic flux. Therefore, they can contribute to $\cS^m$ with non-trivial magnetic background $m\neq 0$. They are characterized by not having a $\bZ_3$ in $C_\lambda$ but an unbroken gauge group $G_\lambda^0$ that has a $\bZ_3$ in its center, which will be quotiented out and become part of $\pi_1(H_\lambda)$ if we switch to the $E_6/\bZ_3$ theory.

The two types are almost matched and listed below:

\begin{center}
    \centering
    \begingroup
    \renewcommand{\arraystretch}{1.15}
    \begin{tabular}{c|c|c|c||c|c|c|c}
        $\lambda$ & $C_\lambda$ & $G_\lambda^0$ & skein & $\lambda$ & $C_\lambda$ & $G_\lambda^0$ & skein \\\hline
        0 & 0 & $E_6$ & 1+{\color{red}26} & $2A_2$ & $\bZ_3$ & $G_2$ & 1+{\color{blue}26}\\\hline
        $A_1$ & 0 & $A_5$ & 1+{\color{red}26} & $2A_2+A_1$ & $\bZ_3$ & $A_1$ & 1+{\color{blue}26}\\\hline
        $3 A_1$ & 0 & $A_1+A_2$ & 1+{\color{red}26} & $A_5$ & $\bZ_3$ & $A_1$ & 1+{\color{blue}26} \\\hline
        $A_2$ & $\bZ_2$ & 2$A_2$ & 8+{\color{red}208} & $E_6(a_3)$ & $\bZ_6$ & 0 & 8+{\color{blue}208} \\\hline
        $D_4$ & 0 & $A_2$ & \cellcolor{orange!25}1+{\color{red}26} & $E_6(a_1)$, $E_6$ & $\bZ_3$ & 0 & \cellcolor{green!25}2$\times$(1+{\color{blue}26})
    \end{tabular}
    \endgroup
\end{center}
One can check that all $G_\lambda$ for the magnetic phases indeed contain the $\bZ_3$ center needed for turning on a non-trivial 't Hooft flux, while every $C_\lambda$ in the electric phases contain a $\bZ_3$ subgroup. One might be concerned that there is one more electric phase as we have grouped two of them together in the last row (green). However, it turns out that they are indistinguishable from the $E_6(a_3)$ phase, and what one should really worry about is the excess of one magnetic phase (orange). 

\item Neutral phases. These are phases that can contribute to neither non-trivial $e$ nor $m$ sectors for the $T^3$ skein module. For $G=E_6$, such phases all flow to a free U(1) theory in the IR. They are all expected to contribute a one-dimensional space to $\cS$ in the $e=m=0$ sector. There are in total nine of them: 
\begin{center}
    \centering
    \begingroup
    \renewcommand{\arraystretch}{1.15}
    \begin{tabular}{c|c|c|c||c|c|c|c}
        $\lambda$ & $C_\lambda$ & $G_\lambda^0$ & skein & $\lambda$ & $C_\lambda$ & $G_\lambda^0$ & skein \\\hline
        $2A_1$ & 0 & $B_3+T_1$ & 1 & $A_4$ & 0 & $A_1+T_1$ & 1\\\hline
        $A_2+A_1$ & 0 & $A_2+T_1$ & 1 &  $A_4+A_1$ & 0 & $T_1$ &1\\\hline
        $A_2+2A_1$ & 0 & $A_1+T_1$ & 1 &  $D_5(a_1)$ & 0 & $T_1$ & 1\\\hline
        $A_3$ & 0 & $B_2+T_1$ & 1 & $D_5$ & 0&$T_1$ &\cellcolor{green!25}1\\\hline
        $A_3+A_1$ & 0 & $A_1+T_1$ & 1
    \end{tabular}
    \endgroup
\end{center}
In the table, $T_1$ represents a $\U(1)$ factor. We keep the same format as the electric and magnetic phases, although the only interesting information for a neutral phase is really just the $H_\lambda$ for each $\lambda$. Also, the $D_5$ phase is expected to be indistinguishable with the $D_5(a_1)$ phase, and we should only count one of them when constructing $\cS$.

    \item A very exceptional hybrid phase. Recall that these are phases with a non-abelian $K^\text{IR}_\lambda$ that can contribute to both non-trivial electric and magnetic sectors. We have seen these before for classical groups, where they always involve O(2) gauge groups in the IR. For $E_6$, one has a very special one---the HE phase $D_4(a_1)$ where the infrared gauge group is $K_\lambda=N_2:=\U(1)^2\rtimes S_3$ being the normalizer of a maximal torus of $A_2$.\footnote{One can similarly define $N_n:=N(T_{A_n})\simeq \U(1)^{n}\rtimes S_{n+1}$ the normalizer of the Cartan of $A_n$. Then $N_1=$ O(2).} Although many aspects of this phase is analogous to the O(2) cases, it is more complicated and more interesting with a non-abelian component group. The group $N_2$ can be viewed as either a subgroup of SU$(3)$ (from $E_6$) or, for the magnetic version, a subgroup of PSU$(3)$ (from $E_6/\bZ_3$) and, just like the O(2) case, it can contribute to $e\neq 0$ sectors in the former case while to $m\neq 0$ sectors in the latter case. The counting of $N_2$-connections on $T^3$ is studied in Appendix~\ref{app:NT}, which gives
\begin{center}
\begin{tabular}{c|c|c|c|c|c}
    \multicolumn{2}{c|}{} & \multicolumn{2}{c|}{$E_6$}
   & \multicolumn{2}{c}{$E_6/\bZ_3$}
   \\\hline
   $\lambda$& $C_\lambda$ & $K_\lambda$ & skein
   & $K_\lambda$ & skein
   \\\hline\hline
   $D_4(a_1)$ & $S_3$ & $N_2$ & 21+{\color{blue}104}
   & $N_2$ & 21+{\color{red}104}
\end{tabular}
\end{center}
\end{itemize}
Now, we can sum them all up to obtain
\begin{equation}
    \dim \sk(T^3,E_6)= 442=52+26\cdot{\color{blue}15},
\end{equation}
while 
\begin{equation}
    \dim \sk(T^3,E_6/\bZ_3)= 468=52+26\cdot{\color{red}16},
\end{equation}
The following table gives the refined count
\begin{center}
    \centering
    \begin{tabular}{c|c|c}
        dim $\cS^{e,m}$ & $e=0$ & $e\neq 0$ \\\hline
         $m=0$& 52 & {\color{blue}15} \\\hline
         $m\neq0$& {\color{red}16}   & 0 
    \end{tabular}
\end{center}

One can attempt to remedy this mismatch by arguing that the $D_4$ magnetic phase is also indistinguishable. In general, the overlapping phenomenon often happens when the same Bala--Carter Levi has several distinguished orbits (i.e.~multiple orbits share the same minimal Levi). For $E_6$, the phases labeled by $\lambda=E_6,\,E_6(a_1)$ and $D_5$ are expected to be of this type, being indistinguishable from $E_6(a_3)$ and $D_5(a_1)$. The first two, $E_6$ and $E_6(a_1)$ are electric phases both described by the same $\bZ_3$ gauge theory in the infrared, making them indistinguishable from a subsector of the $E_6(a_3)$ phase. The next one, $D_5$, is a Maxwell/neutral phase, but no Wilson line can possibly tell it apart from the other neutral phase $D_5(a_1)$. So far, this is in line with the observation that indistinguishable phases are often either electric or neutral.

However, the extra phase $D_4$ is a magnetic phase with $K_\lambda=\SU(3)$, and we have never seen before a magnetic phase being indistinguishable from one that is not magnetic. In general, the confinement of Wilson lines would make it detectable. In fact, even if one can find in the $D_4(a_1)$ phase a vacuum where any non-trivial SU(3) Wilson line will have zero vev and is therefore indistinguishable from the $D_4$ vacua, this does not solve the problem as the mismatch concerns the $m\neq 0$ sectors. The $D_4$ phase can contribute in each of the 26 non-zero $m$ and one can check that it can be distinguished from any linear combination of the vacua in the $D_4(a_1)$ phase with the same magnetic flux. One way to see this is that the $N_2$-holonomy along the 1-cycle dual to $w_2(N_2)$ will be trivial for any flat connections with the same $w_2$, making it very different from the $D_4$ phase. 

Previously, when we cannot match the skein modules on the two sides, we try to match the enriched version $\tilde\cS$. We won't attempt to perform the detailed analysis here, but a quick count suggests that simply separating the quantum vacua cannot restore the duality. Indeed, one would get additional electric contributions from $\lambda=2A_2,\,2A_2+A_1,$ and $A_5$, but these seem to be balanced by the additional magnetic contributions from $0,\,A_1,$ and $3A_1$. Another possibility is that adding back the 't Hooft and dyonic operators can make some of the indistinguishable electric phases distinguishable. Readers who lean toward this option might be encourage by the fact that Wilson lines always separate magnetic phases. Then shouldn't it be natural that 't Hooft/dyonic operators can also separate electric phases? Although this is tempting, it is not guaranteed by any electric-magnetic duality as we are in the $\cN=1$ world. Furthermore, we cannot find a way to separate either $E_6$ or $E_6(a_1)$ from a subsector of the $E_6(a_3)$ phase, because the centralizer for all of the three phases are trivial, and there do not exist co-characters of $G$ compatible with the $\fsl_2$-triples that we can use to define any non-trivial 't Hooft operators. We hope to return to this problem after achieving a better understanding of the ``enrich skein module'' $\tilde \cS$ in a future work.

\paragraph{$E_7$ and $E_8$.} For the remaining exceptional groups, we will not list all the nilpotent orbits, but instead group them according to their $C_\lambda$ and only give a total number per group. The result is in the table below: 
\begin{center}
    \centering
    \begin{tabular}{c|c|c|c|c|c|c}
        $C_\lambda$ & 0 & $S_2$ & $S_3$  &  $S_5$ &total& skein\\\hline
        $E_7$ & 16($-1$)+{\color{blue}16($-$5)} & 8+{\color{blue}3($-$2)} & 1+{\color{blue}1} &  ---  &45 &$\sim504=140+7\cdot{\color{blue}52}$\\
        $E_7/\bZ_2$ & 16($-5$)+{\color{red}16($-$1)} & 9($-2$)+{\color{red}2} & 1+{\color{red}1}  &--- &45 &$\sim588=140+7\cdot{\color{red}64}$\\
        \hline
        $E_8$ & 38($-13$) & 25($-10$) & 6$(-3)$ &   1&70 & $\sim500$\\\hline
        \hline
        $\left|\cM(T^3,C_\lambda)\right|$ & 1 & 8 & 21 &   206 & --- &--- \\
    \end{tabular}
\end{center}
For each type of $C_\lambda$, we have two numbers in the $E_7$ row, with the blue numbers counting the electric phases, whose $C_\lambda$ is the listed one plus the $\bZ_2$ center,\footnote{In other words, what being listed is the number of phases with the same component group $A(\cO_\lambda)$.} For the $E_7/\bZ_2$ row, the red number counts the magnetic phases. By comparing them, one can also get the combined number for neutral and HE phases---in fact all six of them are in the $S_2$ column. For reference, we have included the cardinality of the moduli space of flat $C_\lambda$-connections for the relevant $C_\lambda$ (see Appendix~\ref{app:DW}). The count for $E_7$ and $E_8$ is approximate, as we have made a collection of simplifying assumptions---which can in principle be checked and refined from data in \cite{LawtherTesterman2011}---in the computation for the dimension of skein module.

For $E_8$ the center is trivial, and all the phases are neutral. Some of them are indistinguishable, and the negative numbers in parentheses subtract the number of phases removed due to such (potential) overlap. The overlapping pattern seems somewhat non-trivial, and for the table above, we have assumed the following.
    \begin{itemize}
        \item Every other distinguished orbit in $E_8$ is indistinguishable from $E_8(a_7)$. This removes three phases with trivial $C_\lambda$, four with $C_\lambda=\bZ_2$, and three with $C_\lambda=S_3$.
        \item Every other distinguished orbit in the $E_7$ Levi is indistinguishable from $E_7(a_5)$. This removes three phases with trivial $C_\lambda$ and two with $C_\lambda=\bZ_2$.
        \item $D_7(a_1)$ and $D_7(a_2)$ are assumed to flow to O$(2)$ theories, which we also assume to have mutually indistinguishable identity component for their flat connections (e.g.~the same $\U(1)$ subgroup of $E_8$) but with two non-conjugate reflections. As these orbits are listed under $S_2$, we need to add $21\times2-1=41$ to the naive count of dimensions.
        \item For the rest, we have similarly assume that only the distinguished orbit with the largest $C_\lambda$ in a given Bala--Carter Levi contribute. This additionally removes 7 phases with trivial $C_\lambda$ and 4 with $C_\lambda=\bZ_2$, and most likely results in an undercount. We then attempt to (over-)compensate by assume that O$(2)$ arises in $K_\lambda$ whenever allowed, including for $D_5+A_2,\; A_4+2A_1,\; A_4+A_1,$ and $A_3+A_2$. 
    \end{itemize}

For $E_7$, the two issues are also present, and we have made similar assumptions:
\begin{itemize}
    \item The phases $\lambda=E_7,\,E_7(a_1),\,E_7(a_2),\,E_7(a_3),\,E_7(a_4),\,E_6,\,D_6,$ and $D_6(a_1)$ are indistinguishable from $E_7(a_5),\,E_6(a_3)$ and $D_6$. Among them, all except $E_6$ and $E_6(a_3)$ are electric.
    \item For the six phases with continuous part in $K^\text{IR}_\lambda$, four of them contain O(2) in the IR. This increases the count by $4\cdot21=84$.
\end{itemize}
This gives $504=140+7\cdot{\color{blue}52}$ for $E_7$, while $588=140+7\cdot{\color{red}64}$ for $E_7/\bZ_2$. Again, we see that the dimensions are different for the Langlands-dual pair. Now that there are five extra magnetic phases, this is significantly harder to explain away, compared with some of the previous cases, by amending the overlap analysis. 

Based on this observation, we think it is very likely that the Langlands duality for skein modules also fails to hold for $E_7$, and one again needs to include the 't Hooft operators to restore the duality for the enriched skein modules $\tilde\cS$.

\medskip

 This concludes our discussion about the skein modules for $T^3$. Notice that in the $M_3=T^3$ case, due to topological invariance, the only information we need about each phase is the infrared gauge group $K_\lambda^{\text{IR}}$ (plus information about the fate of the center of $H$ if we want the refined count). However, there is an important subtlety that invalidates this simplification when we generalize the computation for $\sk(T^3)$ to $\sk(\Sigma\times S^1)$, where an additional ingredient is needed.

\section{Cosmic strings}\label{sec:cosmic}

The $\cN=1$ deformation of the Vafa--Witten theory can be turned on for any Kähler four-manifolds locally, but there are two global issues, and both are associated with the fact that, after the topological twist, one of the three scalars---as well as the superpotential---becomes a $(2,0)$-form. 

While the $\cN=4$ superpotential
\begin{equation}
    W=\Tr T[U,V],
\end{equation}
is still well-defined, the $\cN=1^*$ deformation 
\begin{equation}
    W'=-\frac{m}{2}\cdot \Tr \left(T^2+U^2+V^2\right)
\end{equation}
no longer makes sense with a scalar parameter $m$. In fact, one cannot write down a non-degenerate mass matrix with only scalar entries for the three superfields, as its determinant also has to be a $(2,0)$-form. A very economical choice for the mass matrix is the one used in \cite{Vafa:1994tf} of the form,
\begin{equation}
   M= -\begin{pmatrix}
\omega & 0 & 0 \\
0 & 0 & \frac{m}{2} \\
0 & \frac{m}{2} & 0
\end{pmatrix},
\end{equation}
where $m$ remains a scalar while $\omega$ is a holomorphic $(2,0)$-form.

One issue that arises immediately is when the Kähler manifold has vanishing $H^{2,0}$ and hence no global holomorphic $(2,0)$-forms. Another point of view on this scenario is that any $\omega$ on such a manifold will have codimension-2 singularities with the local behavior
\begin{equation}
    \omega\sim z^{-n} dz\wedge dw
\end{equation}
for some positive integer $n$, which makes it hard to justify that correlation functions in the TQFT are continuous under the deformation (though one might still be able to get some mileage by incorporating such singularities as defects in the TQFT). 

When we consider the 4d $\cN=4$ theory on the geometry $\Sigma\times S^1\times \bR$, such an issue only arises when $\Sigma=\CP^1$ has genus one. Furthermore, in such a case, we expect that the skein module is one dimensional for any semisimple $G$, which agrees with the fact that the 4d theory on $S^2$ flows to a symmetric orbifold of a free theory. Although there is still an interesting question about torsions in the skein module of $S^1\times S^2$,\footnote{They are presumably related to twisted-sector states in the 2d theory that can be created by Wilson lines when $q$ is a root of unity. Alternatively, one can first compactify on $S^1$, and the additional point operators in the 3d effective theory that appear when $q$ is a root of unity can create states in the $S^2$ Hilbert space. When the 4d theory with the 3d boundary is lifted to a 6d system compactified on a ``solid torus'' (more accurately cigar times a circle), these operators come from multi-wrapped strings along the longitudinal direction. See \cite{Gaiotto:2023ezy} for a recent study of these operators and the implications for their existence.\label{foot:torsion}} we will not study this case further, but will later try to ``analytically continuate'' the answer for higher genus to the genus-zero case and check it against the expected answer.

For higher genera, the opposite problem arises: $\omega$ can be holomorphic globally, but will have zeroes where the mass matrix $M$ becomes degenerate. For a generic Kähler structure, such loci form codimenisonal-two ``cosmic strings'' (or ``superconducting vortices'') with the profile for $\omega$ given by   
\begin{equation}
    \omega\sim z^{n}dz\wedge dw
\end{equation}
for some positive $n$ in the neighborhood of such a string. In fact, generically, one might expect to be able to split  the charge-$n$ strings to obtain a collection of charge-1 strings.

In our setting, we want to preserve the product structure of the geometry, and construct an $\omega$ given by putting together two holomorphic 1-forms/differentials on respectively $\Sigma$ and $S^1\times \bR$. For the latter, there is a rather canonical choice given by $dw$ in the ``cylindrical coordinate'' $w\sim w+2\pi i$, which is constant and has no zero. For the $\Sigma$ factor, a generic choice of a holomorphic 1-form has $\chi(\Sigma)=2g-2$ simple zeros, and, after the mass deformation, there will be a collection of $2g-2$ points that support cosmic strings on $\Sigma$. This leads to a configuration invariant under the translation in the $S^1$ and $\bR$ directions.

One way to think about a holomorphic 1-form is that it gives $\Sigma$ the structure of a flat surface, i.e.~a flat metric except for conical singularities at its zeros (see \cite{Zorich2006Flat} for a review of this subject). While a physical cosmic string in the universe (if they exist) is often expected to have a deficit angle, the one at an order-$n$ zero of the holomorphic differential on $\Sigma$ actually has a surplus angle $2\pi n$. 

Using topological invariance, we can also work with a highly non-generic holomorphic 1-form/differential on a $\Sigma$ with a special modulus. In fact, we will coalesce all simple cosmic strings into a single one with charge $2g-2$. A theorem of Kontsevich and Zorich guarantees that this is always possible \cite{2003InMat.153..631K}. 

As part of the mass deformation is turned off at such a locus, it is natural to expect that there are massless degrees of freedom supported on the cosmic string, making the Hilbert space $\cH(\Sigma\times S^1)$ infinite dimensional even after turning on the deformation. One might be discouraged by this fact and conclude that the deformation cannot be used to solve the higher-genus cases. 

However, our proposal for the skein module, embedded into $\cH^\vee$ as a subspace $\cS$, can be finite dimensional even when $\cH$ itself is not. In fact, the very first examples we considered, lens spaces $L(p,q)$, are of this type. What concerns us is not the infinite-dimensional Hilbert space on the defect itself, but how Wilson lines in the bulk act on this Hilbert space.

\subsection{Bulk--defect coupling}

Among the two parameters in $M$, the scalar $m$ can still be made large throughout the geometry, and we have effectively a 4d $\cN=2$ theory.  The cosmic strings in such a theory is studied in \cite{Witten:1994ev}, where it was found that each simple string will carry a $(0,2)$ Fermi multiplet in the adjoint representation coupling to the gauge field in the bulk. 

In the bulk, similar to the $T^3$ case, there will be phases labeled by $\lambda$, which classifies conjugacy classes of $\fsl_2$-triples in $\frak g$ that describe the values of the three scalars away from zeros of $\omega$. The phases are again in bijection with nilpotent orbits, and in each phase, we have a collection of cosmic strings with massless fermions on their worldsheet. They transform in the adjoint representation $\frak{g}=\frak{k}\otimes \bC$ of the UV gauge group $K$, which becomes a typically reducible representation of $K_\lambda$. We will denote this representation as $R_\lambda$, even though the underlying vector space is isomorphic to $\frak g$ and independent of $\lambda$.

When we collide them to form a composite string, there will then be $2g-2$ free Fermi multiplet in $R_\lambda^{\otimes (2g-2)}$, and we will argue that its coupling to the bulk can be modeled on a Chern--Simons--WZW system with gauge group $K_\lambda$ at level $(2g-2)\cdot c_\lambda$ where $c_\lambda:=I(R_\lambda)$ is equal to the Dynkin index of $R_\lambda$.\footnote{Our convention is such that the adjoint of a simple group has $I$(adj) $ =h^\vee$ equals the dual Coxeter number. The fundamental representation of SU$(2)$ then has index $\frac{1}{2}$. When $G_\lambda$ is not simple, the index, and hence the level $(2g-2)\cdot c_\lambda$, is interpreted as an element of $H^4(\mathrm{B}G_\lambda,\bZ)$ (or, more precisely, [B$G_\lambda$,P$^4$BO]), which can be often conveniently parametrized by a collection of integers. Each integer is associated with a factor $K_\lambda^{(i)}$ being either a simple subgroup of $H_\lambda$ or a U(1) subgroup of $K_\lambda$. In both cases, the Dynkin index $c_\lambda^{(i)}$ is given by $h^\vee\cdot I({K_\lambda^{(i)}};H)$ with the second factor being the embedding index of $K_\lambda^{(i)}$ into $H$. Alternatively, for a $\U(1)$ subgroup given by a cocharacter $\gamma$, the level is given by $(g-1)\cdot\sum_{\alpha>0}\alpha(\gamma)^2$, with the sum over positive roots of $H$. We will explain this in more detail at a later point.}

The WZW part is not too surprising, as one can often bosonize free fermions into a WZW model. To see how the Chern--Simons theory arises and to determine the level, one can perform a reduction along the ``meridian circle'' $|z|=\varepsilon$.

\begin{figure}[htb!]
        \centering
        \includegraphics[width=0.85\linewidth]{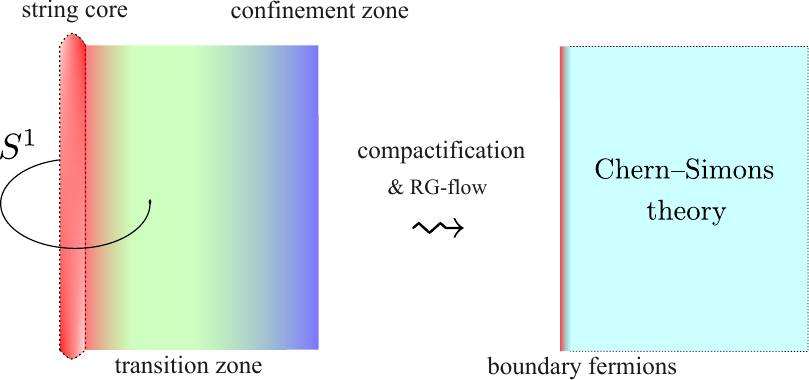}
        \caption{On the left, we illustrate the three regions with qualitatively different dynamics in the presence of a cosmic string. After compactifying on the meridian circle, the system at low energy can be viewed as a bulk Chern--Simons theory coupling to massless chiral fermions on the boundary.}
        \label{fig:string}
    \end{figure}

If we choose $\omega$ to have a steep profile near $z=0$ (i.e.~the coefficient $\omega\sim \kappa \cdot z^{2g-2}dz$ is sufficiently large), there are three qualitatively different regions in spacetime with different range of distances from the string:
\begin{itemize}
    \item Core of the string. This is when $z$ is very small and $\omega$ is approximately zero when compared with the length scale for $z$ due to the $(2g-2)$-power scaling. Another way of stating this is that the coefficient $\kappa$ is a dimensionful parameter that determines the size and ``fuzziness'' of the string. 
    \item Transition zone. This is when $z$ is large enough for us to assume that we are well-separated from the string, but small enough that confinement has not yet set in. The physics in this zone is described by a weakly coupled $\cN=1$ $K_\lambda$ theory. If $K_\lambda$ is already abelian, the coupling constant won't depend on the radius and this zone will actually extend to infinity. However, if $K_\lambda$ has a non-trivial non-abelian part $H_\lambda$, its coupling constant will increase with radius until one enters the next region.
    \item Confinement zone. This is when $z$ is large and we are below the confinement scale. The IR dynamics is then described by an $\cN=1$ $K^\text{IR}_\lambda$ gauge theory. 
\end{itemize}
The three zones are illustrated in the left part of Figure~\ref{fig:string}. 

When reducing along the meridian, we now have a boundary--bulk coupled system, and, in the transition zone, the bulk theory can be thought of as an $S^1$-compactification of the 4d $\cN=1$ theory. 

A usual compactification will not lead to a Chern--Simons term, but, in the present case, the mass parameter has a non-trivial profile along the circle,
\begin{equation}
    \omega(\varphi)\sim e^{2\pi i (2g-2)\varphi}dz,
\end{equation}
making the situation slightly different. One can cancel this phase with a chiral rotation
\begin{equation}
    T(\varphi) \mapsto e^{-2\pi i (g-1)\varphi}\cdot T(\varphi),
\end{equation}
but this has the effect of shifting the theta angle of the theory
\begin{equation}
    \theta\mapsto \theta+2\pi \cdot (2g-2)c_\lambda
\end{equation}
due to the fact that the superfield $T$ transforms in the $R_\lambda$ representation of $K_\lambda$ and the fermions in the multiplet will have $2I(R_\lambda)$ zero modes in an one-instanton background. In other words, once we come back to $\varphi=0$ along the circle, the theta angle of the 4d theory will be shifted. Therefore, the compactification on the circle will generate a Chern--Simons term in 3d from the $\theta$-term in 4d,
\begin{equation}
    \frac{1}{8\pi^2}\int_{M_3\times S^1}\theta\cdot\Tr  F_A\wedge F_A =\frac{\Delta\theta}{8\pi^2}\int_{M_3} \mathrm{CS}(A)=\frac{2(g-1)\cdot c_\lambda}{4\pi}\int_{M_3} \mathrm{CS}(A).
\end{equation}

One can also perform the analysis in the confinement zone. There, as the size of the circle is larger than the confinement scale, the monodromy can be thought of as inserting a domain wall at a point on $S^1$ interpolating between two massive vacua of the 4d $\cN=1$ theory at distance $2(g-1)\cdot c_\lambda$ away. Such walls are discussed in \cite{Acharya:2001dz} and indeed have Chern--Simons theory on the world-volume.\footnote{From this perspective, the cosmic strings are living on the boundary of these domain walls. Notice that, for each simple factor $H^{(i)}_\lambda$ of $H_\lambda$, the shift---hence the charge of the domain wall---is always a multiple of $2h^\vee\big(H^{(i)}_\lambda\big)$. Therefore, these domain walls can be ``absorbed by the rest of $\Sigma$.'' }

To fix an overall shift of the Chern--Simons level, one needs to be slightly more careful and take into account of several factors, including the shift introduced by integrating out the massive fermions in the chiral multiplet and the gauginos with the right choice of the framing of the circle. We claim that the level $2(g-1)\cdot c_\lambda$ is the correct one for the bosonic Chern--Simons theory (hence with the appropriate shifts when viewed as $\cN=1$ or $\cN=2$ super--Chern--Simons theories). This ensures that, for the trivial charge-0 string, we do get a trivial theory in the bulk.

Now we have a picture of how the Wilson lines in the 4d bulk act on the Hilbert space of the cosmic string---it is identical to the action of the Wilson lines in Chern--Simons theory on its WZW/free-fermion boundary. Therefore, the skein module is not rendered infinite dimensional by the massless modes on the string. Instead, one just needs to count the additional states that can be generated by the action of the independent Wilson lines, whose number is bounded by the number of weights for $K_\lambda$ at level $(2g-2)\cdot c_\lambda$.\footnote{We can now see the benefit of coalescing the cosmic strings into a single one---when there are multiple strings, one also needs to take into account configurations with some of the Wilson lines starting on one string and ending on another. In the end, the answer should be the same with an arbitrary complex structure on $\Sigma$ and an arbitrary holomorphic differential, but the present choice seems the simplest.} 

We now proceed to analyze the contribution of different phases, starting from the two extremes. 

\subsection{Extreme phases}

Starting from now, we will assume that $K=H$ is simple so that we don't need to keep track of multiple levels to begin with.

\subsubsection*{Regular orbit}

In the phase given by the principal embedding of the $\fsl_2$-triple, the gauge symmetry is broken to the center $Z(K)$. The representation $R_\lambda$ is trivial, as it comes from the adjoint representation of $K$ with a trivial action of the center. Therefore, one can completely forget about the cosmic strings as no Wilson lines can act non-trivially on their Hilbert spaces. Then one just has to count the flat $Z(K)$-connections on $\Sigma\times S^1$, arriving at $|Z(K)|^{2g+1}$.

The story would be analogous for phases labeled by other non-regular distinguished orbits in $\frak g$, for which we can first count the number of flat $C_\lambda$-connections. And just as in the $M_3=T^3$ case, there can be overlaps with other phases sharing the same Bala--Carter Levi and one would have to avoid overcounts when summing over different phases.

\subsubsection*{Zero orbit}

For the trivial $\fsl_2$-triple, the cosmic string has a more interesting coupling with the bulk, modeled by that between a Chern--Simons theory with gauge group $K_\lambda=K$ at level $k=(2g-2)\cdot h^\vee$ and a boundary WZW model.

Naively, one should count all independent Wilson lines. When $K$ is simply connected, this amounts to counting all integrable weights of $\hat{\frak{g}}_{k}$, and otherwise one should construct the line operators in the Chern--Simons theory from these weights by following the Moore--Segal procedure of discrete gauging \cite{MOORE1989422} (in short, one needs to take a subset, form a quotient, and then add ``twisted lines''). 

However, we are only interested in the line operators via their action on the Hilbert space of the cosmic string, and it is possible that certain non-trivial Wilson lines do not act in an interesting way. In fact, in a Chern--Simons theory, 
Wilson lines are also vortices or flux tubes, and these carrying a flux in the center of $K$ should then act trivially on the Hilbert space of the cosmic string. Therefore, we should only count the orbits under the center action. Similarly, for the non-simply-connected $K$, we should also identify the twisted sector lines, which, before discrete gauging, are junctions with the central flux tubes. In other words, we count modulo the $Z(K)$ 1-form ``fusion'' symmetry and the $\pi_1(K)$ 0-form symmetry acting on line operators. Therefore, in the end, we should just count weights of ${K}_{(2g-2)h^\vee}$ modulo the center action.

\subsubsection*{2d perspective}

This can also be understood more directly from the action of the Wilson lines on the $S^1$ Hilbert space of the string. The Hilbert space has a decomposition,
\begin{equation}
    \cH_\text{string}\simeq \bigoplus_{\nu\in \Lambda_k(\frak g)}V_\nu\otimes M_\nu,
\end{equation}
which is the branching decomposition from the conformal embedding 
\begin{equation}\label{Conformal}
    \hat{\frak g}_{k=(2g-2)h^\vee}\subset \fsl\big((2g-2)\dim\frak g\big)_1.
\end{equation}
The first part, $V_\nu$, is an irreducible module of the affine $\hat{\frak{g}}_{k}$ symmetry associated with the highest weight $\nu$, while $M_\nu$ is a multiplicity space controlled by the coset model of \eqref{Conformal}. As the fermions are in the adjoint, only dominant weights $\nu\in \Lambda_{\mathrm{rt},k}(\frak g)\subset \Lambda_k(\frak g)$ in the root lattice $Q(\frak g)$ can appear.\footnote{We use $P$ and $Q$ to denote the weight and root lattices, and use $P^\vee$ and $Q^\vee$ for the co-weight and co-root lattices. Elements of $Q$ and $Q^\vee$ are referred to as root(-lattice) and co-root(-lattice) vectors, with the ``-lattice'' part often omitted to avoid clutter. The set of level-$k$ weights is the subset $\Lambda_k(\frak g)\subset P(\frak g)$ in the fundamental alcove of the affine Weyl group given by the condition $\langle\mu,\alpha\rangle\ge0$ for any positive root $\alpha$ as well as $\langle\mu,\vartheta\rangle \le k$ for the highest root $\vartheta$. Here, the angular bracket denote the Killing form on $\frak t^\vee$, the dual of the Cartan of $\frak g$.}

\begin{figure}[htb!]
    \centering
    \includegraphics[width=0.9\linewidth]{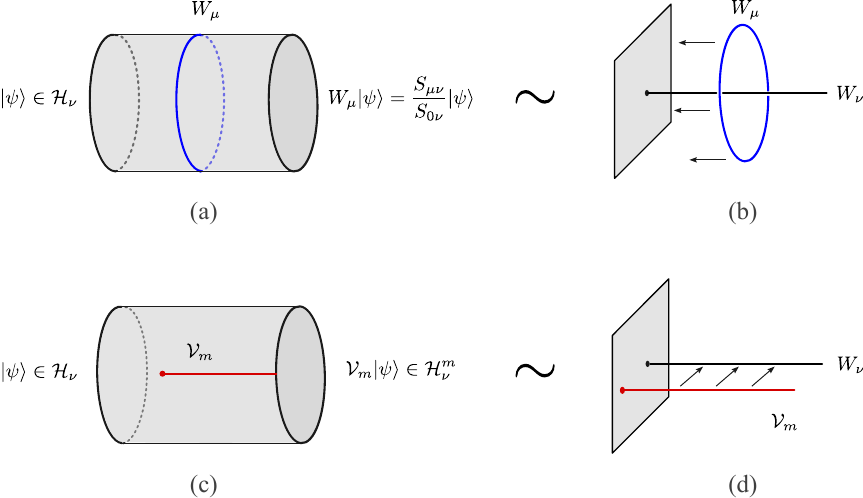}
    \caption{Two actions of operators on the Hilbert space of the cosmic string (grey) on $S^1$. In (a), a Wilson/Verlinde line $W_\mu$ (blue) is inserted along an $S^1$. Its action on any state in the sector $\cH_\nu$ is given by the scalar multiplication with $\frac{S_{\mu\nu}}{S_{0\nu}}$. This can be understood from the bulk perspective illustrated in (b), where, in the radial quantization, a state in $\cH_\nu$ is created by a point operator attached to the bulk Wilson line $W_\nu$. The action of $W_\mu$ is then given by its braiding with $W_\nu$. The sub-figure (c) depicts the action of a boundary monopole operator (red dot), coming from the lasso action of an 't Hooft operator labeled by a co-character $m$ of the gauge group $K$. This operator is on the end point of an ``'t Hooft vortex line''  $\cV_m$ (red) carrying a magnetic flux, mapping the initial state to a twisted sector. From the bulk perspective illustrated in (d), the action can be understood as the fusion of $W_\mu$ with the vortex line $\cV_m$.}
    \label{fig:braiding}
\end{figure}

When a Wilson line $W_\mu$ is inserted along $S^1$, it is a Verlinde line from the 2d perspective, whose action is diagonal on $V_\nu$ by a phase factor \cite{Verlinde1988},
\begin{equation}
    W_\mu \ket{\psi}=\frac{S_{\mu\nu}}{S_{0\nu}}\cdot \ket{\psi}, \quad \text{for $\ket\psi\in\cH_\nu\simeq V_\nu\otimes M_\nu$},
\end{equation}
with $S$ being the modular $S$-matrix acting on $\bC[P_k(\frak g)]$. This action is illustrated in the top part of Figure~\ref{fig:braiding}. When $\eta \in \Lambda_k(\frak g)$ represents a generator of the center 1-form symmetry, $S_{\eta\nu}=S_{0\nu}$ for $\nu \in \Lambda_{\mathrm{rt},k}(\frak g)$ gives a trivial action. Furthermore, for arbitrary $\mu$, we have $S_{\eta\cdot\mu,\nu}=S_{\mu,\nu}$ and we will only be able to distinguish $Z(\frak g)$-orbits in $\Lambda_k(\frak g)$, due to the lack of states in $\cH_\text{string}$ that are sensitive to the action of the center of $K$.

\subsubsection*{The counting problem}

To describe this set of orbits more concretely, one can consider dominant weights of $\frak g$ at level $k$ viewed as the subset $\Lambda_k(\frak g)\subset P(\frak g)$ in the fundamental alcove. The center 1-form symmetry is generated by a subset that we denote as $Z(\frak g)\subset \Lambda_k(\frak g)$. These are weights living on certain corners of the alcove that are mapped to a fundamental co-weight by $\vev{\cdot,\cdot}/k$. The fusion action of $Z(\frak g)$ on $\Lambda_k(\frak g)$ is by shifting (and then bringing back to the alcove by affine reflections). The line operators also carry charges under the 1-form symmetry, given by the braiding phase with central lines in $Z(\frak g)$. The charge can be computed using the pairing with the corresponding fundamental co-weight mod $\bZ$. The level $k$ being a multiple of $h^\vee$ ensures that the fusion preserves the charges, which, in the 3d TQFT, corresponds to the statement that the 1-form symmetry doesn't have an 't Hooft anomaly. 

One last problem to deal with is about remaining linear relations between Wilson lines when they act on $\cH_\text{string}$, since if two lines $W_\mu$ and $W_{\mu'}$ act in the same way, they would not be distinguishable. We claim that there is no additional relations---equivalently, the ``asymmetric $S$-matrix,'' with one index restricted to roots and the other modded out by the $Z(\frak g)$ action is still non-degenerate. The justification for this is that the two operations are $S$-dual of each other, and the asymmetric $S$-matrix can be interpreted as the basis change that diagonalizes the fusion rules in the 2d TQFT obtained from first reducing the 3d Chern--Simons theory with simply-connected gauge group and then only gauge the 1-form (or only the 0-form) symmetry. Notice that the analogous statement also holds when we only gauge a subgroup. See \cite{Gukov:2021swm} for related discussions.

Now we can layout our prescription for finding the contribution of the cosmic string to the skein module in this phase. Given a compact global form $K$ associated with $\frak g$, we take weights that are neutral under $\pi_1(K)\subset Z(\frak g)$, and then count the orbit of the $Z(\frak g)$ action. Equivalently, one can take the ($S$-)dual perspective and count the number of $\pi_1(K)$-orbits of root-lattice vectors in $\Lambda_k(\frak g)$. It would be a consistency check that they give the same count.

For any given $G$ and genus $g$, it is not difficult to count ``by hand'' in simpler cases and by a computer program in more complicated cases. When $\frak g$ is $G_2$, $F_4$ or $E_8$, the center is trivial, and the problem reduces to the counting of level-$k$ weights, which is given by a generating function. For each series of classical groups, the orbit count can be done uniformly. When $\frak g=\fsl_N$, we study this in Appendix~\ref{app:Weights} with the answer given by \eqref{GammaA} as a divisor sum for $N$. However, we also want to obtain the refined count, which we analyze next.

\subsubsection*{Refinements}

For $M_3=T^3$, the confining phase given by the trivial $\fsl_2$-triple only has non-trivial magnetic refinements. However, the situation is now different, as the confinement is ``turned off'' near the cosmic string. In the above procedure, the Wilson lines of $K$ being counted are charged under the $Z(K)$ 1-form symmetry, giving each a character in $Z(K)^\vee\subset Z(\frak g)^\vee$. Therefore, each can contribute to a sector labeled by an electric flux $e\in Z(K)^\vee\subset H_1(\Sigma\times S^1,Z(K)^\vee)$ along the $S^1$ direction.

One can also take the dual perspective and consider the quotient of $\cH_\text{string}$ given by $\bC\left[\Lambda_{\mathrm{rt},k}(\frak g)\right]$. Then $H^1(S^1,Z(\frak g))\simeq Z(\frak g)$ acts on this space by fusion---or equivalently just as multiplication by the center when viewing $\Lambda_{\mathrm{rt},k}(\frak g)$ as a subset of the Cartan of the simply-connected cover of $K$ modulo the Weyl-group action---decomposing it into characters. When $K$ is not simply connected, one should also identify root-lattice vectors up to the action of $\pi_1(K)$. After this, the grading will be given by the characters of $H^1(S^1,Z(K))$, which acts on $\pi_1(K)$-orbits. 

While this takes care of the electric refinements, we still have sectors with non-trivial magnetic fluxes, and we claim that, when the magnetic flux is entirely through $\Sigma$ (e.g.~along the string), the count is unaffected. This is again due to the fact that the degrees of freedom on the cosmic string cannot detect the center of the group. Another way to think about this is that the magnetic flux sectors can be generated by insertion of 't Hooft operators, labeled by co-characters of $K$ that are not in the co-root lattice. In the relative theory, they lead to ``non-genuine'' line operators in the sense that they actually live on the boundary of topological surface operators, which are exactly generators for the $\pi_1(K)$ 1-form magnetic symmetry and carry non-trivial 't Hooft fluxes. Therefore, inserting such a non-genuine line operator at $S^1\times \{t=0\}$ will modify the background magnetic flux by the class dual to the $\pi_1(K)$-valued 1-cycle supported on $S^1$. However, when descending to the Chern--Simons theory, this becomes a line operator with central monodromy, with trivial effect on the cosmic string Hilbert space. We then arrive at the prescription given earlier. This is analogous to the top part of Figure~\ref{fig:braiding} but with the $W_\mu$ replaced with a central vortex line $\cV_m$ (i.e.~$\mu$ is on the corner of the Weyl alcove equal to $k$ times a co-character $m$).

Notice that, in the $e=0$ and $m=0$ sector, as we simply count the number of $Z(\frak g)$-orbits of root vectors, the result does not actually depend on the global form of $K$ or $G$. This ensures the consistency of the lift to the relative theory. However, the prescriptions for sectors with $e\neq 0$ and $m\neq 0$ appear at first to be rather different, making it an interesting question whether adding the cosmic string spoils the electric-magnetic duality.

\subsubsection*{String--bulk selection rule}

On the electric side (simply-connected $G$), we denote this total count as $\Gamma(\frak g;k)$, with $k=(2g-2)h^\vee$, and the refinement by the central characters as $\Gamma^r(\frak g;k)$ with $r \in Z(\frak g)^\vee$. For $r=0$, it gives the number of $Z(\frak g)$-orbits of root vectors. The refined count with any global form $K$ and arbitrary $e$ and $m$ can be constructed from them as a linear combination by summing over $r\in Z(\frak g)^\vee$ in the pre-image of $e\in Z(K)^\vee$, whereas the count doesn't depend on $m\in \pi_1(K)$.

 However, there is one remaining piece that we need to find $\cS$ when $K$ is not simply connected. We have seen in the above that the line operators can contribute to sectors with non-trivial magnetic flux along $S^1$, but how about more general background $m$? Should the cosmic string also modify the story there by also contributing to these sectors? One way to think about this issue is to separate the total flux as $m=m_\text{string}+m_{\text{bulk}}$, with $m_\text{string}$ and $m_\text{bulk}$ the components in the two parts of $H^2(\Sigma\times S^1,\pi_1(K))\simeq H^2(\Sigma,\pi_1(K))\oplus H^1(\Sigma,\pi_1(K))$. Previously, what we have considered is the $m_{\text{bulk}}=0$ case and how the different states generated by the action of Wilson lines on the string contribute to $m$ (there we simply denote $m_\text{string}$ as $m$ when focusing on the cosmic string). Now we need to understand the effect of turning on $m_{\text{bulk}}$.

 We claim that when $m_{\text{bulk}}$ is non-trivial, the allowed states in $\cH_\text{string}$ are significantly reduced, leading to a smaller number of independent Wilson-line insertions along the string. To see this, recall that one way to generate a state in a more general flux sector is to insert an 't Hooft loop along a non-trivial 1-cycle $\gamma$ in $\Sigma\times S^1$ with a topological surface operator stretched from it in the $\bR^+$ direction. 
  
 For a state $\ket{\psi}\in\cH_\nu$ on the string to survive in the presence of such a flux tube, topological invariance should require that the ``lasso action'' of the surface operator on the string state is trivial, as one can move the cosmic string along $\Sigma$ to the other side of $\gamma$ without crossing it (at least when $\gamma$ project to a simple closed curve on $\Sigma$ representing  a non-trivial 1-cycle). 

 This condition, when phrased within the Chern--Simons--WZW system, is that $\ket\psi$ also lives in the twisted sector with the insertion of a central lines, which is the compactification of the topological surface operator on the meridian circle. This requires that $\nu$ is a fixed point of $m_\text{bulk}$. One can understand this statement by generating the state in $\cH_\nu$ in radial quantization via an insertion of a primary at the origin. Then the Wilson line attached to it, which is also labeled by $\nu$, must be invariant under the fusion action of $m_\text{bulk}$. This is illustrated in the bottom part of Figure~\ref{fig:braiding}. 
 
 Notice that this is, in a sense, the $S$-dual of  the effect of turning on $m_\text{string}$. For $m_\text{string}$, the action is via braiding with the flux tube, which is trivial as $\nu$ is a root-lattice vector, allowing $m_\text{string}$ to be turned on without any constraints. However, the lasso action of $m_\text{bulk}$ is via the fusion action, which requires $\nu$ to be a fixed point, leading to a drastic reduction of allowed states in $\cH_\text{string}$.

 Unlike $m_\text{string}$, which is automatically $\pi_1(K)$-valued, $m_\text{bulk}$ takes values in $H^1(\Sigma,\pi_1(K))$, and the action is only defined once we choose a 1-cycle in $H_1(\Sigma,\bZ)$ to pair with $m_\text{bulk}$ via the cap product. However, we want the condition to hold for any choice of the 1-cycle, and this combined condition only depends on the \emph{type} of $m_\text{bulk}$---the minimum subgroup $F_\text{bulk}$ of $\pi_1(K)$ for which $m_\text{bulk}$ is in the image of 
 \begin{equation}
     H^1(\Sigma,F_\text{bulk})\to H^1(\Sigma,\pi_1(K)).
 \end{equation}
 
For generic $m_\text{bulk}$, its type $F_\text{bulk}$ is the entire $\pi_1(K)$, and we should only count root vectors fixed by the entire $\pi_1(K)$-action.  On the other hand, if $m_\text{bulk}$ is non-generic, $F_\text{bulk}$ can be a proper subgroup of $\pi_1(K)$, there can be more weights $\nu$ that survive in $\cH_\text{string}$ and hence $\cS^\vee$, allowing there to be more independent Wilson lines in $\cS$. Although it is rather straightforward to completely classify types of $m_\text{bulk}$ and the corresponding fixed root vectors for simple Lie groups, we will not do it here, as we will be content to be able to deal with either simply-connected gauge groups or PSU$(N)$ with prime $N$. For the latter, only one non-trivial $\nu$ at the center of the Weyl alcove survives when $m_\text{bulk}$ is non-trivial.

\subsubsection*{Asymptotic behavior}

Now, one can in principle already find the contribution of the cosmic string in this phase to skein modules of $\Sigma\times S^1$ for any reductive $G$ by reducing the problem to the counting of weights and their orbits. We give the answer for the $A_n$ case in Appendix~\ref{app:Weights} and will resist the temptation to go through all the other simple groups. In general, the number of weights scales with the volume of the fundamental alcove, and is expected to be a polynomial in $g$ of degree rank$(H)$ (or a ``quasi-polynomial''---a collection of polynomials labeled by congruence classes of $g$),
\begin{equation}
    \Gamma(G;g)\sim \#\cdot g^{\text{rank}(H)}+\ldots
\end{equation}
The refined count $\Gamma^\gamma(G;g)$ will have a similar behavior, but with the leading coefficient divided by $|Z(\frak g)|$.

This polynomial growth is a non-TQFT-like behavior and what exactly prevents the dimensions of skein modules to be expressible as the partition function of a (reasonably well-behaved) TQFT. Notice that the Hilbert space $\cH(\Sigma\times S^1)$ of the topologically twisted $\cN=1$ theory, though infinite dimensional, behaves as the Hilbert space of a TQFT, as the part coming from $2g-2$ cosmic strings grows as $\cH_\text{1-string}^{\otimes (2g-2)}$. In particular, if the single-string Hilbert space $\cH_\text{1-string}$ were finite dimensional, the dimension of the tensor power would be $\big(\!\dim \cH_\text{1-string}\big)^{2g-2}$. However, the way that the skein module is embedded in this Hilbert space is such that the TQFT behavior is lost. We hope that this perspective can shed light on other instances where the Betti version or other versions of the geometric Langlands correspondence involves non-TQFT-like features.

We now illustrate this procedure for incorporating cosmic strings for $G=\SL(2)$ and PSL(2), which can already be solved at this stage before the discussion about intermediate phases, as they are absent in these cases.

\subsection{Example: $\fsl(2)$}

We first start with the $K=\SU(2)$ case.

For the regular embedding ($\bZ_2$ phase), there is almost no modification compared with the $T^3$ case except that the $\bZ_2$ gauge theory now has $2^{2g+1}$ flat connection and they all can be distinguished from each other. Again, as $\bZ_2$ is part of the center of $G$, the degree of freedom on the cosmic string decouples, and won't contribute to dimension of $\cS$. 

To refine this count, notice that such flat connections are in bijection with $H^1(M_3,\bZ_2)$, and the collection of Wilson lines has to populate all electric-charge sectors in $H_1(M_3,\bZ_2)$ to separate them. Therefore, this phase contributes a one-dimensional space to $\cS^e$ for all $e$.

For the confinement phase given by the zero $\fsl_2$-triple, the entire SU$(2)$ is unbroken classically, and the degrees of freedom on the cosmic string are coupled to the bulk, with non-trivial actions of Wilson lines. Although the Hilbert space $\cH(\Sigma\times S^1)$ becomes infinite dimensional due to the massless modes on the cosmic string, we only need to count Wilson lines of the SU$(2)_{4g-4}$ theory modulo the center action. Weights of $\fsl(2)$ are labeled by integers, and the level-$k$ dominant weights are these in the Weyl alcove $[0,k]$. 

\begin{figure}[htb!]
    \centering
    \includegraphics[width=0.85\linewidth]{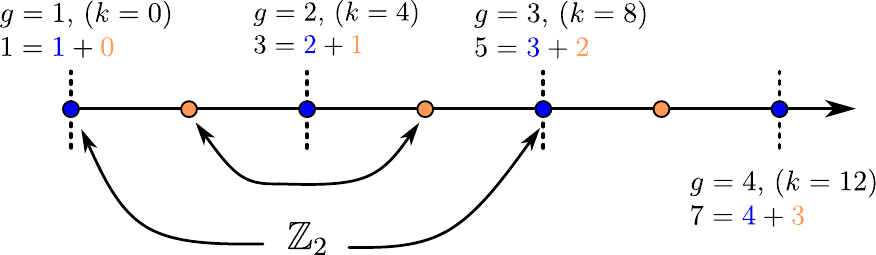}
    \caption{For $\fsl(2)$, the dominant weights at level $k$ are labeled by non-negative integers $\{0,1,\ldots , k\}$. The two colors for the weights in the figure reflect whether the central character is trivial (blue) or non-trivial (orange). The $\bZ_2$ center acts by a reflection, along axes depicted by dashed lines for various $k$ in the figure. The action for the $g=2$ (hence $k=4$) case on weights is explicitly illustrated. To obtain $\Gamma^0$ (or $\Gamma^1$) for a given $k$, one just counts blue (or orange) dots from zero until the reflection axis.}
    \label{fig:su2}
\end{figure}

The $\bZ_2$ center action on a weight $\mu$ is given by the reflection along the point $k/2$, sending $\mu\mapsto k-\mu$. As a consequence of this symmetry, we should only count weights from 0 to $k/2=2g-2$, giving us in total $2g-1$ weights. This is illustrated in Figure~\ref{fig:su2}.

Putting the two phases together, we have 
\begin{equation}
    \dim \cS=2^{2g+1}+2g-1,
\end{equation}
recovering the results of \cite{gilmer2019skein,detcherry2021basis}.

To obtain the refined count, notice that a weight $\mu$ contributes to a sector with $e$ along the $S^1$ direction given by $\mu \bmod 2$. Then the $2g-1$ weights will be separated into two classes, with $g$ weights in the neutral sector and $g-1$ weights in a non-trivial sector.

For $K=\SO(3)$, the principal embedding of the triple gives a trivial vacuum, while the magnetic $\bZ_2$ gauge symmetry in the confinement phase giving the $2^{2g+1}$ generators for $\cS$, one for each value of $m$. When $m_\text{bulk}\neq 0$, the string--bulk selection rule is only satisfied by the unique $\bZ_2$-invariant root vector (on the dashed line in Figure~\ref{fig:su2}). Therefore, there are no additional states from the cosmic string in these $2^{2g+1}-2$ sectors. For the two sectors with $m_{\text{bulk}}=0$, we also include the additional string states given by the SO(3) integrable weights (blue in Figure~\ref{fig:su2}) for both the $m_\text{string}=0$ and $1$. This adds additional $g-1$ states to each of them, making each $g$ dimensional.

We organize the refine counts for the two cases in the table below:
\begin{center}
    \centering
    \begin{tabular}{c|c|c|c|c|c|c}
       & phases & $(0,0)$ & $(\alpha,0)$ & $(0,\beta)$ & $(\alpha,\beta)$ & phase contribution \\\hline
       \multirow{2}{2.3em}{SL(2)} & E & 1 & 1 & 1 & 1 & $2^{2g+1}$\\
        &M & $g$ & $g-1$ & 0 & 0 & $2g-1$\\\hline
         \multirow{2}{3em}{PSL(2)} & E & 1 & 0 & 0 & 0 & 1\\
        &M & $g$ & $g$ & 1 & 1 & $2^{2g+1}+2g-2$\\\hline\hline
        \multicolumn{2}{c|}{classes}  & 1 & 1 & $2^{2g}-1$ & $2^{2g}-1$ & $2^{2g+1}$\\\hline
        \multicolumn{2}{c|}{total} & $g+1$ & $g$ &$2^{2g}-1$  & $2^{2g}-1$ & $2^{2g+1}+2g-1$ \\
    \end{tabular}
\end{center}
Here, $\alpha$ and $\beta$ denote non-zero elements in the two parts of the decomposition 
\begin{equation}
    H_1(\Sigma\times S^1,\bZ_2)\simeq H_0(\Sigma,\bZ_2)\oplus H_1(\Sigma,\bZ_2).
\end{equation}
We use them to label both electric and magnetic fluxes. The count doesn't depend on the choice of $\beta$ as any non-zero values of $\beta$ are related by the Sp$(2g,\bZ_2)$ action. The ``classes'' row gives the number of homological classes for each type.

Although the phase-wise duality observed for the $M_3=T^3$ case stops to hold, we have equalities between SL(2) and PSL(2) for each charge sector once the two phases are combined. It is interesting that the difference in the prescriptions for counting the electric and magnetic contributions of the cosmic string ``cleverly'' compensates for the difference of the bulk contributions to preserve the Langlands duality.\footnote{A useful way to think about the duality is by regarding one of the $g+1$ cosmic-string states in the $e=0$ sector on the SL(2) side as the bulk contribution in the magnetic phase. (On the PSL(2) side, we in fact have already taken this perspective, only viewing the $g-1$ states in the each of the two $m_\text{bulk}=0$ sectors as contributions of the cosmic string, even though there are actually $g$ states in total.) After this reshuffling, the bulk contributions enjoy phase-wise duality while the contribution of the cosmic string is self-dual.\label{foot:reshuffle}}

\subsection{Example: $\fsl(3)$}

For SU(3) or PSU(3), there is an additional phase labeled by the sub-regular/minimal orbit $\lambda=[21]$, with the IR dynamics given by a free U(1) theory. This is the simplest ``intermediate phase,'' which we will try to better understand in order to tackle more general ones. Before that, we first go through the two phases given by the regular and zero triples.

\subsubsection*{Two extreme phases}

The story for the bulk contributions is very similar to the $T^3$ case (or, in fact, also to the SL(2) and PSL(2) cases), with the electric phase $\lambda=[3]$ giving $3^{2g+1}$ generators for $\cS(\Sigma\times S^1,\SL(3))$ but only 1 for $\cS(\Sigma\times S^1,\mathrm{PSL}(3))$, while the magnetic phase $\lambda=[1^3]$ giving 1 generator for SL(3) while $3^{2g+1}$ for PSL(3).

\begin{figure}[htb!]
    \centering
    \includegraphics[width=0.85\linewidth]{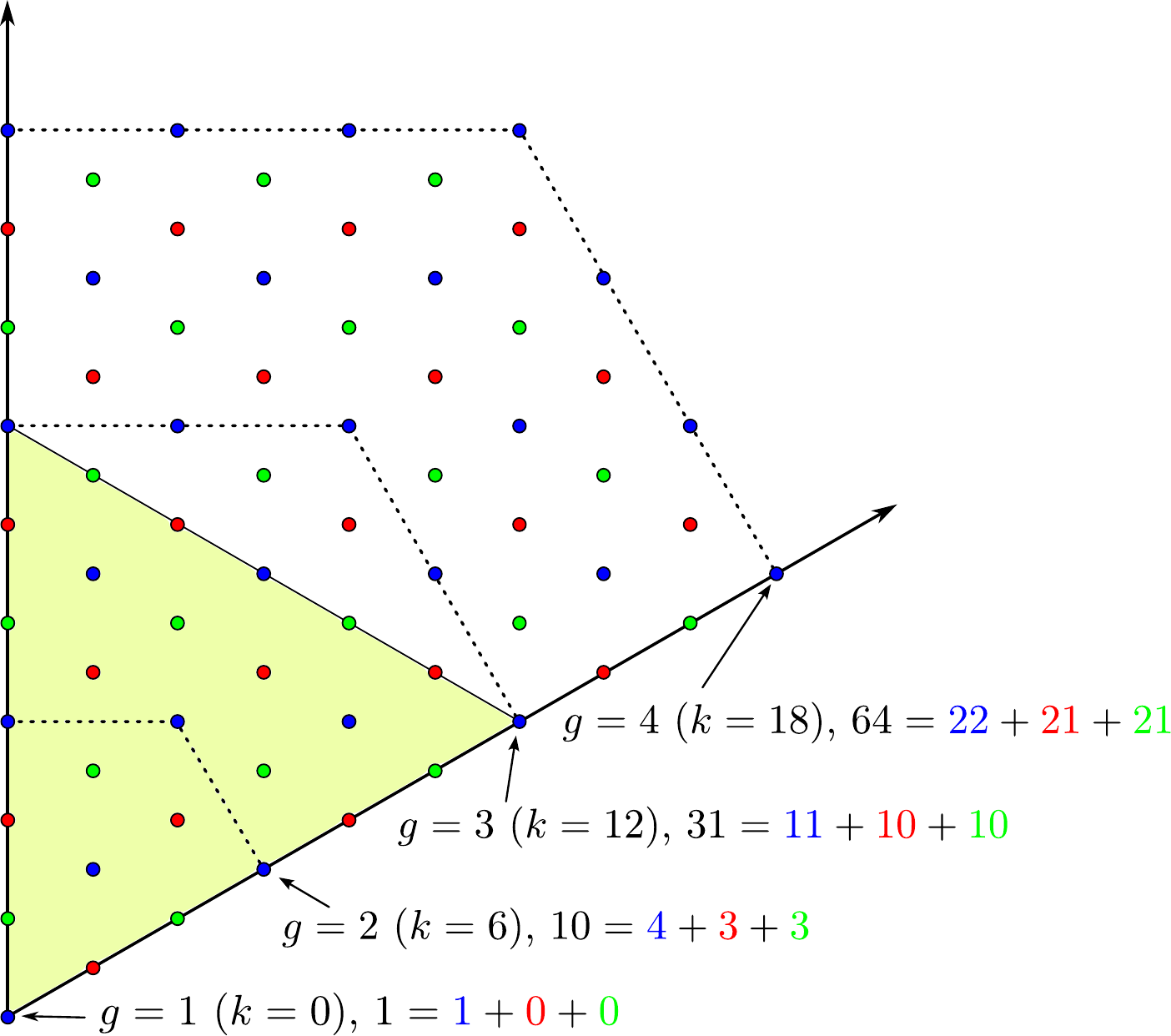}
    \caption{In the Weyl chamber of $\fsl(3)$ depicted above, the level-$k$ weights are in the equilateral triangle known as the level-$k$ Weyl alcove (yellow region in the figure for $k=6$). The $\bZ_3$ center acts on the alcoves by rotations, and the fundamental domains are the kite-shaped regions between the two dashed lines (shown for $k=6(g-1)=6,12,18$) and the axes. The $\bZ_3$-classification of the weights by the central characters are illustrated by colors: blue ($r=0$), green ($r=1$) and red ($r=2$). We give the counts, with the $\bZ_3$ refinement, for the number of orbits for $g=1,2,3$ and $4$. For the neutral phase, the weights of the U(1)$_{3k}$ Chern--Simons theory is on the boundary of the alcove, causing them to be counted again. The corners are in a sense counted one more time in the electric phase.}
    \label{fig:su3}
\end{figure}

As for the cosmic string in the magnetic phase, we then need to count weights of $\fsl(3)$ at level $k=6(g-1)$ up to the action of the center. This is done systematically for general $N$ in Appendix~\ref{app:Weights}, but the present case is simple enough and one can count ``by hand,'' illustrated in Figure~\ref{fig:su3}.

At level $k$, the weights can be labeled by a triple $(\mu_0,\mu_1,\mu_2)\in\bZ^3_{\ge0}$ that sums to $k$. There are in total
\begin{equation}
    \binom{k+2}{2}=\frac{(k+2)(k+1)}{2}=(3g-2)(6g-5)
\end{equation}
such triples. The center $\bZ_3$ acts as cyclic permutations,
\begin{equation}
    c: \quad (\mu_0,\mu_1,\mu_2) \mapsto (\mu_2,\mu_0,\mu_1).
\end{equation}
The action is almost free, except for the weight $(\frac k3,\frac k3,\frac k3)$. Therefore, the total number of orbits are
\begin{equation}
    \Gamma(\fsl(3);6g-6)=\frac{(3g-2)(6g-5)-1}{3}+1=3(2g-1)(g-1)+1.
\end{equation}
To have a refined count, notice first that the central character is
\begin{equation}
    r=\mu_1+2\mu_2 \bmod 3,
\end{equation}
which is preserved by $c$ as $3|k$. It is easy to see that there are equal numbers of weights with $r=1$ and $r=2$ but one more for $r=0$. Therefore we have 
\begin{equation}
    \Gamma^{r}(\fsl;6g-6)= \begin{cases}(2g-1)(g-1)+1, & \text{$r=0$,}\\
    (2g-1)(g-1), & \text{$r=1,2$.}\end{cases}
\end{equation}
Again, after substracting the 1 in $\Gamma^0$ (cf.~Footnote~\ref{foot:reshuffle}), the cosmic string gives the same contribution to SL(3) and PSL(3), becoming independent of both $e$ and $m$ along the string. Although their numbers agree, notice that conceptually there is an asymmetry: for SL(3), the electric background $e_{\text{string}}$ along $S^1$ comes from the electric charge (the central character $r$) of the Wilson line inserted along the string, while for PSU(3) the magnetic flux $m_\text{string}$ is related to the $S^1$ holonomy of the $\bZ_3$ magnetic gauge field which lives in the bulk.

\subsubsection*{Neutral phase}

The phase labeled by $\lambda=[21]$ is a neutral phase according to our classification, with seemingly uninteresting physics. However, the cosmic string is non-trivially coupled to the bulk gauge field, as the adjoint of $\fsl(3)$ is charged under this $K_\lambda=\U(1)$ subgroup. For $K= \SU(3)$, this subgroup is parametrized by $\theta\in \bR/2\pi\bZ$ as diag$\{e^{i\theta},e^{i\theta},e^{-2i\theta}\}\in \SU(3)$ while the periodicity is reduced by a factor of 3 if we consider the $\mathrm{PSU}(3)$ version of the story. 

For such a subgroup, one has the relation between Chern--Simons levels $k_{\U(1)}=3k_{\SU(3)}=18(g-1)$ with the ratio 3 being the embedding index.\footnote{This factor can be worked out by taking the 8-dimensional adjoint representation of SU$(3)$ and decomposing it as a representation $R_\lambda$ of U(1). It then becomes two charge-3, two charge-$(-3)$ and four trivial representations of this U(1) subgroup. The embedding index is then given by sum of squares of the charges divided $4h^\vee_{\SU(3)}$.} As a consequence, the string--bulk coupling is modeled on that between a U(1) Chern--Simons theory at level $18(g-1)$ and free fermions on the boundary. 

This Chern--Simons theory has $18(g-1)$ independent Wilson lines, and they are exactly the weights on the boundary of the $\fsl$(3) Weyl alcove for level $k_{\SU(3)}=6(g-1)$. To see that they all come from line operators in the UV theory---aside from the fact that we have literally embeded all of them in the $\fsl(3)$ weight lattice---it might be easier to take the perspective of vortex lines, as they will not become a superposition when breaking SU(3) to U(1).\footnote{Here, we implicitly used the fact that we can turn on the vevs of scalars to go between the phases. In other words, the zero orbit is in the closure of the minimal one.} To generate all vortex lines of the U(1) theory, we only need to consider lines in the SU(3) theory with monodromies in the U(1) subgroup. Using $k^{-1}$ times the Killing form $\vev{\cdot,\cdot}$, we can map them to $\frak t^\vee$ and they are exactly the weights on the boundary of the Weyl alcove. For SU(3) at level $k>0$, there are $3k$ of them indexed by elements of $\bZ_{3k}$, and the center $\bZ_3$ of SU(3) is now a subgroup generated by a $k$-shift. However, the level $k=0$ is special, as one still has the trivial weight (in other words, $\bZ_0$ still has one element and is not empty). Therefore, the total contribution of this phase is $k=6(g-1)$ for $g>1$, but $1$ for $g=1$, which indeed reduces to the $T^3$ case.

Putting the three phases together, we have
\begin{equation}
    \dim \sk (\Sigma\times S^1;\SL(3))=3^{2g+1}+3(2g+1)(g-1)+1+\delta_{g,1}.
\end{equation}
Interestingly, even though our computation does not cover the $g=0$ case, the formula still correctly gives
\begin{equation}
    \dim \sk (S^2\times S^1;\SL(3))=1.
\end{equation}
When $g=0$, the ``contribution'' of the neutral phase is $-6$, which indeed indicates that the $\cN=1$ deformation in the $g=0$ case is a large deformation, in the sense that the Hilbert space of the deformed theory is not smoothly connected with the original one, but perhaps related by certain differential or spectral sequence.

\subsubsection*{Refinement}

For the neutral phase, the refinement is again given by the $\bZ_3$ central characters of the U(1) weights, and we have $2(g-1)$ generators for each charge except that when $g=1$ we only have one generator in the neutral sector. 

Combining the other phases then gives the following refined counts for SL(3):
\begin{center}
    \centering
    \begin{tabular}{c|c|c|c|c|c}
        phases & $(0,0)$ & $(\alpha,0)$ & $(0,\beta)$ & $(\alpha,\beta)$ & phase contribution \\\hline
        E & 1 & 1 & 1 & 1 & $3^{2g+1}$\\\hline
        N & $2g-2+\delta_{g,1}$ & $2g-2$ & 0 & 0 & $6g-6+\delta_{g,1} $\\\hline
        M & $2g^2-3g+2$ & $2g^2-3g+1$ & 0 & 0 & $6g^2-9g+4$\\\hline\hline
        classes  & 1 & 2 & $3^{2g}-1$ & $2(3^{2g}-1)$ & $3^{2g+1}$\\\hline
        total & $2g^2-g+1+\delta_{g,1}$ & $4g^2-2g$ &$3^{2g}-1$  & $2(3^{2g}-1)$ & $3^{2g+1}\!+\!6g^2\!-\!3g\!-\!2+\!\delta_{g,1}$ \\
    \end{tabular}
\end{center}
Here, $\alpha$ and $\beta$ denote arbitrary non-zero elements in the two parts of (the dual of) $H_1(\Sigma\times S^1,\bZ_3)\simeq H_0(\Sigma,\bZ_3)\oplus H_1(\Sigma,\bZ_3)$. The count doesn't depend on the choice as $\alpha=1$ and $2$ are related by charge conjugation, while any non-zero values of $\beta$ are related by the Sp$(2g,\bZ_3)$ action on the first homology. 

For PSL(3), we have
\begin{center}
    \centering
    \begin{tabular}{c|c|c|c|c|c}
        phases & $(0,0)$ & $(\alpha,0)$ & $(0,\beta)$ & $(\alpha,\beta)$ & phase contribution \\\hline
        E & 1 & 0 & 0 & 0 & 1\\\hline
        N & $2g-2+\delta_{g,1}$ & $2g-2$ & 0 & 0 & $6g-6+\delta_{g,1} $\\\hline
        M & $2g^2-3g+2$ & $2g^2-3g+2$ & 1 & 1 & $3^{2g+1}+6g^2-9g+3$\\\hline\hline
        classes  & 1 & 2 & $3^{2g}-1$ & $2(3^{2g}-1)$ & $3^{2g+1}$\\\hline
        total & $2g^2-g+1+\delta_{g,1}$ & $4g^2-2g$ &$3^{2g}-1$  & $2(3^{2g}-1)$ & $3^{2g+1}\!+\!6g^2\!-\!3g\!-\!2+\!\delta_{g,1}$ \\
    \end{tabular}
\end{center}
Here, we keep the same label for $H^2(\Sigma\times S^1,\bZ_3)\simeq H_1(\Sigma\times S^1,\bZ_3)$. The counting is mostly straightforward, with the only point worth explaining about being the fact that the neutral phase can also contribute to non-trivial $m_\text{string}$ sectors but not to non-vanishing $m_\text{bulk}$. This is because, although turning on a central flux carrying a non-zero first Chern class causes energy in the 4d theory with an abelian gauge group, it becomes a topological line operator in the 3d Chern--Simons theory when reducing on the meridian $S^1$ of the cosmic string, and therefore can be turned on along (and only along) the cosmic string. Note that the contribution of the cosmic string is independent of $m_\text{string}$ in both the neutral and magnetic phase, except for the $g=1$ case where it doesn't actually exist. 

Again, the duality is not completely phase-wise (i.e.~between $\lambda$ and $\lambda^T$), but (after a slight reshuffling) phase-wise for the bulk part, while the cosmic string contributions are self-dual in both the neutral and magnetic phases.

For $K=\SU(3)$, notice that the total contribution of the string is given by first, in the magnetic phase, counting the $\bZ_3$-orbits for all weights in the alcove and then counting, in the neutral phase, the orbits of weights on the boundary. In a sense, the trivial sector for the electric phase can be viewed as the contribution of the vertices of the alcove which has only one orbit. Therefore, the total cosmic string contribution can be reorganized by weights (as opposed to by phases), but counted with multiplicities. This perspective is particularly useful for $A_{N-1}$ with $N$ a prime, which we will study later. Before that, we will first give a general description for the ``intermediate phases'' (e.g.~all non-extreme ones).

\subsection{Intermediate phases}

For a general intermediate phase, we will have a gauge group $K_\lambda\subset K$. There are two ways to determine the Chern--Simons level(s) that controls the bulk--string coupling:
\begin{itemize}
    \item One can start in the phase with $K_\lambda=K$, where the Chern--Simons level is $(2g-2)h^\vee$, and turn on the vev of the scalars to arrive at the phase $\lambda$. After breaking $K$ to $K_\lambda$, to express the Chern--Simons term in the usual normalization for $K_\lambda$, we should compare the Killing form on $\frak{k}\subset \frak g$ with that for each simple and $\U(1)$ factor of $K_\lambda$. 

    \item In the phase labeled by $\lambda$, the 4d $\cN=1$ chiral multiplet couples to $K_\lambda$ via the representation $R_\lambda\simeq\frak{g}$. The Chern--Simons level is determined by $(g-1)$ times the number of fermionic zero modes from the chiral multiplet on the one-instanton background. 
\end{itemize}
The two methods give the same answer, as they are essentially just two different ways of computing the Dynkin index of the embedding $K_\lambda\subset K$. For both approaches, in the end, we just need to rewrite the bilinear form given by $\Tr_{\!\frak g}$ in terms of factors of $K_\lambda$. For a simple factor, the two are related by the embedding index,
\begin{equation}
    \frac{1}{2h^\vee}\left.\Tr_{\!\frak{g}}\right|_{H_\lambda^{(i)}}=\frac{{I(H,H_\lambda^{(i)})}}{2h^\vee(H_\lambda^{(i)})}\cdot \Tr_{\frak{h}_\lambda^{(i)}}.
\end{equation}
And the Chern--Simons level is then
\begin{equation}
k_{H_\lambda^{(i)}}= I(K|H_\lambda^{(i)})\cdot k_K,
\end{equation}
where we have taken into account that the trace in the adjoint gives a level equal to twice the dual Coxeter number.

The story is slightly more subtle for U(1) factors, as the canonical Killing form is degenerate, and we need to pick a bilinear form and be consistent with the normalization. With the choice compatible with the ``usual'' normalization of the Chern--Simons level (i.e.~integrating out a massive fermion shift the U(1) Chern--Simons level by half of the charge squared and U(1)$_\text{odd}$ are spin TQFTs), we claim
\begin{equation}
    k_{\U(1)}=I(U(1);K)\cdot k_K
\end{equation}
with the embedding index
\begin{equation}
    I(U(1)_\gamma;K)=\frac1{4h^\vee}\sum_{\alpha\in \Phi(\frak g)} \alpha(\gamma)^2.
\end{equation}
Here, $\gamma$ is a cocharacter of $K$ that specifies the U(1) subgroup, and we used the fact that the weights of $\frak{g}$ as a U(1)$_\gamma$-representation are given by the pairing $\alpha(\gamma)$ between $\gamma$ and the roots of $\frak g$. With $k_K=2(g-1)h^\vee$, we have 
\begin{equation}
    k_{\U(1)_\gamma}=(g-1)\cdot\sum_{\alpha>0} \alpha(\gamma)^2.
\end{equation}
This is the same as the Chern--Simons level for this U(1) generated from integrating out $g-1$ copies of 3d $\cN=2$ chiral multiplet in the representation $R_\lambda$. 

Putting all of these together, we write the level for $K_\lambda$ as $k_\lambda=2(g-1)\cdot c_\lambda$ and refer to $c_\lambda$ as the Dynkin index of $R_\lambda$, even though they are in fact both collections of numbers.

Now one can compute the contribution of the cosmic string, $\Gamma(K_\lambda;2(g-1)c_\lambda)$ and its refinement. In general, this seems to be a complicated task as one needs to pay attention to many details, such as the global form of $K_\lambda$, to get the correct count. However, the weights being counted in each phase $\lambda$ form a subset of these of $K$---we are interested in Wilson lines for the $K$ gauge theory after all---and it is useful to organize the counting using geometry of Weyl alcove of $K$ by giving it a stratification according to multiplicities of weights.  

After completing this count, one needs to combine the cosmic-string contribution with the bulk contribution, and then finally combine all the phases together. In this process, one needs to take into account the following issues:
\begin{itemize}
    \item When we organize the count by weights, we also need to remember the bulk factor $|\pi_0\cM(\Sigma\times S^1; K^\text{IR}_\lambda)|$.
    \item For non-simply-connected $K$, the string--bulk selection rule is in effect for the intermediate phases as long as $\pi_1(H_\lambda)\to \pi_1(K)$ is non-trivial, similar to the $\lambda=0$ phase studied before. It selects a subset of root-lattice vectors fixed by $F_{\text{bulk}}$---the type of $m_\text{bulk}$. An alternative way to think about this is that, for a given phase $\lambda$ and a root vector $\nu$ representing a $\pi_1(K)$-orbit of root vectors, there is a maximal $F_{\text{bulk}}\subset \pi_1(K)$ in the image of $\pi_1(H_\lambda)$ that is compatible with $\mu$, and this orbit contributes to sectors labeled by $m_{\text{bulk}}\in H^1(\Sigma, F_{\text{bulk}})$. 
     
    \item We also need to ``duplicate'' the cosmic-string contribution for non-trivial values of $m_\text{string}$ in the image of $\pi_1(K_\lambda)$ in $\pi_1(K)$. Notice that the abelian part of $K_\lambda$ can give rise to non-trivial $m_\text{string}$ but not non-trivial $m_{\text{bulk}}$ as we have explained in the PSU$(3)$ case.

    \item In addition, both components of $m$ can arise from the non-abelian nature of $K^\text{IR}_\lambda$ or from the non-abelian central extension of $C_\lambda$. For the latter, the computation in Appendix~\ref{app:DW} about flat connections of D$_8$ and Q$_8$ should serve as the first step toward a more in-depth understanding of this issue for $B$- and $D$-series, with the next step being to refine the computation to obtain the counts in each flux sector. In particular, the total magnetic flux $m_\text{string}$ along $S^1$ actually has two parts which are combined via the group extension
    \begin{equation}
        \mathrm{im}(\eta_\lambda)\to \pi_1(K)\to \mathrm{coker}(\eta_\lambda),
    \end{equation}
    obtained via the map
    \begin{equation}
        \eta_\lambda:\quad \pi_1(H_\lambda)\to \pi_1(K).
    \end{equation}
    
    \item For phases that are indistinguishable, they should remain indistinguishable after including the cosmic strings. Notice that such phases would necessarily have the same $H_\lambda$, as otherwise they can be distinguished by confinement patterns. This might be viewed as another advantage of organizing the count by weights, as it might offer an easier way to avoid overcounts.

    \item When the holonomy of the continuous part of $K^\text{IR}_\lambda$ (e.g.~when it is O(2) or $N_2$) is non-trivial along $S^1$, only certain compatible $\cH_\nu$ subsectors survive. These are labeled by $\nu$ which, when viewed as a $K^0_\lambda$-valued monodromy of the corresponding vortex line, agrees with the $K^\text{IR,0}_\lambda$-holonomy after passing to its abelianization. 

    \item For the discrete part $C_\lambda$ of $K^\text{IR}_\lambda$, the holonomy in the subgroup $\mathrm{im}(\iota_\lambda)$ coming from the center $Z(K)$ via
    \begin{equation}
        \iota_\lambda:\quad  Z(K)\to C_\lambda
    \end{equation}
    is shifted by $H^1(S^1,Z(K))$ and contributes to sectors with $e_\text{string}\in H_1(S^1,Z(K)^\vee)$. This is in addition to the action of $\ker(\iota_\lambda)\subset K_\lambda^0$ on root-lattice vectors. Therefore, just like $m_\text{string}$, the electric flux $e_\text{string}$ along $S^1$ also has two parts, one in $\mathrm{im}(\iota_\lambda)^\vee$ and another in $\ker(\iota_\lambda)^\vee$  combined via the dual of the group extension
    \begin{equation}\label{iotaExtension}
        \ker(\iota_\lambda)\to Z(K)\to \mathrm{im}(\iota_\lambda).
    \end{equation}
This might sound complicated, but as the vortex lines relevant for the phase $\lambda$ is a subset of these for $K$, they are represented as root-lattice vectors for $K$ with the action of the entire $Z(K)$. Therefore, one does not actually need to compute and combine the two parts---the action of $Z(K)$ on this set already contains all the information that one needs.    
    On the other hand, there can be a more interesting interplay between the component group $A(\cO_\lambda)$ with $\cH_\text{string}$ similar to the previous point. 
    
     \item One can ask whether there is a similar selection rule on the electric side. When the holonomies of $C_\lambda$ along $\Sigma$ are central, they are not expected to modify $\cH_\text{string}$. However, when $A(\cO_\lambda)$ has a non-trivial holonomy, one needs to take into account the non-trivial action of $A(\cO_\lambda)$ on the Hilbert space of the string. It is natural to expect that, for a sector labeled by $\nu$ in $\cH_\text{string}$, it can only survive in the presence of non-trivial holonomies if it is fixed by the action of $A(\cO_\lambda)$ (or, in fact, the subgroup generated by all holonomies) on the root vectors. These can include ``complex conjugation'' for $A$-type or $E_6$ subalgebras of $\frak g_\lambda$, spinor--co-spinor duality for $D_{n\ge5}$ factors, triality for $D_4$ factors, permutation of isomorphic subalgebras, etc. On the other hand, the holonomy of $A(\cO_\lambda)$ along $S^1$ does not seem to have a similar effect and can be included without any modification for $\cH_\text{string}$. 
\end{itemize}

It would be interesting to revisit the last point (after also including 't Hooft and dyonic operators when $H_\lambda$ is non-trivial) to see whether this system naturally relates to a myriad of interesting topics in geometric representation theory centered around the action of the component group. However, our final goal for this section is to understand the SL$(N)$ case, where $C_\lambda$ is always a quotient of $Z(K)$ with $A(\cO_\lambda)=0$ and this problem does not arise. 

\subsection{Generators}

This algorithm gives us a set of generators for (the dual of) the skein module of $\Sigma\times S^1$ as triples $(\lambda,\alpha,\nu)$, with:
\begin{itemize}
    \item $\lambda$ an nilpotent orbit that labels an IR phase of the theory. We remove phases that are indistinguishable from subsectors of other phases.
    \item $\alpha\in \pi_0(\cM(\Sigma\times S^1; K^\text{IR}_\lambda))$ a component of the moduli space of flat $K^\text{IR}_\lambda$-connections. When $K$ is not simply connected, we need to additionally remember the value of $m_\text{bulk}$, which we view as part of the data of $\alpha$ (e.g., when $K^\text{IR}_\lambda=0$, $\alpha$ is just $m_\text{bulk}$). We would not need to keep this datum if we were simply listing generators for a given sector $\cS^m$. However, we are constructing generators for $\cS=\bigoplus_m\cS^m$, many of which are identical except for the $m_\text{bulk}$ part. 
    \item A root-lattice vector $\nu$ at level $k_\lambda=2(g-1)\cdot c_\lambda$ (or, dually,  a central orbit of weights) of $\frak g_\lambda$. It has to be compatible with the $K^\text{IR}_\lambda$-holonomy along $S^1$. When $K$ is not simply connected, $\nu$ represents a $\pi_1(K)$-orbits and has to be compatible with $m_\text{bulk}$, satisfying the string--bulk selection rule. Also, we need to remember the value of $m_\text{string}$, which we view as part of the data of $\nu$. In the end, the triple is a generator for $\cS^m$ with $m=m_\text{bulk}+m_\text{string}.$
\end{itemize}

The first two of the three labels are not of the usual type used in skein theory, but it is not hard to convert them by choosing a set of Wilson lines that distinguish different components $\alpha$ in different phases. This part is similar to the discussion for $M_3=T^3$. Now, to combine with the label $\nu$ one needs to add Wilson lines in the $S^1$ direction to some of the previous generators, and it is not immediately clear which of the previous generators one should choose to ``enlarge.'' 

This difficulty reflects the fact that the above generators actually live in the dual of $\cS$. After all, our present approach starts by embedding skein modules into the dual of the Hilbert space of the 4d $\cN=4$ theory. Therefore, the following procedure is more natural. One starts with a larger set of generators on the skein side. This can be taken to be the set that distinguishes all $(\lambda,\alpha)$ pairs, and for each configuration in this set, one enlarges it to a set of $|\frak g_{(2g-2)\cdot h^\vee}|$ ones by including a Wilson line along $S^1$ with all possible weights of $\frak g$ at this level. They can be chosen to be of the form that some Wilson lines wind cycles along $\Sigma$ but are constant on $S^1$, while the rest wind the $S^1$ fiber alone. Many of these configurations are redundant, since only in the $\lambda=0$ phase does one need all these Wilson lines labeled by the weights of $\frak g$. Therefore, the last step is to eliminate such redundancy by pairing with the generators for $\cS^\vee$ given above, for which we need to be slightly more explicit about the pairing. 

Due to confinement, many of the configurations give 0 in a phase with non-trivial $H_\lambda$, and it is reasonable to expect that a sufficient condition for this to happen is the following. Consider a simple closed curve $\gamma$ in $\Sigma$. It determines an embedded $T^2$ given by $\gamma\times S^1$, which, in the presence of Wilson lines, has a collection of marked points colored by representations of $\frak g$. The tensor product of them is also a representation of $\frak {g}$, and hence $\frak h_\lambda$, and one can ask whether it has an invariant subspace. And the condition is that, for some choice of $\gamma$, there is no $\frak h_\lambda$-invariant subspace. When this happens, due to confinement and topological invariance, the correlation function is expected to be zero, as one can stretch the neighborhood of the $T^2$ into a long cylinder. Notice that the same condition does not apply for the surface $\Sigma$ when $g>2$, or any other embedded higher-genus surfaces, due to the cosmic strings. 

For a configuration whose pairing with $(\lambda,\alpha,\nu)$ is not automatically vanishing due to confinement, one can then evaluate it by keeping only the $\frak h_\lambda$-invariant part along $\Sigma$, which is a representation of $K^\text{IR}_\lambda$. This part is not too different from the $T^3$ case. What is different now concerns Wilson lines along the $S^1$ direction.\footnote{One can also introduce Wilson loop that wrap $S^1$ and some cycle in $\Sigma$ at the same time. However, using skein relations, one expects that we do not need to consider them. 
}

In the $\lambda=0$ phase, we discussed this evaluation which is given by the ratio $\frac{S_{\mu\nu}}{S_{0\nu}}$ for a Wilson line label by the level-$k$ weight $\mu$ on a state in $\cH_\nu\subset\cH_\text{string}$. This still makes sense in a more general phase labeled by $\lambda$. However, we also need to take into account of the holonomy of $C_\lambda$ along the circle. For this purpose, it is useful to view $\nu$ as describing the monodromy of a vortex line. Concretely, one can first view it as an element of $\frak{t}$ via the bilinear form $\vev{\cdot,\cdot}/k$, which can then be exponentiated to obtain a conjugacy class of $K_\lambda^0$. This combined with the data of the flat $C_\lambda$-connection determines a conjugacy class of $K_\lambda$-holonomy on the circle, which can be paired with the representation of $K$ (and hence also of $K_\lambda$) to produce a number via the trace map.

We now get the full pairing, which we can use to find a complete basis for $\cS$ via linear algebra by identifying two skeins $\bra L\sim \bra{L'}$ if they pair identically with all $(\lambda,\alpha,\nu)$. Notice that, when $K$ is not simply connected, the Wilson lines will have to be restricted to these of $K$, but one also needs to keep the information about $m_\text{string}$ and $m_\text{bulk}$. The linear relations to be imposed are within each $m$ sector only.

The above generators labeled by the triple are expected to span the dual of $\cS$ and, if no surprises arise---some of the possible scenarios for them to appear were mentioned previously---one should also expect that they are independent and form a basis for the dual space. We will first assume that there are no additional relations, but will revisit this assumption in Section~\ref{sec:checks} when comparing with the geometry of Hitchin moduli spaces. 

Now, we can determine the dimension of $\cS$---or at least provide an upper bound---by simply counting the number of dual generators, which has been implicitly our strategy used in many of the previous examples for $M_3=T^3$. Generalizing them to $\Sigma\times S^1$ is not difficult for $G$ with smaller ranks but becomes quickly tedious and complicated once the rank becomes larger.

Our focus later will be on the $A$-series, where it is possible to get some mileage even for general rank. But before that, we will first work out one more example with the group being $G_2$, where we will see some features not present for the $A$-series.

\subsection{Example: $G_2$}

\begin{figure}[htb!]
    \centering
    \includegraphics[width=0.5\linewidth]{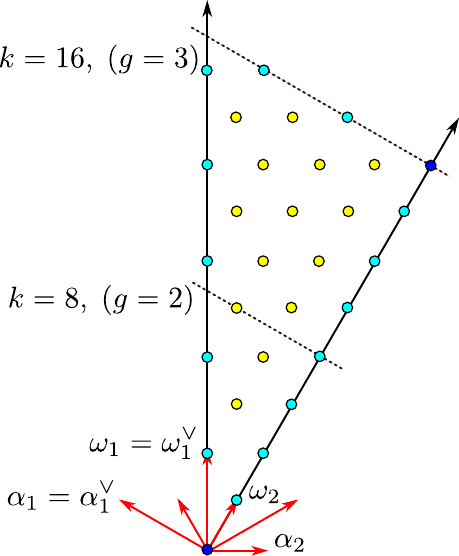}
    \caption{The Weyl alcove of $G_2$ and the integrable weights are illustrated for level $k=8$ and $16$. The red arrows denote positive roots, with the dominant long and short roots coinciding with the fundamental weights. One simple co-root and one fundamental co-weight are not labeled in the figure, but they are given by $\omega_2^\vee=3\omega_2$ and $\alpha_2^\vee=3\alpha_2$. The highest root and co-root decomposes as $\theta=\omega_1=2\alpha_1+3\alpha_2$ and $\theta^\vee=\omega_2^\vee=3\alpha_1^\vee+2\alpha_2^\vee$. If one combines the cosmic-string contribution of the three phases with non-trivial $H_\lambda$, weights in the interior of the alcove (yellow) is counted once, while those one the edges are counted twice (teal) with two special ones counted three times (blue).}
    \label{fig:g2}
\end{figure}

The Weyl alcove for $G_2$ is half of an equilateral triangle shown in Figure~\ref{fig:g2}. The highest co-root decomposes as $\theta^\vee=3\alpha_1^\vee+2\alpha_2^\vee$, and the integrable weights at level $k$ are parametrized by $(\mu_1,\mu_2)\in\bZ_{\ge0}^2$ satisfying
\begin{equation}
    \vev{\mu,\theta^\vee}=3\mu_1+2\mu_2\le k.
\end{equation}
The dual Coxeter number is $h^\vee=4$, and, therefore, we are interested in the cases with the level $k=8g-8$ divisible by 8.

The phases labeled by the principal ($\lambda=G_2$) and the sub-regular ($\lambda=G_2(a_1)$) orbits do not have a non-trivial $K_\lambda^0$ and hence no contribution from cosmic strings. Furthermore, the $\lambda=G_2$ phase is indistinguishable from the trivial connection in the  $\lambda=G_2(a_1)$ phase. Therefore, the combined contribution of the two phases is given by the number of flat $C_{G_2(a_1)}\simeq S_3$ connections on $\Sigma\times S^1$, which is found to be 
\begin{equation}
    |\cM(\Sigma\times S^1,S_3)|= 2\cdot 6^{2g-2}+13\cdot 3^{2g-2}+ 6\cdot 2^{2g-2}
\end{equation}
 in Appendix~\Ref{app:DW}.

 For the other three phases, the discrete part of the gauge group is trivial. Therefore, we only need to take into account the contribution of the cosmic string.  

\paragraph{Phase with $\lambda=0$.} One sums over all the level-$k$ weights of $G_2$. A simple computation yields the answer as a quasi-polynomial, 
\begin{equation}
    \Gamma(G_2;k)=\begin{cases}\frac{(k+2)(k+4)}{12} & \text{if }24\nmid k\\\frac{k(k+6)}{12}+1
    &\text{if } 24|k.
    \end{cases}
\end{equation}
In term of $g$, this is 
\begin{equation}
    \Gamma(G_2;8g-8)=\begin{cases}\frac{(4g-3)(4g-2)}{3} & \text{if } g\equiv 0,2 \pmod 3\\\frac{(4g-4)(4g-1)}{3}+1
    &\text{if } g\equiv 1 \pmod 3,
    \end{cases}
\end{equation}
or, more compactly, $\frac{1}{3}(16g^2-20g+6+\delta_{g-1\bmod 3})$, where the last term is a delta function that is 1 if $3|(g-1)$ and zero otherwise.

\paragraph{Phases with $\lambda=A_1$ and $\lambda=\tilde A_1$.} For both of them, $K_\lambda=\SU(2)$, but they are non-conjugate subgroups of $G_2$ given by either a long or short root. The embedding index can be computed to be 1 and 3 respectively. Therefore, the SU(2) Chern--Simons theory in the two cases has level $8g-8$ and $24g-24$. The Wilson lines of the two theories are on the boundary of the Weyl alcove, with one set aligned with the dominant short root, while for the other phase they are aligned with the dominant long root until hitting the corner and taking a turn to continue along the affine wall. Although they have different levels, they contain the same number of $G_2$ weights: $\frac k2+1=4g-3$ for each factor. The two sets overlap at two points, $\mu=(0,0)$ and $\mu=(0,4g-4)$.

\medskip
Combining all phases, we have
\begin{equation}
    \dim \cS(\Sigma\times S^1;G_2)=2\cdot 6^{2g-2}+13\cdot 3^{2g-2}+ 6\cdot 2^{2g-2}+\frac{1}{3}(16g^2+4g-12+\delta_{{g-1\bmod 3}}).
\end{equation}
Similar to the SU(3) case, we have a ``correction term,'' but it is now a different type with a different origin. The expression can be more concisely written in terms of $\chi=2g-2$ as 
\begin{equation}
\dim \cS(\Sigma\times S^1;G_2)=2 \cdot 6^{\chi} + 13 \cdot 3^{\chi} + 6 \cdot 2^{\chi} +
\frac{1}{3}(4\chi^2 + 18\chi + 
8+\delta_{{\chi} \bmod 3}).
\end{equation}

For small values of $g=0,1,2,3$ and $4$, we have this formula giving $-1$, $24$, $233$, 3789 and 103260, with exponential growth dominated by the $6^{2g-2}$ term for larger $g$. When $g=1$, this gives back our previous result for $M_3=T^3$. 

What is more curious is the value for $g=0$---it no longer gives 1 (which is the expected value). Of course, there is no reason for the formula to be valid for $g=0$, where the deformation cannot be turned on, and unlike the $A_1$ and $A_2$ cases, naively extrapolating it does not give the correct answer anymore for $G_2$. Perhaps one should be encouraged by the fact that $-1$ is quite close to the expected answer, as the discrepancy could be due to certain minor corrections either to our analysis or in the special situation of $g=0$.\footnote{When comparing the formula for $\dim \cS$ with geometry of Hitchin moduli spaces in Section~\ref{sec:checks}, we observe that phases with $H_\lambda$ that has a non-trivial center not part of $H$ can lead to mismatches. Here, both $\lambda=A_1$ and $\tilde A_1$ can potentially have this problem. If we are allowed to use the center of $K_\lambda=\SU(2)$ in one of the two phases to eliminate $2g-2$ generators, then we will have the dimensions given by 
\begin{equation}
    1,\; 24,\; 231,\; 3785,\; 103254,\; ... 
\end{equation}
instead, with expected answer for $g=0$. However, we do not see how the additional relations for the generators arise in this case.} 

\subsection{$\SL(N)$}

There are various simplifications when $G=\SL(N$), making the counting problem more tractable compared to other groups. We will first analyze the case of general $N$ to outline the algorithm, which we then carry out in examples.

\subsubsection{General structures}

The level-$k$ weights of $\SL(N)$ are parametrized by an $N$-tuple
$\mu=(\mu_0,\mu_1,\ldots,\mu_{N-1})\in \bZ_{\ge0}^N$ satisfying $\sum_i\mu_i=k$. 

The fundamental Weyl alcove is an $N$-simplex with $M$-cells in bijection with subsets of size $N-M-1$ of the nodes of the affine Dynkin diagram of $A_{N-1}$. The weights living in a cell have zeroes at places corresponding to this subset (i.e.~a zero at the $j$-th place if the $j$-th node is chosen).  The cells are then classified by partitions of $N$ which is determined by the sub-diagram. More precisely, the sub-diagram is a collection of $A$-type Dynkin diagrams, and each $A_{i-1}$ component gives an $i$ in the partition, with 1 added to bring the sum to $N$.

For each partition $\lambda$, one then has a collection of cells. It is clear that the orbits of the $\bZ_N$ center action live within this collection. Furthermore, the central character of $\bZ_N$ on the weight space $r(\mu)\in\bZ_N$ is given by 
\begin{equation}
    r(\mu)=\sum_{j=0}^{N-1}j\cdot\mu_j \bmod N.
\end{equation}
As $N|k$ in our setting, the $\bZ_N$-action preserves the central characters, making the refined count well defined.

Now, one can count from the perspective of either vortex lines (root-lattice vectors equipped with $\bZ_N$ action) or Wilson lines ($\bZ_N$-orbits of weights refined by central characters). The two methods are dual to each other and similar in complexity. We used the latter perspective when studying the $N=2$ and $N=3$ cases, but the former perspective seems slightly simpler conceptually when $\gcd(\lambda)>1$. 

When a partition $\lambda$ has $\gcd(\lambda)>1$, the collection of cells will have connected components indexed by $\bZ_{\gcd(\lambda)}$. This can be argued by analyzing the connectivity of vertices via this collection. It is easy to see that the $i$-th vertex $v_i=(0\ldots,0,k,0,\ldots,0)$ with only non-zero entry at the $i$-th place can be connected to $v_j$ via this collection if and only if \begin{equation}
    i-j\equiv 0\pmod{\gcd(\lambda)}.
\end{equation}  
To give an example, consider $N=6$ and $\lambda=[42]$. The two connected components have weights of the types \[(*,0,0,0,*,0), \;\;(0,0,*,0,*,0), \;\;(*,0,*,0,0,0)\] and \[(0,*,0,0,0,*),\;\; (0,0,0,*,0,*), \;\;(0,*,0,*,0,0).\]

To obtain the refined count, one looks at the $\bZ_N$ action on root-lattice vectors in this phase. The orbits are classified by their lengths, which are divisors of $N$. A length-$d$ orbit will give rise to $d$ states, one for each $e_\text{string}$ in the subgroup $\bZ_{d}^\vee\subset \bZ_N^\vee$. Notice that $d$ must be divisible by $\gcd(\lambda)$, as the action of $\bZ_{\gcd(\lambda)}$ permutes the components. The group extension
\eqref{iotaExtension} is now 
\begin{equation}
    \bZ_{N/\!\gcd(\lambda)}\to \bZ_N\to \bZ_{\gcd(\lambda)},
\end{equation}
where the subgroup $\bZ_{N/\!\gcd(\lambda)}$ acts on each component internally. The group $\bZ_d$ acting on a particular orbit can be decomposed with respect to the short exact sequence above, and it always contains the $\bZ_{\gcd(\lambda)}$ part. Notice that once we know the length $d$, we already know everything we need to incorporate this orbit. 

In contrast, if we take the Wilson-line/weight-orbit perspective, we do need to first work out the two parts of $e_\text{string}$. For a weight $\mu$, the part in $\bZ_{N/\!\gcd(\lambda)}^\vee$ is given by $r(\mu)/\!\gcd(\lambda)$, which is an integer as the central character is always a multiple of $\gcd(\lambda)$. One then takes into account all the $\bZ_{\gcd(\lambda)}$-connections along $S^1$ and get one state for each value of $e_\text{string}$ in the coset $r(\mu)/\!\gcd(\lambda)\cdot \bZ_{\gcd(\lambda)}^\vee\subset \bZ_N^\vee$. 

In the end, one multiplies the total number by the multiplicity $\gcd(\lambda)^{2g}$ from $\bZ_{\gcd(\lambda)}$-connections on $\Sigma$. For the refined count, one instead duplicates the answer for any given $e_\text{string}$ sector to each $e_\text{bulk}\in H_1(\Sigma,\bZ_{\gcd(\lambda)}^\vee)\subset H_1(\Sigma,\bZ_N^\vee)$. In the $\lambda=[42]$ example, a sector receiving a non-trivial contribution from this phase can have $e$ with an arbitrary component $e_\text{string}$ in $H_1(S^1,\bZ_6)$ but only an order-2 component $e_\text{bulk}$ in $H_1(\Sigma,\bZ_6)$. Finally, adding all the phases $\lambda$ together then gives the final answer.

Now, in principle, we are well-equipped to tackle any specific case with a given $N$ and $g$ by hand. To obtain a formula for arbitrary $g$ and fixed $N$, we only need to work out the counting function $\Gamma^r(\lambda;k)$ that counts $\bZ_N$-orbits with character $r$ in the collection of cells given by $\lambda$, or equivalently an analogous function that counts the number of orbits of root vectors with a given length $d$. 

If one actually wants to count phase-by-phase, the starting point is $\lambda=[N]$, and one counts orbits of weights in the entire $N$-simplex. This is carried out in Appendix~\ref{app:Weights}, where the resulting $\Gamma^r$ is expressed as a double summation. As any $M$-cell is itself an $(M+1)$-simplex, this counting function is basically the only thing that one needs, except for the counts of orbits under subgroups of the center, as the center might map a cell to other ones. To count the orbits for a given phase/partition, one can combine all the cells that it contains. One technical difficulty that one needs to overcome is to correctly take into account of overlaps. In other words, if one simply adds together the weights on each of the $M$-cells associated to a given partition, one would overcount the weights on $(M-1)$-cells where different $M$-cells meet. Therefore, one should subtract these, but then one would undercount the weights on certain $(M-2)$-cells. One can try to be very careful to get the correct answer in simpler cases, but it seems rather complicated when $N$ becomes large.

Working this out in greater generality seems to involve a level of technical details well beyond what we intend to go into in this paper. Also, as we will discuss in the next section, there is a connection between this counting problem with the one about counting fixed points in the moduli space of Higgs bundles, where results in many cases are available in the literature. Therefore, our goal will be to establish the connection instead of actually counting.

In the remainder of this section, we will study the $N=4$ and $5$ cases to illustrate the general process mentioned above.

\subsubsection{Example: $\SL(4)$}

In the $N=4$ case, the alcove is a tetrahedron, with a 3-cell, four 2-cells (faces), six 1-cells (edges) and four vertices. There is almost a one-to-one correspondence between the dimensions of cells and phases, except that two of the edges are associated with the $\lambda=[22]$ phase, while the other four are associated with $\lambda=[31]$.

The contribution of the phase $\lambda=[1^4]$ is given by counting orbits of the weights in the 3-cell. The refined count is given in Appendix~\ref{app:Weights}, with 
\begin{equation}
    \Gamma^0(A_3;k)=\frac{k^3+6k^2+20k+96}{96},
\end{equation}
while $\Gamma^2(A_3;k)=\Gamma^0(A_3;k)-1$ and $\Gamma^1(A_3;k)=\Gamma^3(A_3;k)=\Gamma^2(A_3;k)-\frac{k}{8}$. Notice that $k=4\chi=8g-8$ is always divisible by 8. From the vortex-line perspective, \begin{equation}
\Gamma^1(A_3;k)=\Gamma^3(A_3;k)=\frac{k^3+6k^2+8k}{96}
\end{equation} 
is the number of length-4 orbits of root-lattice vectors in the alcove. The difference $\Gamma^2(A_3;k)-\Gamma^1(A_3;k)=\frac{k}{8}$ is the number of length-2 orbits, which are of the form 
\begin{equation}\label{lengthTwo}
    \left(\frac k2-a,a,\frac k2-a,a\right)\sim \left(a,\frac k2-a,a,\frac k2-a\right)
\end{equation}
with even integer $a$ between 0 and $\frac{k}{4}-2$. On the other hand, the difference $\Gamma^0(A_3;k)-\Gamma^2(A_3;k)=1$ counts the only length-1 orbit given by $a=\frac k4$.

For $\lambda=[211]$, we should count the orbits of root vectors on the faces. There is only one length-$2$ orbit given by $a=0$ in \eqref{lengthTwo}, while there are  $\frac {k^2}{8}$ length-4 orbits. When $g>1$, there will not be any length-1 orbit, but there will be one for $g=1$. However, as $C_\lambda$ is still $\bZ_2$ and can have a holonomy along $S^1$, the $g=1$ case is equivalent to having a length-2 orbit.

Both the phases $\lambda=[31]$ and $[22]$ involve 1-cells. For the former, there are four edges forming a circle with $k$ root vectors in total. The $\bZ_4$ action is generated by a ``$\frac \pi2$-rotation.'' Therefore, there are $\frac k4$ length-4 orbits (except for $g=1$ when there is still a length-1 orbit).  For the $\lambda=[22]$ phase, the root vectors are on two disconnected edges, related to the fact that $K_\lambda=\SU^\pm(2)$ is disconnected. There is one length-2 orbit and $\frac k 4$ length-4 orbits. We also need to remember the $2^{2g}$ values of possible $e_\text{bulk}\in H_1(\Sigma,\bZ_2)$.

Finally, the four vertices corresponding to the $\lambda=[4]$ phase form a single length-4 orbit. Combined with the choice of $e_\text{bulk}\in H_1(\Sigma,\bZ_4)$, this phase contributes to every electric sector once.

Combining them, we have the total count given by 
\begin{equation}\label{SL4}
    \dim \cS(\Sigma\times S^1;\SL(4))=4^{2g+1}+ (4g-3)\cdot 2^{2g+1}+\frac{1}{3}(64g^3-48g^2-58g+51).
\end{equation}
The delta function is canceled out in the sum. One can check that this recovers the expected answer 75 when $g=1$, while naively extrapolating it to $g=0$ gives 15.

One can also find the refined counts from the information about the numbers of orbits of different types above. The results are given below, with the abbreviations $\delta$ for $\delta_{g,1}$ and $F:=\frac{1}{3}(2\chi^3+9\chi^2+7\chi)+1=\frac{16}{3}g^3-4 g^2-\frac{10}{3}g+3$.
\begingroup
\setlength{\tabcolsep}{0.3em}
\begin{center}
    \centering
    \begin{tabular}{c||c|c|c|c}
        $a$ & $b=0$ & $b\neq0$ even & $b\neq0$ odd & total \\\hline
       0  & $F\!+\!g\!+\!2\!+\!\delta$ & $2g$ & 1& $4^{2g}\!+\!(2g\!-\!1)2^{2g}\!+\!F\!-\!g\!+\!2\!+\!\delta$  \\\hline
       1 or 3 & $F$ & $2g-1$ & 1  & $4^{2g}\!+\!(2g\!-\!2)2^{2g}+F-2g+1 $\\\hline
       2 & $F\!+\!g\!+\!1\!-\!\delta$ & $2g$ & 1  &  $4^{2g}\!+\!(2g\!-\!1)2^{2g}\!+\!F\!-\!g\!+\!2\!-\!\delta$\\\hline\hline
        classes  & $1\times 4$ & $(2^{2g}-1)\times 4$ & $(4^{2g}\!-\!2^{2g})\!\times\!4$ & $4^{2g}\times 4$ \\\hline
        total & $4F+2g+3$& $(4g-1)(2^{2g+1}-2)$& $4^{2g+1}\!-\!2\!\cdot\! 2^{2g+1}$ & $4^{2g+1}\!+\!(4g\!-\!3)2^{2g+1}\!+\!4F\!-\!6g\!+\!5$
    \end{tabular}
\end{center}
\endgroup
\noindent 
Here, $a$ and $b$ are the two components of $H_1(\Sigma\times S^1,\bZ_4)\simeq H_1(S^1,\bZ_4)\oplus H_1(\Sigma\times S^1,\bZ_4)$, and being even means that $b$ is of order two. The total count in the bottom-right corner agrees with~\eqref{SL4} after substituting in the expression for $F$.

Notice that, for this computation, much of our effort is spent on understanding the special shorter orbits, which only affect sectors with even $e$. If we are only interested in the ``co-prime case'' with $a=1$ or 3, the computation would be much easier.

\subsubsection{Example: $\SL(5)$}

When $N$ is prime, there is much simplification compared with the general case. From the vortex-line point of view, the only short orbit of $\bZ_N$ is the one at the center of the alcove, while all other orbits are of length $N$. From the Wilson-line perspective, which we will take in this subsection, the refined count in every phase is the same for all $r$ except that there is one extra $r=0$ state in $\lambda=[1^N]$.

For $N=5$, the unique fixed point of the $\bZ_5$ center action on weights of $\fsl(5)$ at the level $k=10g-10$ is $\mu=(\frac k5,\frac k5,\frac k5,\frac k5,\frac k5)$. This enables us to easily find the number of orbits from the total number of weights.

Also, when $N$ is small, it is slightly easier to use a different way of organizing the count, by not proceeding phase by phase, but instead first counting the $(N-1)$-cell and $(N-2)$-cells. The overlaps between the $(N-2)$-cells automatically take care of some of the lower-dimensional cells. 

We start by summing together the weights in the 4-cell and 3-cells, leading to
\begin{equation}
    F_0=\binom{k+4}{4}+5\binom{k+3}{3}.
\end{equation}
This counts a weight in the interior of an $i$-cell with multiplicity $5-i$, but due to the overlap between $\lambda=[221]$ and $[311]$, the weights on 1-cells (edges) needs to be counted one more time, while the weights on 0-cells (vertices) needs to be counted two more times due to the further overlap between $\lambda=[41]$ and $[32]$.

To compensate for this, one first adds a correction term
\begin{equation}
    F_1=10(k+1)
\end{equation}
to $F_0$, to take into account the 10 edges, each with $k+1$ weights. However, this overcounts each vertex twice, and one needs to add another correction term $F_2=-10$. Therefore, the total cosmic-string contribution is\footnote{As mentioned previously, in $\Gamma_{\text{total}}$, it is natural to either include or not include the ``trivial'' contribution of the cosmic string in the $\lambda=[5]$ phase. For the former choice, one would add an additional +1 in the formula and need to remember that the bulk contribution is then $5^{2g+1}-1$ in the $\lambda=[5]$ phase to avoid an overcount (or, equivalently, multiplying the +1 with, as opposed to adding it to, the $5^{2g+1}$ factor). But one benefit would be that $\Gamma_{\text{total}}^{0}=\dim\cS^0$ directly gives the dimension of the sector with trivial flux when $N$ is prime. We went with the latter convention by not including it, which seems to make various other formulae slightly nicer-looking. }
\begin{equation}
    \Gamma_\text{total}=\frac{F_0+F_1+F_2-1}{5}+5\delta_{k,0}=\frac{1}{5}\left[\binom{k+4}{4}\!-\!1\right]+\binom{k+3}{3}+2k+5\delta_{k,0}.
\end{equation}
Here, the $5\delta_{k,0}=5\delta_{g,1}$ term takes into account the fact that, when $k=0$, the overlap between 3-cells does not actually incorporate the other phases. Notice that if one counts phase by phase, there will be one Kronecker delta for each intermediate phase with abelian factors in $K^0_\lambda$. Therefore, the coefficient agrees with the five intermediate phases for $N=5$, which also generalizes to all prime $N$.

Adding the contribution of the $\bZ_5$ flat connections in the $\lambda=[5]$ phase gives
\begin{equation}
    \dim \cS=5^{2g+1}+\Gamma_\text{total}=5^{2g+1}+\frac{1}{5}\left[\binom{10g\!-\!6}{4}\!-\!1\right]+\binom{10g\!-\!7}{3}+20(g-1)+5\delta_{g,1}.
\end{equation}

To get the refined count, we need to either count the $r=0$ sector of the cosmic string or that with an $r\neq0$, but they just differ by $1+5\delta$. Therefore, the latter is simply given by 
\begin{equation}
\Gamma_{\text{total}}^{r\neq 0}=\frac{1}{5}\left(\Gamma_{\text{total}}^{r\neq 0}-1-5\delta\right)=\frac{F_0+F_1+F_2-6}{25}.
\end{equation}

If one wants to be more concrete, one can repeat the procedure for the total count but with refinement. The number of $\bZ_5$-orbits of weights with central character $r$ on the 4-cell is
\begin{equation}
    \Gamma^{r\neq0}(A_{4};k)=\frac{1}{25}\left[{\binom{k + 4}{4}-1}\right].
\end{equation}
For a 3-cell, the number of such weights is given by
\begin{equation}
        \Gamma^{r\neq0}_{(3)}(A_{4};k)=\frac{1}{5}\left[{\binom{k + 3}{3}-1}\right],
\end{equation}
 where the subscript indicates that this is a count over a 3-cell. For a 1-cell, we have
\begin{equation}
        \Gamma^{r\neq0}_{(1)}(A_{4};k)=\frac{k}{5},
\end{equation}
but we also need to remember that there are 10 edges, forming two orbits.

Putting them together gives the total cosmic-string contribution in a sector with $r\neq 0$, 
\begin{equation}
    \Gamma_{\text{total}}^{r\neq 0}=\frac{1}{5}\left[\frac{1}{5}{\binom{k + 4}{4}}+\binom{k + 3}{3}+2k-\frac{6}{5}\right],
\end{equation}
which also agrees with our earlier prediction. 
Then the contribution of the $r=0$ sector is
\begin{equation}
    \Gamma_{\text{total}}^{0}=\Gamma_{\text{total}}^{r\neq 0}+1+5\delta_{g,1}=\frac{1}{5}\left[\frac{1}{5}{\binom{k + 4}{4}}+\binom{k + 3}{3}+2k+\frac{19}{5}\right]+5\delta_{g,1}.
\end{equation}
The refined count is then
\begin{center}
    \centering
    \begin{tabular}{c||c|c|c}
         & $b=0$ & $b\neq0$  & total \\\hline
       $a=0$  & $\Gamma_{\text{total}}^{0}+1$ & $1$ & $\Gamma_{\text{total}}^{0}+5^{2g}$ \\\hline
       $a\neq 0$ & $\Gamma_{\text{total}}^{a}+1$ & 1 & $\Gamma_{\text{total}}^{a}+5^{2g}$   \\\hline\hline
        classes  & $1\times5$ & ${(5^{2g}-1)\times5}$ & $5^{2g}\times5$ \\\hline
        total & $\Gamma_{\text{total}}+5$ & $5^{2g+1}-5$ &$\Gamma_{\text{total}}+5^{2g+1}$
    \end{tabular}
\end{center}
Replacing $5$ with an arbitrary prime $p$, one gets the general shape of the table for the $N=p$ case. There, one also have a universal structure for the cosmic string contributions
\begin{equation}
    \Gamma_{\text{total}}^{r}=\frac{1}{p}\left[\frac{1}{p}{\binom{k + p-1}{p-1}}+\binom{k + p-2}{p-2}+(p-3)\cdot \binom{k + p-4}{p-4}+\ldots\right].
\end{equation}
The coefficients $c_m$ for $\binom{k-p-m}{p-m}$ is expected to be a polynomial of $p$ of degree at most $m-3$ when $m\ge 4$. It starts with $c_4=p-3$, which comes from the overlap between $\lambda=[31^{p-3}]$ and $[221^{p-4}]$ requiring us to count some $(p-4)$-cells one more time. Using the $\bZ_p$ action, they can be put into the form $(*,0,0,\ldots)$ with another zero in the ``$\ldots$'' region, which has $p-3$ places. Systematically working out the subsequent coefficients seems to be a rather complicated task.

\section{Checks}\label{sec:checks}

By now, we have made various predictions for skein modules of 3-manifolds, especially about their dimensions on $T^3$ and $\Sigma\times S^1$. As we have wandered far from the realm of established results, it would be valuable to have some reassurance by testing our predictions in some way. In this section, we perform two tests: the first demonstrates that the dimensions exhibit Langlands duality, at least when $N$ is prime, and the second compares $\cS^{e,m}$ with the structure of the fixed loci in the moduli spaces of Higgs bundles.

\subsection{Electric-magnetic duality}\label{sec:duality}

The $S$-duality of the 4d $\cN=4$ theory with a gauge group of type $ADE$ can be understood from the point of view of the relative theory, whose Hilbert space is doubly graded by both electric and magnetic fluxes, and the $S$-duality at the level of the Hilbert space becomes an electric-magnetic duality,\footnote{In this paper, we try to use three different terminologies for three related but slightly different dualities. We use ``$S$-duality'' for the duality at the level of 4d $\cN=4$ theories, the topologically twisted theories, and the twisted $\cN=1$ deformed theory. We use ``electric-magnetic duality'' to refer to the symmetry of the doubly graded Hilbert space and skein module under the exchange of electric and magnetic fluxes, while ``Langlands duality'' is used for the isomorphism between $\sk(M_3;G)$ and $\sk(M_3;^\mathrm{L}\!G)$ conjectured in \cite{jordan2022langlands}.}
\begin{equation}
    \cH^{e,m}\simeq \cH^{m,e}.
\end{equation}
See e.g.~\cite{Witten:2000nv,Kapustin:2006pk,Gukov:2025dol} for related discussions. As the two gradings come from 1-form symmetries of the relative theory and the inner product respects the symmetry, $\cH^\vee$ (of either the physical theory or the topologically twisted one), and hence $\cS$ by applying our construction for the relative version of the theory, also naturally become doubly graded. If this embedding of the doubly graded skein module $\sk^{e,m}(M_3)$ into $\cH^\vee$ respects the $S$-duality of the relative theory, one recovers a conjecture by Ben-Zvi--Gunningham--Jordan--Safronov which states that $\sk^{e,m}(M_3)$ has the same electric-magnetic duality under the exchange of $e$ and $m$ \cite{jordan2022langlands}.

We have seen in Section~\ref{sec:deform} some evidence that the Langlands duality for the skein modules does not really hold outside the $A$-series (except for the $D$-series which we haven't tested beyond $D_4$) even for $M_3=T^3$. While it might still be interesting to understand the failure of the duality for $B$--$C$ pairs and $E$-type groups on $\Sigma\times S^1$ to have a better diagnosis and pinpoint the ``root cause,'' our focus will be instead on the $A$-series, where we test the electric-magnetic duality as a refinement of the Langlands duality and a check for our proposed algorithm. 

Again, there is a simplification for $\fsl(p)$, which will serve as our starting point.

\subsubsection*{Duality for $\fsl(p)$}

With $p$ being a prime, there are only two global forms, $\SL(p)$ and PSL$(p)$. The doubly graded Hilbert space of the relative theory is empty unless either $e=0$ or $m=0$, as otherwise the electric and magnetic fluxes are not ``mutually local,'' and an argument analogous to Dirac's charge quantization condition tells us that a state in such a sector is unphysical due to the phase ambiguity. The Hilbert spaces of the SL and PSL theories are respectively the summation of the different $e$- and $m$-sectors. 

For the relative theory, one only has bulk contributions to non-trivial electric sectors from the phase given by the regular $\fsl(2)$-triple, whose infrared $\bZ_p$ gauge theory contributes a one-dimensional space to $\cS^{e,0}$ for each $e$. On the other hand, the bulk contributions to the non-trivial magnetic sectors are only from the phase given by the trivial triple, which generates a one-dimensional space for $\cS^{0,m}$ for each $m$. Therefore, the bulk contributions are invariant under the duality by themselves, and we only need to show that the cosmic-string part also satisfies the electric-magnetic duality.

For the relative theory, the bulk--string coupling is modeled on the coupling between boundary free fermions and a relative version of the Chern--Simons theory, which has both electrically charged Wilson/vortex line operators and magnetically charged boundary monopole operators from 't Hooft lines of the 4d theory.\footnote{It would be interesting to better understand this relative version of the coupled system, which seems to be related to the compactification of the 6d $(2,0)$ theory on a 2-handlebody with boundary being $L(k,1)$ and can be studied along the line of \cite{Gukov:2025dol}. On the other hand, the bulk theory might also be engineered by $T[M_3]$ with $M_3$ being the mapping torus of $T^k\in\SL(2,\bZ)$, with this monodromy leading to the identification of the Wilson and vortex lines in the infrared.} As the topological operators of both kinds can end, they can contribute to both $\cS^{e,0}$ and $\cS^{0,m}$. 

The analysis for the contributions of the cosmic string is very analogous to the one in the previous section: 
\begin{itemize}
    \item Weights of $K_\lambda$ at level $(2g-2)\cdot c_\lambda$ contribute to $\cS^{e,0}$ with $e=e_{\text{string}}=r\in\bZ_p$ along the $S^1$ direction given by the central character, with the entire $\bZ_p$ populated by cosmic-string states except for $\lambda=[p]$ (where the $\bZ_p$ is populated by bulk states).
    \item The $r=0$ sector given by the root-lattice vectors is duplicated in $\cS^{0,m}$ with $m=m_\text{string}\in\bZ_p$ in the image of $\pi_1(K_\lambda)$, which is always the entire $\bZ_p$ except for $\lambda=[p]$. 
\end{itemize}
For the intermediate phases, $K_\lambda$ always has an abelian factor. Using a translation by its weights, one can show that the numbers of weights for any $r$ are the same and, dually, the $\bZ_p$ action on root vectors is free. For the $\lambda=[1^p]$ phase, there is one more element in the $r=0$ sector (which can be thought of as the unique fixed point of the $\bZ_p$-action on level-$k$ weights), but we are also supposed to subtract one from each $\cS^{0,m_{\text{string}}}$ as we have counted these as the bulk contribution of the magnetic $\bZ_p$ gauge theory in this phase. 

After this, we see that the cosmic-string contributions are symmetric under $e\leftrightarrow m$ in each phase. This symmetry combined with the electric-magnetic duality between bulk contributions in the $\lambda=[p]$ and $[1^p]$ phases gives
\begin{equation}\label{skeinIso}
    \cS^{e,m}\simeq  \cS^{m,e}.
\end{equation}
This is actually a natural isomorphism between vector spaces (as opposed to just an equality of dimensions). For the bulk contribution, the isomorphism sends the generators associated with the $\bZ_p$ electric and magnetic gauge theories to each other. As for the cosmic-string states, the isomorphism is given by the ``asymmetric'' (but still non-degenerate) $S$-matrix relating $\bZ_p$-orbits of $K_\lambda$-weights with central character $r$ with the character-$r$ space of the $\bZ_p$-action on the $\bC$-span of root-lattice vectors. Even for  $e=m=0$, the isomorphism
\begin{equation}
    \cS^{0,0}\stackrel{\sim}\longrightarrow \cS^{0,0},
\end{equation}
when restricted to the cosmic-string part, involves a ``discrete Fourier transform'' given by a scalar multiple of the $S$-matrix $S_{\mu\nu}$ where both $\mu$ and $\nu$ represent $\bZ_p$-orbits of root-lattice vectors.

The isomorphism \eqref{skeinIso} also gives an isomorphism between SL- and PSL-skein modules,
\begin{equation}
    \cS(\Sigma\times S^1;\SL(p))\simeq  \cS(\Sigma\times S^1;\mathrm{PSL}(p)),
\end{equation}
as vector spaces graded by $H_1(M_3,\bZ_p)\simeq H^2(M_3,\bZ_p)$.

\subsubsection*{General rank}

For $\fsl(N)$ with $N$ not necessarily prime, there can be various global forms of the gauge group labeled by divisors of $N$. Correspondingly, $\cS^{e,m}$ can be non-empty even when $e$ and $m$ are both non-trivial, as long as they remain mutually local, in the sense that they come from a isotropic subgroup of $Z(\frak g)\times Z(\frak g)$ where the pairing is trivialized. In the present case, each divisor $d$ of $N$ labels a ``Lagrangian'' (i.e.~maximal isotropic) subgroup  $\bZ_{N/d}\times \bZ_{d}\subset\bZ_N\times \bZ_N$ generated by $(d,0)$ and $(0,N/d)$, while a general Lagrangian subgroup is a ``tilted'' version of such standard subgroups that corresponds to turning on discrete theta angles of the theory. (See \cite{Gukov:2020btk} and references therein for more details on this perspective as well as the classification of Lagrangian subgroups for various $\frak g$.)

\paragraph{Bulk contributions.} For the bulk parts, the phase-wise duality via transposition remains valid, as we have both $\bZ_{\gcd{(\lambda)}}$ electric and $\bZ_{\gcd{(\lambda^T)}}$ magnetic 1-form symmetry in a phase $\lambda$, and they are exchanged by switching from $\lambda$ to the dual phase $\lambda^T$. The former phase will give rise to states in (the dual of) $\cS^{e,m}$ with $e\in H^2(M_3,\bZ_{\gcd{(\lambda)}})$ and $m\in H^2(M_3,\bZ_{\gcd{(\lambda^T)}})$, which come from an isotropic subgroup $\bZ_{\gcd{(\lambda)}}\times \bZ_{\gcd{(\lambda^T)}}$ of $\bZ_N\times \bZ_N$ (it only becomes Lagrangian when $\lambda$ is an equal partition). On the other hand, the bulk gauge fields in the $\lambda^T$ phase will contribute in exactly the dual way.

However, we now also have cosmic strings in all the phases except $\lambda=[N]$, and, a priori, they can spoil the duality as they interact with the electric and magnetic symmetries in different ways, and their contributions are different in dual phases.

\bigskip

The situation is significantly simplified once we look at the ``co-prime cases'' with $(e,m)$ being of order $N$. This also forces one of them to vanish. As the ``bulk'' and ``string'' components play quite different roles in the analysis, there are several cases.

\paragraph{With order-$N$ $e_\text{bulk}$ or $m_\text{bulk}$.} If the non-trivial components of $e$ or $m$ in $H^1(\Sigma,\bZ_N)$ is of order $N$, they can only come from the phase associated with, respectively, the principal or zero $\fsl(2)$-triple. As the cosmic-string contribution becomes trivial in both---by the decoupling of the string in the former phase and the string--bulk selection rule in the latter phase---this part satisfies the electric-magnetic duality. 

\paragraph{With $e_\text{bulk}=m_\text{bulk}=0$.} In this case, either $e_\text{string}$ or $m_\text{string}$ will be of order-$N$.
When there is only a component of $e$ in $H^2(\Sigma,\bZ_N)\simeq\bZ_N$, states in $\cS^{e,0}$ can come from all the phases, including these with $\gcd(\lambda)>1$ as we have discussed previously when studying the $N=4$ case. On the other hand, for $\cS^{0,m}$ with such a co-prime $m\in H^2(\Sigma,\bZ_N)$, although it is not subject to the selection rule since $m_\text{bulk}=0$, it cannot receive contributions from phases with $\gcd(\lambda)>0$, as only the $\bZ_{N/\gcd(\lambda)}$ subgroup of $\bZ_N$ is in $K_\lambda$. Therefore,  unlike the case with a prime $N$, the matching now has to involve ``conspiracies'' between different phases, and the cosmic string in some other phases has to contribute more to sectors with non-trivial $m$ instead. Recall that, for an $e_\text{string}$ co-prime with $N$, one simply counts the number of length-$N$ orbits of root-lattice vectors belonging to a given phase. For a non-trivial $m$, one is supposed to count all orbits, but should remove the states already counted as parts of the bulk contribution. When $N$ is prime, this yields an equality between the two sides, as there is a unique short orbit (actually a fixed point), which can be regarded as the bulk contribution. However, when $N$ is not a prime, there are other orbits of reduced size (e.g.~\eqref{lengthTwo} for $N=4$), leading to a mismatch for the cosmic-string contributions on the two sides for $\lambda=[1^N]$. Now one can hope that the two mismatches exactly offset each other. We will not carry out this analysis in general, since we find that this is already not the case for $N=4$, where they do not cancel each other.

\paragraph{Potential mismatch for $\fsl(4)$.} Indeed, we computed previously that the number of order-4 orbits of root vectors belonging to the $\lambda=[22]$ phase is $k/4$, but, on the other hand, the number of length-2 orbits in the $\lambda=[1^N]$ phase is $k/8$. They are proportional and both count root vectors on a line segment. However, one is half of an edge connecting the vertex $(k,0,0,0)$ and the mid-point $(\frac k2,0,\frac k2,0)$ (itself not included) of an edge, while the other is between the latter point and the center of the alcove $(\frac k4,\frac k4,\frac k4,\frac k4)$ (not included as counted as part of the bulk contribution), and the numbers of root-lattice vectors on them differ by a factor of two. As a consequence, for $e$ and $m$ being $1,3\in \bZ_3\simeq H^2(\Sigma,\bZ_4)$, one has
\begin{equation}
    \dim \cS^{e,0}(\Sigma\times S^1;\fsl(4))=\frac{2}{3}\chi^3+3\chi^2+\frac{7}{3}\chi+1,
\end{equation}
while 
\begin{equation}
    \dim \cS^{0,m}(\Sigma\times S^1;\fsl(4))=\frac{2}{3}\chi^3+3\chi^2+\frac{11}{6}\chi+1.
\end{equation}

\bigskip

Given that the mismatch is just by a factor of 2 in from of $\frac k8$, one might hope that it is due to a minor oversight in our analysis and proceed to study cases with more general $N$ as well as more general $e$ and $m$. There, one would also need to incorporate the $\gcd(\lambda)^{2g}$ sectors with non-trivial $e_\text{bulk}$ on the electric side. The same factor also appears in the magnetic side, but in the dual phase labeled by $\lambda^T$ and it is not clear how one can match them once the cosmic-string contributions are also taken into account on both sides. Furthermore, there are also sectors with both $e$ and $m$ being non-trivial, and more care is needed to correctly generalize the previous analysis. We will not pursue this here but would encourage the interested reader to work out some lower-rank examples (in particular, revisit our analysis for $N=4$) to see whether the electric-magnetic duality actually holds beyond prime $N$.

\subsection{Fixed points in Hitchin moduli spaces}

In a sense, the electric-magnetic duality for $\fsl(p)$ is not a strong check on our results, as the computations on the two sides are largely similar, and what we argued is that their difference is zero. It is desirable to have a stronger check against results computed through completely independent methods, and the geometry of the Hitchin moduli space provides such an opportunity.

The Hitchin moduli space has a close connection with 4d physics since its inception in \cite{hitchin1987self}, with the link to the twisted $\cN=4$ theory established in \cite{Bershadsky:1995vm}, where it was found that the twisted compactification of the 4d theory with gauge group $K$ on $\Sigma$ gives a sigma model with target the Hitchin moduli space $\cM_H(\Sigma,K)$. 

If we consider the Hilbert space of the 2d theory on another $S^1$, a sector of BPS states (or states in the topological sigma model) is given by cohomology classes of the target space. However, the Hitchin moduli space in general has singularities, which can give rise to additional states in the Hilbert space. To avoid such complications, one can first focus on the co-prime case, with either non-trivial $e$ or $m$.

From the point of view of the moduli space of the 4d theory on $M_3$, turning on $e$ corresponds to using the action of $H^1(M_3,Z(K))$ to decompose the Hilbert space into characters and picking the sector labeled by $e\in H_1(M_3,Z(K)^\vee)$, while turning on $m$ is about fixing the second Stiefel--Whitney class of topological type of the bundle to be $m\in H^2(M_3,\pi_1(K))$. We will use the relative version of the theory, which can be thought of as having $K$ being simply-connected, but still allows 't Hooft flux $m$ valued in the center of $K$. 

When $M_3=\Sigma\times S^1$ and $K=\SU(N)$, the case of $m\in H^2(\Sigma,\bZ_N)\subset H^2(\Sigma \times S^1,\bZ_N)$ being co-prime with $N$ leads to a smooth moduli space, which, via the Kobayashi--Hitchin correspondence \cite{hitchin1987self,Donaldson1987,Simpson1988,Corlette1988}, can be identified with the moduli space of rank-$N$ poly-stable Higgs bundles with a fixed twisted determinant bundle of degree $m$, and is very well studied (see e.g.~\cite{hausel2013global} for a nice review). 

One can also work with the ``full'' moduli space, which additionally takes into account of the holonomy along the $S^1$ factor. The full space contains $N$ copies of the Hitchin moduli space, connected via lower-dimensional strata. Only in the full moduli space, one can talk about turning on the component of $e$ along the $S^1$ direction (i.e.~$e_{\text{string}}\in H_1(S^1,\bZ_N)\subset H_1(\Sigma\times S^1,\bZ_N)$).\footnote{Notice that $e_{\text{string}}$ (and also $m_\text{string}$) can be represented by a 1-cycle in $H_1(S^1)\subset H_1(\Sigma\times S^1)$ or 2-cocycle in  $H^2(\Sigma)\subset H^2(\Sigma\times S^1)$, with the two perspectives related by the Poincaré duality. When using the former perspective, we often say that the flux (tube) is along $S^1$, while when using the latter perspective, we say the flux is through $\Sigma$.} Also, if one goes beyond the $A$-series, it seems rather important to work with the full space, as it will contain Hitchin moduli spaces associated with different gauge groups. See \cite{Gukov:2025dol} for a more detailed discussion about this and closely related moduli spaces. To avoid further complicating the analysis, we will work with the usual Hitchin moduli space in the remainder except for the $N=2$ case where we utilize the geometry of the full moduli space toward the end. This is mostly fine in the $A$-series, provided that we do not turn on $e$ along the $S^1$ direction. For related reasons, we will also not turn on other components of $m$ other than $m_\text{string}$ along $S^1$---otherwise, it would be more natural to work with variants of the full moduli space.\footnote{Because of this, it is not possible to check the electric-magnetic duality at the level of Hitchin moduli spaces, which can only see $m_\text{string}$ and $e_\text{bulk}$.}

For the sigma model, the full Hilbert space is infinite-dimensional due to the non-compactness of the target, but the skein module of $\Sigma \times S^1$ is expected to be given by the middle-dimensional cohomology from the point of view of quantization \cite{ben2018integrating,jordan2022langlands,gunningham2023deformation}. This can also be understood via ``brane quantization'' \cite{Gukov:2008ve} along the line discussed in \cite{Gukov:2022gei} (see also \cite{kapustin2003remarks,oblomkov2004double,kapustin2005branes,Nekrasov:2010ka,Yagi:2014toa,Allegretti:2024svn,Allegretti:2024idu,Huang:2024mtw}).

When the moduli space is smooth, the middle-dimensional cohomology is generated by downward Morse flows from the fixed loci of Hitchin's $\bC^*$-action. Therefore, one can simply classify and count the number of components of the fixed loci. This was studied by Hitchin for the $N=2$ case in \cite{hitchin1987self} and by Gothen for the $N=3$ case \cite{Gothen}. We will mainly focus on comparing with their results but will also briefly discuss $N=4$ and  the more general case.

\subsubsection*{Rank-2 case}

For $G=\SL(2)$, there are two moduli spaces labeled by, respectively, $m=0$ and $1$, with the latter being the co-prime case. 

There are $g$ components of fixed loci in the $m=1$ moduli space, which exactly matches with the dimension of $\cS^{0,m}$. The ``lowest'' component, which can be identified with the moduli space of twisted rank-2 bundles---corresponds to the ``bulk contribution'' in the confining phase with non-trivial $m_\text{bulk}$, while the other components are in one-to-one correspondence with $\bZ_2$-orbits of weights of $\SU(2)_{4g-4}$ with odd central characters. In other words, they are related to the cosmic-string states. In fact, these weights/orbits are labeled by positive odd numbers smaller than $2g-2$, and they give exactly the values of the moment map of the $\bC^*$-action at the fixed loci, with a coefficient $\frac{\pi}2$. The action of $H^1(\Sigma,\bZ_2)$ on them is trivial, compatible with the fact that the corresponding states are all in the $e=0$ sector.

The $m=0$ moduli space has $2^{2g}+g-1$ fixed loci. They can be classified by an even integer $\mu$, which ranges from 0 to $2g-2$ and corresponds to $\bZ_2$-orbits of even weights in SU(2)$_{4g-4}$. For $\mu=2g-2$, there are $2^{2g}$ isolated fixed points, forming a torsor of $H^1(\Sigma,\bZ_2)\simeq \bZ_2^{2g}$. Therefore, they contribute one state to $\cS^{e,0}$ for each $e\in H_1(\Sigma,\bZ_2)$. There is one more state in the $e_{\text{string}}=0$ subspace, whose dimensions is $2^{2g}+g$, compared with the number of components of fixed points. The additional states come from the singular loci of the lowest fixed components, which is the moduli space of semi-stable bundles and has singularities along the strictly semi-stable ones.

\begin{figure}[htb!]
    \centering
    \includegraphics[width=0.8\linewidth]{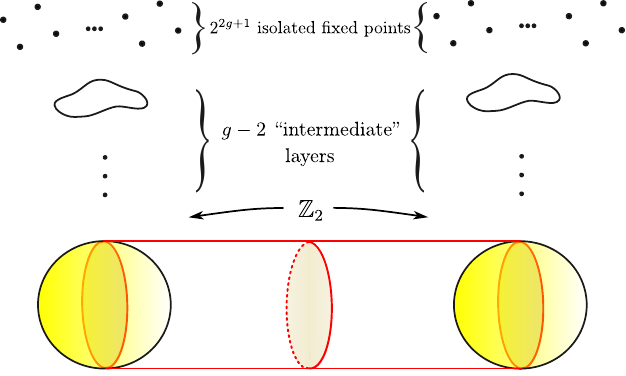}
    \caption{Illustrated on the two sides are fixed loci of the two copies of moduli space of SL($2,\bC$)-Higgs bundles. Each has $2^{2g}$ isolated fixed points at the top layer (highest value of the moment map). The lowest components (yellow) have singularities (red), which are connected in the full moduli space---approximated by the SL($2,\bC$)-character variety of $\Sigma\times S^1$---via abelian connections on $\Sigma\times S^1$ (red). The $H^1(S^1,\bZ_2)\simeq \bZ_2$ electric symmetry invisible for a single copy of the Hitchin moduli space acts on the full moduli space by ``flipping the interval.'' }
    \label{fig:full}
\end{figure}

To access the $e_{\text{string}}=1$ subspace, one needs to use the full moduli space of the 4d $\cN=4$ theory on $\Sigma\times S^1$. For our purposes, it can be approximated by the character variety of $\Sigma\times S^1$. This full moduli space in the SL(2) case contains two copies of the Higgs-bundle moduli space, connected along a $(2g+1)$-dimensional component which can be thought of as the flat $\bC^*$-connections on $\Sigma\times S^1$. It intersects the lowest fixed loci of the Higgs-bundle moduli space along the strictly semi-stable bundles. This is illustrated in Figure~\ref{fig:full}.

Therefore, almost all of the $2^{2g}+g$ states with $e_{\text{string}}=0$ are doubled, except the one coming from the strictly semi-stable bundles, as they are connected in the full moduli space. Collectively, they are also preserved by the action of $H^1(\Sigma\times S^1,\bZ_2)$, and only contribute to the $e=0$ sector. On the other hand, all of the other fixed loci are doubled in the full moduli space. There are now $2^{2g+1}$ isolated fixed points, with the transitive action of $H^1(\Sigma\times S^1,\bZ_2)$, and their contribution can be identified with the one from the $\bZ_2$ phase. The other components are also doubled, but they only have a non-trivial action of the subgroup $H^1(S^1,\bZ_2)$, and give rise to states with $e=e_\text{string}$ along the $S^1$ direction. And they can be identified with the cosmic-string contributions in the confinement phase.\footnote{To ``distribute'' the middle-dimensional cohomology to the two phases, it is in a sense more natural to put the $H^1(\Sigma,\bZ_2)$-invariant two-dimensional subspace from the $2^{2g+1}$ isolated fixed points in the confinement phase, while the two lowest fixed loci in the electric $\bZ_2$ phase. Also, to access $\cS^{0,m}$ with more general $m$, one can turn on a more general twist on $\Sigma\times S^1$. It is natural to expect that the moduli space will become a point, similar to how flat O$(2)$- and $N_2$-connections behave on $T^3$ when a non-trivial $w_2$ is turned on.  } 

Notice that the reasons for the ``truncation'' are different on the two sides but nonetheless lead to the same condition. On the cosmic-string side, it is related to the Chern--Simons theory only having a finite number of independent line operators, while on the Higgs-bundle side, it is due to the stability condition. More precisely, fixed points with non-zero values of the moment map are described by a pair $(E,\Phi)$ with $E$ a rank-two vector bundle (with a fixed determinant) that decomposes into two line bundles $L_1,L_2$ of degrees $\mu/2$ and $-\mu/2$. Stability requires that the Higgs field is lower triangular $\Phi=\begin{pmatrix}
    0&0\\\phi&0
\end{pmatrix},$ so that it leaves invariant the line bundle $L_2$. The field $\phi$ is then a holomorphic section of the line bundle $L_1^\vee\otimes L_2\otimes K_\Sigma$, which has degree $2g-2-\mu$. Demanding that this be non-negative to allow a non-zero $\phi$ to exist leads to the condition $\mu\le 2g-2$.

\subsubsection*{Rank-3 case}

For the $N=3$ case, one has fixed loci of three different types, depending on how the vector bundle $E$ decomposes, given exactly by the partition of $N$. For the partition $2+1$, there are two sub-types with the decomposition $(2,1)$ and $(1,2)$. They become  distinct once we demand that the Higgs field $\Phi$ is again lower triangular.

In the $m=1$ (or 2) moduli space, the middle-dimensional cohomology has dimension $2g^2-g$, which is exactly in agreement with the dimension of $\cS^{0,m}$. In fact, there is a finer agreement between different types of fixed points and the contribution from different phases, which we now check.\footnote{There are typos in \cite{Gothen} rendering the count to be different from ours. The correct total count is given in \cite{hausel2003mirror,hausel2008mixed}. We will redo the count in a convention that is slightly easier to connect with the weights of $\fsl(3)$.}

Recall that there are $(2g-1)(g-1)$ states from the magnetic phase $\lambda=[1^3]$, which are $\bZ_3$-orbits of level-$(6g-6)$ weights of $\fsl(3)$ with central character $m$ (green or red dots in Figure~\ref{fig:su3}). The condition can be written as $\mu_1+2\mu_2\equiv m\pmod 3$ using the affine Dynkin label $(\mu_0,\mu_1,\mu_2)$ for weights. Using the $\bZ_3$ cyclic rotation, one can put the weights between the dotted lines in Figure~\ref{fig:su3}, satisfying $\mu_1+2\mu_2\le 6g-6$, and $2\mu_1+\mu_2\le 6g-6$. The inequalities are actually strict in the co-prime case with $m\neq 0$.

For fixed points of type $(1,1,1)$ on the Higgs-bundle side, we have $E=L_1\oplus L_2\oplus L_3$ decomposes into three line bundles whose degrees $l_i$ sum to $m$. With a lower-triangular
\begin{equation}
    \Phi=\begin{pmatrix}
        0&0&0\\
        \phi_1&0&0\\
        0&\phi_2&0
    \end{pmatrix}, 
\end{equation}
the (semi-)stability condition requires the slopes of $L_2\oplus L_3$ and $L_3$ to be less than (or equal to) $\frac m3$, which leads to $2l_1-l_2-l_3\ge 0$ and $l_1+l_2-2l_3\ge 0$. Another constraint comes from the condition for the existence of non-zero $\phi_1$ and $\phi_2$. As they are holomorphic sections of $L_1^\vee\otimes L_2\otimes K_\Sigma$ and $L_2^\vee\otimes L_3\otimes K_\Sigma$, the non-negativity of their degrees gives
\begin{equation}
    a:=2g-2+l_2-l_1\ge0 \quad\text{and}\quad b:=2g-2+l_3-l_2\ge 0.
\end{equation}
Rewriting the stability conditions using $a$ and $b$, these are exactly
\begin{equation}
    a+2b\le6g-6 \quad\text{and}\quad 2a+b\le6g-6.
\end{equation}
Furthermore, $a+2b\equiv m \pmod 3$, making them subject to exactly the same conditions as $\mu_1$ and $\mu_2$. Therefore, we have explicitly established the one-to-one correspondence.

Notice that the matching is ``order-reversing'' in the sense that the higher critical points in the moduli space correspond to the weights closer to the origin. (In the previous $N=2$ example, it is hard to tell whether the identification reverses the order or not, as both sides are labeled by points on an interval.) This is a good feature if we think about the $m=0$ moduli space, which has $3^{2g}$ fixed points with the largest value of the moment map corresponding to the contributions of the flat $\bZ_3$-connections on $\Sigma$ in the electric phase with $\lambda=[3]$. Previously, we associate them with the trivial weight, and, via the present correspondence, they are naturally mapped to the critical points with highest value of the moment map. 

For critical points of the $(2,1)$ and $(1,2)$ type, they will be mapped to weights belonging to the neutral phase $\lambda=[21]$, which live on the two longer sides of the kite-shaped region in Figure~\ref{fig:su3}. We now see this more explicitly.

Consider a critical point of type $(1,2)$, which is given by a pair with $E=L\oplus V$ and $\Phi$ lower triangular with the only vanishing component being $\phi\in H^0(\Sigma,L^\vee\otimes V\otimes K_\Sigma)$. The degree $l$ of the line bundle then satisfies 
\begin{equation}
    \frac m3 \le l\le \frac m3 +g-1
\end{equation}
from stability and the condition for non-vanishing Higgs fields. One can then map this to a weight with $\mu_1=3g-3 + m-3l$ and $\mu_2=0$. This indeed lives on a long edge of the kite-shaped regions. The type $(2,1)$ fixed points are similar, but with $\phi\in H^0(\Sigma,L\otimes V^\vee\otimes K_\Sigma)$, and the conditions for $l$ become
\begin{equation}
     \frac m3 -g+1\le l\le \frac m3 .
\end{equation}
One can view this as continuing in the previous case to negative values of $l$ and, by a $\bZ_3$ rotation, mapped to weights on the other long edge of the fundamental domain of the $\bZ_3$ action.\footnote{To have ``continuity'' for this line, one can pick the fundamental domain to be the isosceles triangle between an edge and the center of the alcove so that this line is not ``broken.''} More concretely, one has $\mu_1=0$ but $\mu_2=l-\frac m3 +g-1$. Notice that they have the correct central character as $\mu_1+2\mu_2\equiv m \pmod3$ in both cases.

In the co-prime case, the above inequalities are strict, but when $m=0$, the moduli space is singular exactly due to the ``boundary cases,'' which one needs to be more careful with to recover the result on the skein side. 

First of all, the highest critical points are $3^{2g}$ isolated points acted upon by the electric symmetry $H^1(\Sigma,\bZ_3)$. They match the contribution of the electric phase to $\cS^{e,0}$ with $e_{\text{string}}=0$.\footnote{Similar to the $N=2$ case, one might want to associate the invariant part with $\lambda=[1^3]$, but add back the lowest fix loci of type $(3)$ to $\lambda=[3]$. Also similar is that, in the full moduli space, one sees all the $3^{2g+1}$ fixed points with transitive action of the $H^1(\Sigma\times S^1,\bZ_3)$ electric symmetry.} Toward the other end, at $l=0$, the type $(2,1)$ and $(1,2)$ components become identical and is part of the moduli space of rank-3 bundles as its singular loci. Therefore, they should still be counted but only once. Inside this component, there are singular loci of even higher co-dimensions given by another $(1,1,1)$-type fixed loci, which is labeled by the weight at the center of the alcove and should be counted by itself. There are also the components with $l=g-1$ and $l=-g+1$. They are connected through strictly poly-stable Higgs bundles with $E$ of the form $K^{1/2}\oplus K^{-1/2}\oplus \cO$ and a Higgs field taking the first summand into the second. There are $2^{2g}$ choices for $K^{1/2}$, and one can then view them as of either type $(2,1)$ or $(1,2)$ by grouping $\cO$ with $K^{1/2}$ or $K^{-1/2}$. The fact that the two types are connected explains why they are only counted once on the skein-module side. 

Another interesting phenomenon on the skein side is the identification of the weights on the two shorter edges of the kite-shaped region due to the $\bZ_3$ symmetry. These weights correspond to Higgs bundles with either $l_1=0$ or $l_3=0$. Poly-stability requires the component of the Higgs field mapping out of $L_1$ or $L_1\oplus L_2$ to vanish and the rank-3 Higgs bundle to decompose as a direct sum of a rank-1 and a rank-2 Higgs bundles. Therefore, the two cases with $l_i$'s given by $(0,\mu,-\mu)$ and $(\mu,-\mu,0)$ are related by a gauge transformation, which exactly matches the identification given by the $\bZ_3$ action on weights.

Notice that (poly-)stability plays several rather important roles in the above analysis in achieving precise agreement with the skein side. This observation is consistent with the perspective that the natural moduli space associated with the 4d theory---and hence the skein module---is the one given by BPS equations, which in the present case is approximated by the Hitchin moduli space or, equivalently, via the Kobayashi--Hitchin correspondence, by the moduli space of poly-stable Higgs bundles.

\subsubsection*{Rank-4 case}

In the $N=4$ case, we found previously that $\cS^{1,0}(\Sigma\times S^1) \not\simeq \cS^{0,1}(\Sigma\times S^1)$, with their dimensions differing by $\frac k8=g-1$. To compare with the moduli space of Higgs bundles, one needs to first pick one among the two. 

It is natural to expect that the former is related to the SL(4) moduli space, while the latter is associated with the PSL(4) space, since the Hilbert space of the gauge theory with $\SU(4)$ gauge group is obtained by summing over all sectors with $m=0$ and arbitrary $e$, while the Hilbert space for the PSU$(4)$ theory is constructed by summing over all $m$ with $e=0$.

This also offers a potential way to explain the difference between the two skein modules. Indeed, the PSL(4) moduli space has orbifold singularities, and only the stringy version of the Hodge numbers is expected to match that of the SL(4) moduli space \cite{hausel2003mirror}. Therefore, if the relation between the skein module and the middle-dimensional cohomology of the moduli space requires a different method for incorporating singularities, this may reconcile the difference in the dimensions of $\cS$ for SL and PSL with the topological mirror symmetry between the two moduli spaces.

\medskip

The prediction of our analysis for the dimension of the SL(4)-skein module in the co-prime case (i.e.~$e=e_\text{string}=1$) for $N=4$ is
\begin{equation}
    \dim \cS^{e=1}(\Sigma\times S^1; \SL(4))=\frac23\chi^3+3\chi^2+\frac{7}3\chi+1.
\end{equation}
However, this formula does not match the result on the Higgs-bundle side, which was first conjectured to be $\frac23\chi^3+\frac52\chi^2+\frac{17}6\chi+1$ in \cite{hausel2008mixed} and proved later in  \cite{garcia2014motives,mellit2020poincare}. One can break down this dimension formula into contributions of different types of $\bC^*$-fixed points via the $A_\lambda$-polynomial of \cite{villegas2011refinement}  at $q=1$ for the quiver with one vertex and $g$ loops.\footnote{It is interesting that such a quiver is utilized, which, when interpreted as a gauge theory in 3d, can flow to a Chern--Simons theory in the infrared, but perhaps should be better interpreted as either the 2d or 1d quiver (cf.~\cite{benini2010mirrors} as the 3d lift) describing the 2d theory or quantum mechanics obtained by compactifying the 4d theory on $\Sigma$ or $\Sigma\times S^1$.} One has
\begin{align}
A_{[1111]}(1) &= \frac{2 }{3}\chi^3+\chi^2+\frac{\chi}{3}, \\
A_{[211]}(1) &= \frac{3}{2} \chi^2 +\frac12\chi , \\
A_{[22]}(1) &=A_{[31]}(1) =\chi,\\
A_{[4]}(1)&=1.
\end{align}
They all agree with the contribution of the cosmic string in the corresponding phase except for $\lambda=[211]$, where we found the string to contribute $\frac{k^2}{8}=2\chi^2$ states.

It is not clear what causes this mismatch, as the link between our physics computation and the middle-dimensional cohomology involves several conjectures and assumptions. One special property of the phase $\lambda=[211]$ is that $K_\lambda=\SU(2)\times \U(1)$ has both a non-abelian and an abelian part, with the center of $H_\lambda$ being ``emergent'' (i.e.~not part of $Z(K)$). One should therefore first revisit our analysis for such cases and ask, for instance, whether the Chern--Simons level being used for $K_\lambda$ is indeed the correct one and whether the generators of $\cS$ can develop linear relations in this setting. We will leave a more systematic investigation of this issue for future work and, for now, will instead try to gather more data from higher-rank cases.

\subsubsection*{Higher ranks}

The contribution of the fixed point of type $(1,1,\cdots,1)$ to the middle-dimensional cohomology of the co-prime moduli space is expected to be $A_{[1^N]}(1)$, which we find to always agree with the count of $\bZ_N$-orbits of SL($N$)-weights at level $k=\chi N$ with central character $r$ co-prime with $N$,
\begin{equation}
    A_{[1^N]}(1)=\Gamma^{r}(A_{N-1};k=\chi N).
\end{equation}
For example, when $N=5$, both sides are given by
\begin{equation}
A_{[1^5]}(1)=\Gamma^{r\neq0}(A_{4};5\chi)=\frac{5}{24}\left(5 \chi^4+10 \chi^3+7 \chi^2+2 \chi\right).
\end{equation}
For a given $N$, the asymptotic behavior for large $\chi$ is given by 
\begin{equation}
    \frac{1}{N^2}\binom{\chi N + N-1}{N-1}\sim \frac{N^{N-3}}{(N-1)!}\cdot\chi^{N-1}+\ldots
\end{equation}

Similar to the $N=4$ case, there is a mismatch for $\lambda=[21^{N-2}]$ already at the leading coefficient. The $A_\lambda$-polynomial evaluates to 
\begin{equation}
    A_{[21^{N-2}]}(1)=\frac{1}{N-1}{\binom{\chi(N-1) + N-3}{N-2}}\sim \frac{(N-1)^{N-3}}{(N-2)!}\chi^{N-2}+\ldots
\end{equation}
On the other hand, to leading order, the contribution of $\lambda=[21^{N-2}]$ is given by the number of weights on a single face divided by $N$,
\begin{equation}
    \Gamma^{r}([21^{N-2}];k=\chi N)= \frac{1}{N}{\binom{\chi N + N-2}{N-2}}+\ldots\sim \frac{N^{N-3}}{(N-2)!}\chi^{N-2}+\ldots
\end{equation}
The polynomial $A_{[21^{N-2}]}(1)$ can be rewritten as 
\begin{equation}
    A_{[21^{N-2}]}(1)=\frac{\chi}{N-2}\cdot {\binom{\chi(N-1) + N-3}{N-3}}
\end{equation}
which, aside from the $\frac{1}{N-2}$ factor, is compatible with the counting of $\bZ_N$-orbits of weights in a Chern--Simons theory with gauge group $\SU(N-2)\times\U(1)$, but at the ``wrong'' level $(\chi\cdot (N-1),\chi N)$. From the point of view of reducing from $\SU(N)_{\chi N}$, this level is somewhat unnatural (e.g., it could be fractional if the level of $\SU(N)$ were not divisible by $N$). It would be interesting to understand the mechanism that modifies the natural level compatible with the conformal embedding to the one above. Is it an actual physical effect at the level of quantum field theory (e.g.~non-compactness issues mentioned in Section~\ref{sec:SkeinDeform})? Is it due to additional relations among the generators of the skein module related to the emergent center of $\SU(N-2)$ (with the $\frac{1}{N-2}$ factor possibly hinting at a non-trivial role played by this center)? Or does it instead arise at a later stage, when passing to the cohomology of the Hitchin moduli space?

We hope that, once they are better understood, the connections between $\bC^*$-fixed loci, Chern--Simons theories, and the structure of the Weyl alcove can be useful for the study of the geometry of the moduli space of Higgs bundles, especially beyond the $A$-series. It would be interesting to compare and relate them to other works in the literature on the topology of the nilpotent cone, such as \cite{schiffmann2016indecomposable,bozec2022irreducible,li2022nilpotent} (see also \cite{bu2025cohomology} for related discussions of character varieties of 3-manifolds).

\section{Discussion}\label{sec:future}

In the last section of the paper, we will first discuss new ways to look at some earlier results through the lens of our construction, and then mention several interesting future directions. 

\subsection{New perspectives}\label{sec:other}

\paragraph{On the Gilmer--Masbaum map.} In \cite{gilmer2019skein}, the authors use SO(3) WRT invariants to establish a lower bound on the dimension of the Kauffman-bracket skein module for $\Sigma\times S^1$. Rephrased in our setup, their work can be viewed as choosing the test state $\ket{\psi}$ in the Hilbert space $\cH(M_3)$ to be a ``Chern--Simons state'' $\ket{\text{WRT}_{r,s}}$, labeled by a root of unity $e^{i\pi s/r}$ specified by a pair of integers $r$ and $s$, which then sends a class in the skein module to its WRT invariant. In other words, a natural class of states in the quantum mechanics is given by flat connections on $M_3$, and an appropriate linear combination of them gives the WRT invariant once paired with $\bra{L}$ \cite{Witten:2010zr,witten2011analytic,Witten:2011zz}. The linear combination depends on $(r,s)$, and the Gilmer--Masbaum evaluation map can be interpreted as using such a family of test states to distinguish different skein classes.

\paragraph{Relation to $\hat Z$.} Another set of natural linear combinations is related to the $\hat Z_a$ invariant \cite{Gukov:2016gkn,Gukov:2017kmk}, which, rather than being specified by a root of unity, is labeled by $a\in \mathrm{Tor}\, H_1(M_3,Q(\frak{g}))/W$. These labels can be thought of as $S$-dual to abelian flat connections and are related to spin$^c$ structures for $G=\SL(2)$ \cite{gukov2021two} (see also \cite[Sec.~5.7]{Gukov:2025dol} for a recent discussion of the nature of the $a$ label). Therefore, they also define evaluation maps on skein classes, 
\begin{equation}
    \vev{L|a}\leadsto\hat Z_a(M_3,L),
\end{equation}
giving the generalization of the $\hat Z_a$ invariants to the pair $(M_3,L)$ studied in \cite[Sec.~4]{Gukov:2017kmk}. This set of states labeled by $a$ seems to have a more natural interplay with skein modules, as $q$ can take arbitrary values, unlike the WRT case where it is restricted to be a root of unity. This may provide a productive framework to study both $\hat Z$ and skein modules.\footnote{An interesting work in this direction, which appeared around the same time as the first version of the present paper, is \cite{Park:2026jis}.}

\paragraph{Unifying homological invariants of 3-manifolds.} At this point, it might be illuminating to compare several vector-space-valued invariants associated with 3-manifolds and discuss how they are related to each other through the two-dimensional $T[M_3\times S^1]$ theory.
\begin{itemize}
    \item \textbf{Khovanov homology.} The Khovanov homology \cite{khovanov2000categorification} and its generalizations (e.g.,~\cite{khovanov2008matrix,10.2140/gt.2008.12.1387,gukov2005khovanov,rozansky2010categorification,Witten:2011zz,Gukov:2016gkn,Gukov:2017kmk}) can be obtained from the BPS Hilbert space of the $T[M_3\times S^1]$ theory on an interval (or half-line) with a special ``tip-of-the-cigar'' boundary condition given by the shrinking of the internal $S^1$. When $M_3=S^3$, there is a canonical choice of the other boundary condition given by the unique flat connection on $S^3$. For more general $M_3$, a natural set of boundary conditions is given by the $a$-labels for $\hat Z$. Links in $M_3$ can be incorporated by either codimension-two or codimension-four defects in the 6d theory, leading to operators that are either space-filling from the 2d perspective or localized along the tip-of-the-cigar boundary condition, both of which modify the Hilbert space. The latter approach is closely related to our construction upon compactifying the time circle, while viewing the interval direction as the new time direction.
    \item \textbf{Hilbert spaces of the $\hat Z_a$-TQFT.} As $\hat Z$ is expected to define a 3d TQFT (though it can only satisfy a generalized version of the Atiyah--Segal axioms), one can study its cutting-and-gluing behavior along Riemann surfaces. The Hilbert spaces on $T^2$ and higher-genus surfaces have been studied in \cite{gukov2021two,Jagadale:2022abr,Guicardi:2025kqo}. Using the functoriality of compactification, these spaces are expected to be related to the Khovanov homology of $\Sigma\times S^1$ discussed in the previous point. It would be interesting to compare them and make this connection more precise.
    
    \item \textbf{Floer-like homologies.} The instanton Floer homology \cite{cmp/1104161987} and its variants and generalizations are related to Hilbert spaces of 4d $\cN=2$ theories on 3-manifolds with the Donaldson--Witten twist \cite{witten1988topological}. One way to construct such theories is by compactifying the 6d $(2,0)$ theory on a sphere with two irregular singularities \cite{witten1997solutions}. As this geometry can be viewed as an $S^1$-fibration over an interval, one should be able to recover the Floer homology from the $T[M_3\times S^1]$ theory on an interval with two boundary conditions similar to the tip-of-the-cigar one but with poles/singularities.
        
    \item \textbf{Vafa--Witten homology.} The Hilbert space $\cH(M_3)$ of the Vafa--Witten theory on $M_3$ is closely related to the BPS Hilbert space $\cH_{T[M_3\times S^1]}(S^1)$ of $T[M_3\times S^1]$ on $S^1$. They are not identical due to the difference between the 4d $\cN=4$ theory and the $T[T^2]$ theory. As  $\cH(M_3)$ can be recovered from the larger space by a ``moduli-space flow'' (i.e.~zooming in to the neighborhood of an SCFT singularity and flowing to the IR), one might expect that $\cH(M_3)$ is a quotient of $\cH_{T[M_3\times S^1]}(S^1)$, though it is not entirely clear whether any BPS states of the $\cN=4$ theory on $M_3$ can be extended to the 6d theory on $M_3\times T^2$ (i.e.~whether there are ``emergent states'' in $\cH(M_3)$). It would be interesting to better understand the exact relation between the two Hilbert spaces. 
    
    \item \textbf{Skein module.} Using our construction, the skein module of $M_3$ is embedded into the dual of the Vafa--Witten homology $\cH(M_3)^\vee$. However, as the construction can be uplifted to 6d, we expect that it can be further embedded into $\cH_{T[M_3\times S^1]}(S^1)^\vee$. In other words, $\cS$ does not contain ``emergent states'' since any $\bra L$ can be lifted to the 6d theory. One interesting consequence is that the TQFT structure of the 2d theory might lead to a non-trivial interplay between skein modules and Floer-like homologies via the open-closed map. More concretely, any state in the open Hilbert space on an interval with two boundary conditions $\cB$ and $\cB^\vee$ can propagate to a state in the $S^1$ Hilbert space, becoming a functional on skein classes in $\cS$. This connection between the open and closed sectors of the 2d TQFT is also related to the next point.
\end{itemize}

\paragraph{On skein-valued curve counting.} In the work of Ekholm and Shende \cite{Ekholm:2019yqp}, it was argued that the counting of holomorphic curves with boundaries on a Lagrangian $M_3$ in a Calabi--Yau three-fold $X$ is naturally valued in the HOMFLY-PT skein module of $M_3$. If one truncates to the SL($N$)-skein module, then the M-theory lift of our construction provides a natural physics realization for their work. The M-theory setup is 
\begin{equation}
    \begin{matrix}
{\mbox {\textrm{$N$ fivebranes:}}}~~~\qquad & \bR & \times & M_3& \times & \textrm{``cigar''}  \\ \qquad & &
&   \cap &  & \cap \\
{\mbox{\rm space-time:}}~~~\qquad & \bR&  \times  & X & \times & $Taub--NUT$   \\ \qquad & &
&   \cup &  & $\rotatebox[origin=c]{90}{$\in$}$ \\
{\mbox{\rm M2-brane:}}~~~\qquad &\bR & \times & C & \times  & O   \\
\end{matrix}
\label{CatGeom2}
\end{equation}
with the condition that $C$ is a holomorphic curve with $\partial C\subset M_3$. Here, ``cigar'' is topologically an $\bR^2$ subspace of the Taub--NUT space, but should be better thought of as a circle-fibration over $\bR^+$ with a smooth tip and approaching constant size at infinity. So far, this is exactly the same as the M-theory definition of $\hat Z$, which counts BPS states given by such M2-branes. To relate to our construction, we compactify the first $\bR$ factor into a circle and make it---as well as the fiber $S^1$ of the cigar---small compared to the size of the rest of the geometry. This leads to almost the same system used in our construction with a small difference---the 4d theory is now the ``$T[T^2]$ theory,'' containing the super--Yang--Mills theory as a subsector at low energy that we can restrict ourselves to. An M2-brane wrapping $S^1\times C$ with $\partial C=\gamma\subset M_3$ then generates a state $\bra{\gamma}$ with $\gamma$ colored by the defining representations of SL$(N)$, naturally living in $\cS\subset \cH(M_3)^\vee$. One can also fix $\gamma$, and consider the superposition of all possible configurations of M2-branes ending on $M_3$ along $\gamma$. This ``Hartle--Hawking'' state is then $\#\cdot\bra{\gamma}$ with the coefficient counting such holomorphic curves. The invariance of this state under skein relations then follows from the choice of the boundary condition of the 4d theory at $x=0$. It would be interesting to better understand this physical system to give interpretations for some of the mysterious features of \cite{Ekholm:2019yqp}, such as certain phase factors dependent on a choice of a 4-chain bounding $M_3$ in $X$.

\subsection{Future directions}

We have already mentioned several interesting directions to explore, ranging from a better understanding of the physical details of the construction\footnote{For example, in this paper, we do not fully justify the invariance of $\cS$ under the deformation and RG flow. This is a point that merits further investigation. In particular, it would be interesting to see how the ghost number argument can be utilized to impose non-trivial constraints on the behavior of the BPS Hilbert space under the RG flow. See also \cite{neitzke2020q,neitzke2022quantum} for related studies on the relation between line operators in the UV and IR.} to testing Langlands duality for $B$--$C$ pairs and to comparisons with the geometry of the moduli spaces of Higgs bundles. We will suggest several additional interesting problems for future exploration below. 

\paragraph{The dual computation.} As expected from $S$-duality, one should be able to reproduce our results in the dual theory with gauge group $^\mathrm{L}\!K$ and with insertions of 't Hooft operators along a Nahm-pole boundary condition at $x=0$. One can also consider this boundary condition, as well as its $(p,q)$-generalizations, on the original side with gauge group $K$, which would be needed for the enriched skein module $\tilde \cS$. One starting point for this computation is to determine whether such a boundary condition is compatible with the $\cN=1$ deformation  in the bulk (or an analogous one) that we used, so as to allow the system to be studied in a similar manner as a collection of different IR phases when $M_3$ has reduced holonomy.

\paragraph{Skein algebra and category.} One omission in this work is the treatment of skein algebras and skein categories \cite{walker2006tqfts} associated with a Riemann surface $\Sigma$. To study these, one can simply replace the circle factor with either the interval $I$ or $\bR$, which still allows one to turn on the $\cN=1$ deformation. A natural place to start is $\Sigma=T^2$, for which cosmic strings are not present after the deformation. It would be interesting to reproduce the description of the skein categories in \cite{gunningham2024skeins} from boundary conditions of the 2d theory, which should be closely related to the category of branes and $\bC^*$-fixed points on the target of the 2d theory viewed as a sigma model (cf.~\cite{Gukov:2022gei}). 

\paragraph{More general three-manifolds.} As the $\cN=1$ deformation requires a Kähler structure, our focus in this paper is mostly on the case of $M_3=\Sigma\times S^1$. However, there are many interesting results about the compactification of 4d or even higher-dimensional theories on various three-manifolds (see \cite{dimofte2011vortex,terashima2011text,Cecotti:2011iy,dimofte2014gauge,Dimofte:2011py,Gadde:2013sca,pei20163d,chung20163d,Chun:2019mal} for a small sample). One potentially useful perspective is to consider the 2d $(2,2)$ $T[M_3\times S^1]$ theory obtained from compactifying 5d $\cN=2$ super--Yang--Mills theories on $M_3$. Such theories in the IR are often described (possibly after deformations) by a twisted effective superpotential with finitely many discrete critical points and the computation can simply reduce to counting them, perhaps with multiplicities and some equivalence relations.\footnote{One example where this can be carried out is when $M_3=L(k,1)$, and the skein module can indeed be viewed as the space of functions on the massive vacua of the 2d theory. This system can be understood from various angles. See \cite{Gukov:2015sna,Gukov:2016lki,Andersen:2016hoj,Gukov:2021swm}.   
One feature that arises when the 2d theory has only massive vacua is that one can ``diagonalize the fusion rule'' in the theory $T[M_3\times S^1]$. It would be interesting to understand this from the skein module side.} Some three-manifolds that are ``one step away'' from $\Sigma\times S^1$ are circle bundles over $\Sigma$ and mapping tori. For the latter, it would be interesting to connect the physics understanding of the $T[M_3]$ theory in \cite{Chun:2019mal} with recent results on the skein modules of such three-manifold in \cite{bierent2025skein}. Another potentially interesting way to deal with more general $M_3$ is to use the fact that every oriented smooth 4-manifold can be decomposed into two Kähler manifolds \cite{baykur2006kahler}. If $M_3\times \bR$ can be similarly decomposed, one can apply the $\cN=1$ deformation on the two parts and glue together the skein modules.

\paragraph{Torsion in sk$(M_3)$.} Another interesting question concerns detecting torsion elements in the skein module. Naively, this seems to be a blind spot in our approach, as torsion should be related to jumps when $q$ is a root of unity, but the information about $q$ seems to disappear once we flow to the infrared phases after the deformation. To access roots of unity, one needs to consider strong-coupling limits with Im$\,\tau\to 0$. It would be interesting to understand whether the $\cN=1$ deformation can still be exploited, even in this limit, to give control over torsion in the skein module of $\Sigma\times S^1$.\footnote{From this point of view, it might not be a coincidence that the $M_3=S^1\times S^2$ case---for which the deformation cannot be applied---is also the case where we know that there are complicated torsions.} Another way to study such limits is by working in the weakly coupled $S$-dual frame. Moreover, at special values of the coupling constant of the UV theory, there might be higher-group/non-invertible symmetries or special types of defects, which could also help to understand the torsion part of the module (cf.~Footnote~\ref{foot:torsion}).

\paragraph{Large-$N$ limit and gravity dual.} When $M_3$ is $S^3$ or a lens space $L(k,1)$, the large-$N$ limit of Chern--Simons theory can be described by a topological string theory on a local Calabi--Yau threefold obtained from $T^*M_3$ via the ``conifold transition''  \cite{Gopakumar:1998ki,Ooguri:1999bv,Aganagic:2002wv}. Although it is not clear how this process can be carried out for more general $M_3$, there are hints that the systems we considered, either in four dimensions or higher, have well-defined and interesting large-$N$ limits. For example, the skein-valued curve counting of \cite{Ekholm:2019yqp} concerns HOMFLY-PT skein modules, and the skein relations can be explained from the perspective of topological strings \cite{vafa2025chern}. On the other hand, the $\cN=1$ deformation of the maximally supersymmetric 4d Yang--Mills theory does have a well-studied string-theory dual due to Polchinski and Strassler \cite{Polchinski:2000uf}, and it would be interesting to see whether such an embedding into string theory sheds further light on the properties of skein modules, skein algebras, and skein categories.

\bigskip

We hope that the present work provides a useful starting point for further investigations along these lines.

\section*{Acknowledgment}

We thank Peter Gothen, Tamas Hausel, Maxim Kontsevich,  Qiongling Li, Satoshi Nawata, Pavel Putrov, Fernando Rodriguez-Villegas, Nicolas Seroux, Vivek Shende, Pavel Safronov, and Cumrun Vafa for helpful discussions. We are especially grateful to Sergei Gukov, David Jordan, and Penghui Li for various illuminating conversations over the years that sparked and sustained the author's interest in this topic. The author also thanks the organizers and participants of the ReNewQuantum Closing Conference and the ``Gauge Summer with BV 2025'' workshop, where parts of this work were presented, as well as the Yau Center for Mathematical Sciences at Tsinghua University and the Shanghai Institute for Mathematics and Interdisciplinary Sciences for hospitality during the author's visits. The work of the author is partly supported by Research Grant 42125 from Villum Fonden, the ERC-SyG project No.~810573 ``Recursive and Exact New Quantum Theory,'' and the Simons Collaboration on ``New Structures in Low-Dimensional Topology.''

\appendix

\section{Some background on the construction}\label{app:Details}

In this appendix, we briefly review some technical background relevant for the embedding of $\sk(M_3)$ into the gauge theory.

\subsubsection*{The topological twist}

To study the Hilbert space of the twisted 4d $\cN=4$ super--Yang--Mills theory with gauge group $K$ on a spatial three-manifold $M_3$, we consider the theory on the geometry $M_3\times \bR$. The topological twist is classified by homomorphisms of the Euclidean group Spin$(3)_E$ to the R-symmetry group SU$(4)_R\simeq\mathrm{Spin}(6)_R$ such that at least one supercharge becomes a scalar over $M_3$ under the twisted Euclidean group $\Spin(3)'_E$. 

 In 4d, there are three classes of twist: the Donaldson--Witten twist (and its orientation reversal) \cite{witten1988topological}, the Vafa--Witten twist \cite{Vafa:1994tf} (see also \cite{yamron1988topological}), and the GL twist \cite{Kapustin:2006pk}, which is parametrized by a variable $t\in \CP^1$ and, for $t= \pm i$, it is also sometimes known as the Marcus twist \cite{Marcus:1995mq}.  

On the geometry $M_3\times \bR$, which has reduced holonomy, the backgrounds for the Vafa--Witten (and its orientation reversal) and the GL/Marcus twists become identical, both described by the natural embedding Spin$(3)_E\to\mathrm{Spin}(6)_R$. An equivalent way to state this is that the spinor representation $\mathbf{4}$ of $\mathrm{Spin}(6)_R\simeq \SU(4)_R$ pullbacks to $\mathbf{2}\oplus \mathbf{2}$ of $\Spin(3)_E\simeq \SU(2)_E$. As a consequence, four out of the 16 supercharges now become scalars,
\begin{equation}
    2\times(\mathbf{2},\mathbf{4})\text{ under $\Spin(3)_E \times \mathrm{Spin}(6)_R$} \quad \leadsto \quad 4\times \mathbf{1}\oplus 4\times \mathbf{3} \text{ under $\Spin(3)'_E$}.
\end{equation}
Therefore, with this partial twist along $M_3$, one obtains an $\cN=4$ supersymmetric quantum mechanics (SQM) along $\bR$, and different topological theories correspond to picking a nilpotent linear combinations of the four supercharges. 

\subsubsection*{Supercharges and R-symmetry}

The SUSY algebra of the 1d $\cN=4$ theory has an SO$(4)$ R-symmetry, but in general it is not a symmetry of the full theory. A genuine symmetry is the centralizer $\SU(2)_R\simeq\Spin(3)_R \subset\mathrm{Spin}(6)_R$ of the image of $\Spin(3)_E$, which is already present in the Vafa--Witten theory or as an enhancement of the U$(1)$ ``ghost-number'' symmetry in the Kapustin--Witten theory. 

To describe its action on the supercharges, we first label them as $Q_{1,\ell}$, $Q_{2,\ell}$, $Q_{1,r}$ and $Q_{2,r}$. The notation reflects the fact that two among the four preserved supercharges of the 1d theory were left-moving in 4d before the twist, while the other two were right-moving. 

In the Lorentzian signature, they would satisfy reality conditions $Q_{1,\ell}^\dagger=Q_{1,r}$ and $Q_{2,\ell}^\dagger=Q_{2,r}$, and one can think of them as linear combinations of the four real supercharges $Q_{\alpha,i}$ of the 3d $\cN=2$ ``$T[M_3]$ theory'' obtained by reducing a 6d $(2,0)$ theory. From this perspective, the first index $\alpha=\{1,2\}$ denotes the two components of $Q$'s as Majorana spinors in 3d, which form a vector representation of SO$(2)_E$ of the 3d Minkowski space, while $i=\{1,2\}$ is the index for the two-dimensional vector representation of the SO$(2)_R$ symmetry. These two abelian symmetries each pick out some special complex combinations, with
\begin{equation}
    \cQ_\ell:=Q_{11}+iQ_{12}-iQ_{21}+Q_{22}\quad \text{and} \quad \cQ_r:=Q_{11}+iQ_{12}+iQ_{21}-Q_{22}
\end{equation}
being the two ``eigenstates'' of SO$(2)_R$ with charge $1$, and
\begin{equation}
    \cQ_1:=Q_{11}+iQ_{21}\quad \text{and} \quad \cQ_2:=Q_{12}+iQ_{22}
\end{equation}
giving two weight-1 irreducible representations of SO$(2)_E$. The latter two are in fact just $Q_{1,r}$ and $Q_{2,r}$ (modulo normalization factors), while the first two can also be written as $\cQ_\ell=Q_{1,\ell}+iQ_{2,\ell}$ and $\cQ_r=Q_{1,r}+iQ_{2,r}$, which explains the choice for their subscripts. To make the two groups more symmetric, one can instead choose the latter two to be $\cQ_r=\cQ_1+i\cQ_2$ and $\bar\cQ_\ell=\cQ_1-i\cQ_2$, which are simultaneous eigenstates under both SO(2)$_R$ and SO(2)$_E$. 

When reducing to 1d, SO(2)$_R$ and $\SO(2)_E$ will be part of the SO(4) R-symmetry of the $\cN=4$ supersymmetric quantum mechanics, enhanced to the two subgroups $\SU(2)_R$ and $\SU(2)_{R'}$ that respectively act on $(\cQ_1,\cQ_2)$ and $(\cQ_\ell,\cQ_r)$ as doublet. The actions of their centers are identical, generated by $-I\in \SO(4)$ that flips all real supercharges.

Another nice property of these linear combinations is that they are automatically nilpotent in the 1d theory, $\cQ_\ell^2=\cQ_r^2=\cQ_1^2=\cQ_2^2=0$. In fact, any complex linear combinations $u\cQ_\ell+v\cQ_r$ or $u'\cQ_1+v'\cQ_2$ are nilpotent. Conceptually, this can be understood as a consequence of the absence of suitable operators with weight two under either of the two U$(1)$ factors in the Cartan of the SO(4) R-symmetry. More concretely, this follows from the fact that $Q_{\alpha,i}^2\propto H$ is proportional to the Hamiltonian while anticommuting with each other in the SQM.

Although it is perfectly justified to work with the Minkowski spacetime in our setup, to relate the supercharges of the 1d $\cN=4$ theory with nilpotent supercharges after topological twists on general four-manifolds (i.e.~not assumed to be just $M_3\times\bR$), it is convenient to work in the Euclidean signature. There, each of the four real supercharges can be regarded as complex, or, equivalently, the complex supercharges can be treated as independent of their complex conjugates.

Then $\cQ_\ell$ and $\cQ_r$ are the two complex supercharges preserved in the Kapustin--Witten theory, while $\cQ_1$ and $\cQ_2$ are preserved in the Vafa--Witten theory. From the 6d perspective, they generate respectively a $(1,1)$ and a $(0,2)$ supersymmetry algebra in the remaining two spacetime dimensions after a partial twist on $M_4$. When $M_4=M_3\times S^1$ (i.e.~with the Euclidean time $\bR$ compactified), the two twists involves the same supersymmetric background, but the choice of the supercharges is still different, living inside respectively a $(1,1)$ and $(0,2)$ subalgebra of the 2d $(2,2)$ theory $T[M_3\times S^1]$.

The $\SU(2)_R$ symmetry preserves ``chirality'' (in either the 4d sense or the 2d/SO(2)$_E$ sense, which are equivalent) and, as it is a genuine symmetry, can be used to relate $\cQ_1$ and $\cQ_2$ (or any complex linear combinations of them). The quotient of the naive SO(4) R-symmetry by this genuine symmetry is the ``wrong'' SO(3)$_{R'}$, which can be thought of as the $\bZ_2$-quotient of the SU$(2)_{R'}$ that acts on the doublet $(\cQ_\ell,\cQ_r)$ with different chiralities. From the point of view of the twisting parameter $t$, this SO(3) acts transitively on the $\CP^1\simeq S^2$ where $t$ takes values via rotations and therefore relates different topological twists given by $\cQ_t:=\cQ_{\ell}+t\cQ_{1,r}$.

Since we have the linear relation
\begin{equation}
    \cQ_r=\cQ_1+i\cQ_2,
\end{equation}
the Kapustin--Witten theory on $M_3\times \bR$ at $t=\infty$ is equivalent to the Vafa--Witten theory.

\subsubsection*{Hilbert space} 

The 1d theory on $\bR$ has a Hilbert space $\cH_{\text{full}}$, from which one can construct the Hilbert space $\cH(M_3)$ of the twisted theory by taking the $\cQ_t$-cohomology. The space $\cH(M_3)$ naively depends on both the coupling constant $\tau$ of the 4d theory as well as the twist parameter $t$. However, the twisted theory only depends on them via a combination---referred to as the ``canonical parameter'' in \cite{Kapustin:2006pk}---given by 
\begin{equation}
    \Psi=\frac{\tau+\bar\tau}{2}+\frac{\tau-\bar\tau}{2}\left(\frac{t-t^{-1}}{t+t^{-1}}\right).
\end{equation}
It is a complex variable, but the theory is invariant under  $\Psi\mapsto \Psi+1 $, and one can use its exponential $q=e^{2\pi i\Psi}$ to parametrize $\cH(M_3)$. 

The action of $\SU(2)_R$ on $\cH_{\text{full}}$ is expected to be a symmetry. However, as it transforms $\cQ_t$, in general it is not a symmetry of the Hilbert space of the twisted theory. Instead, it guarantees that a family of twists given by the orbit of $\cQ_t$ under the $\SU(2)_R$ action all lead to equivalent theories. For the Cartan of SU$(2)_R$, it acts on $\cQ_t$ via a phase, thus becoming a ghost-number symmetry on the space of $\cQ_r$-closed states in $\cH_{\text{full}}$. On the other hand, while the action of the SO(3) quotient still preserves nilpotency of the supercharge, it doesn't preserve the canonical parameter $\Psi$, which is one way to see that it is not a true symmetry of the theory as the twisted theories can be inequivalent for different values of $\Psi$.\footnote{We thank Pavel Safronov for pointing out an erroneous claim regarding the R-symmetry action in the first version of the manuscript as well as providing other helpful comments.}

It is also easy to see that there cannot be more general nilpotent supercharges beyond the SO(4)-orbits of $\cQ_t$. In fact, any complex linear combinations of the four real supercharges can be transformed to $Q_{11}+c\cdot Q_{21}$ up to SO$(4)$ rotations and a complex scalar. Nilpotency then requires $c=\pm i$, but the two choices are also related by the SU(2)$_{R'}$ action. Therefore, all nilpotent supercharges are in the SO$(4)$ orbit of $Q_{11}+iQ_{21}=\cQ_1$, which also include $\cQ_\ell$, $\cQ_r$ and their linear combinations.

In our construction of the embedding of the skein module into the dual of $\cH(M_3)$, we can consider arbitrary values of $t$ in the beginning. However, to use the deformation of the Vafa--Witten theory, we have to restrict to $t=\infty$ and use the right-moving supercharge $\cQ_r$. Then we have $\Psi=\tau$, and $q=e^{2\pi i\tau}$ takes values inside the unit circle. To access points on the unit circle, one needs to consider the strong-coupling limit with Im$\,\tau\to 0$. It might be desirable to map it to a theory with weak couplings using $S$-duality, which is our next topic.

\subsubsection*{$S$-duality}

Under SL$(2,\bZ)$, the canonical parameter transforms as 
\begin{equation}
    \Psi\mapsto \frac{a\Psi+b}{c\Psi+d}.
\end{equation}
The generator $S\in\SL(2,\bZ)$ sends the coupling constant $\tau$ to $-\frac{1}{n_{\frak g}\tau}$ with $n_{\frak g}$ being the lacing number of $\frak g$ (in general a collection of numbers, one for each simple or abelian summand of $\frak g$). Then one can check that the canonical parameter also transforms as 
\begin{equation}
    S: \quad \Psi\mapsto -\frac{1}{n_{\frak g}\Psi}.
\end{equation}
The $S$-duality acts on supercharges and, consequently, on the $t$-parameter as
\begin{equation}
    t\mapsto \frac{|\tau|}{\tau}\cdot t.
\end{equation}
Notice that the action is by a phase and has $t=0$ and $\infty$ fixed. Therefore, one can consistently use $\cQ_r$ in the construction of $\cH(M_3)$ on both sides of the duality.

Many aspects of the $S$-duality of the 4d $\cN=4$ super--Yang--Mills theory have been studied in the literature, and one fact we used in the paper is that the relative theory---the one with mutually non-local Wilson and 't Hooft operators but being consistent as a boundary condition of a 5d TQFT---also has $S$-duality. One way to construct this theory is by compactifying a relative 6d $(2,0)$ theory on $T^2$ (possibly with an automorphism twist in order to obtain non-simply-lace factors) and restrict oneself to the neighborhood of an SCFT point of the moduli space. The $\SL(2,\bZ)$ is then interpreted as the mapping class group of $T^2$ (when an automorphism twist is turned on, the true dualities live in a subgroup that preserves the twist). The relative theory can have both electric and magnetic symmetries, giving rise to a doubly graded Hilbert space on $M_3$.

\subsubsection*{Boundary conditions and line operators}

In \cite{Witten:2011zz}, a Neumann-like boundary condition is constructed to realize the analytically continued Chern--Simons theory on the boundary of the topologically twisted $\cN=4$ theory. This boundary condition can be lifted to 6d (or M-theory for $A$-type gauge groups) as the tip of a Melvin cigar. 

On the boundary, one can construct supersymmetric Wilson lines that are compatible with the nilpotent supercharge. After the topological twist, they then become topological line operators in the analytically continued Chern--Simons theory. One crucial property we used throughout the paper is that they obey skein relations. Another way to state this is that, for any state $\ket\psi\in\cH(M_3)$ (e.g.~viewed as an ``incoming'' one from $x=\infty$), it gives a functional on skein classes of line operators on the $x=0$ boundary. Turning this around and assuming that there are enough ``test states,'' one obtains an embedding of sk($M_3$) into $\cH(M_3)^\vee$.

While the above boundary condition is realized by the D3-NS5 system (potentially at a non-trivial theta angle), its $S$-dual is realized by the D3-D5 system, where there can be supersymmetric 't Hooft line operators. The dual boundary condition and the 't Hooft operators are used in the paper to enlarge the skein module to the enriched version $\tilde\cS$.

To be more precise, both $K$ and $^\mathrm{L}\!K$ theories have the two types of boundary conditions and they are dual to each other by switching the type and changing $K$ to its Langlands/GNO dual,
\begin{equation}
    \cB_{\text{D5}}(K)\mapsto \cB_{\text{NS5}}(^\mathrm{L}\!K),\quad \text{and}\quad \cB_{\text{NS5}}(K)\mapsto \cB_{\text{D5}}(^\mathrm{L}\!K).
\end{equation}
To generate $\tilde\cS$, we need to use $\cB_{\text{D5}}(K)$ and $\cB_{\text{NS5}}(K)$ decorated with, respectively, Wilson and 't Hooft lines, as well as more general dyonic lines living on $(p,q)$-type boundary conditions. Notice that, although we say that $\cB_{\text{D5}}(K)$, $\cB_{\text{NS5}}(K)$ and $\cB_{(p,q)}(K)$ are dual boundary conditions, they are actually on the ``same side'' of the $S$-duality.

\section{Finite-group gauge theory}\label{app:DW}

In this appendix, we derive some results on finite-group gauge theories used in the paper. They are sometimes referred to as Dijkgraaf--Witten theories \cite{dijkgraaf1990topological}, although we will not turn on a non-trivial Dijkgraaf--Witten twist/cocycle. Many aspects of the theory has been studied in \cite{Freed:1991bn} with various results scattered in the mathematical and physics literature.

We are interested in 4d theories with gauge groups $G=S_3, S_4, S_5$, D$_8$, or Q$_8$. In particular, we want to compute the partition function on $\Sigma \times T^2$ expressed as a function of the genus $g$ of $\Sigma$. As this is the partition function of a semisimple 2d TQFT, it takes the general form
\begin{equation}
    Z_g(G):=Z(\Sigma\times T^2;G)=\sum_{i\in\Lambda} \lambda^{g-1}_i, 
\end{equation}
and our task is to determine the set of labels $\Lambda$ and the ``fusion eigenvalues'' $\lambda_i$.

\subsection{Compactification}

One can use the two circle factors to reduce the 4d theory to three and two dimensions. Such a finite-group gauge theory on a circle will have holonomy sectors labeled by conjugacy classes $C(G)$ of $G$. In the sector given by a holonomy $[x]\in C(G)$, the gauge group is broken to the stabilizer subgroup $G^x$ in lower dimensions. The 3d theory in each sector counts the flat $G^x$-connections over $\Sigma$, leading to  
\begin{equation}
    Z_g(G)=\sum_{[x]\in C(G)} \left|\cM(\Sigma,G^x)\right|,
\end{equation}
where we have used the notation $\cM(\Sigma,G^x):=\Hom(\pi_1(\Sigma_g),G^x)/G^x$ for the moduli space of flat $G^x$-connections. 

The expression above is manifestly integral, but is not of the ``canonical form.'' To obtain a TQFT-like formula, one can perform this operation one more time to get to 2d, 
\begin{equation}
    Z_g(G)=\sum_{(x,y)\in \cM(T^2,G)} \left|\Hom(\pi_1(\Sigma_g),G^{x,y})\right|/|G^{x,y}|.
\end{equation}
Here, the sum is over the commuting pairs $(x,y)$ in $G$ modulo conjugation, and $G^{x,y}$ is the centralizer of such a pair. Now, each summand is the partition function of a 2d TQFT, 
\begin{equation}
\left|\Hom(\pi_1(\Sigma_g),G^{x,y})\right|/|G^{x,y}|=\sum_{R\in\mathrm{Irr}(G^{x,y})} \left(\frac{|G^{x,y}|}{\dim R}\right)^{2g-2},
\end{equation}
where the sum is over irreducible representations (``irreps'') of $G^{x,y}$. This procedure also gives a way to think of $\Lambda$ as the union over irreducible representations of $G^{x,y}$ for all commuting pairs, with the fusion eigenvalues given by $\left(|G^{x,y}|/\dim R\right)^{2}$. It is a well-known result in representation theory that this ratio is an integer, which guarantees that the entire partition function is always integral when $g\ge 1$.

We now work out these data explicitly for the finite groups of interest.

\subsection{$S_3$}

There are three types of commuting pairs, the trivial pair $(e,e)$, these containing $a=(12)$, and these with $b=(123)$. For the second type, there are three classes: $(e,a)$, $(a,e)$ and $(a,a)$, while for the third type there are four classes: $(e,b)$, $(b,e)$, $(b,b)$, and $(b,b^2)\sim(b^2,b)$. The centralizers for the three types are respectively $S_3$, $\bZ_2$ and $\bZ_3$, and except for a two-dimensional representation of $S_3$, all the other irreps are one dimensional. Therefore, we have the following list of fusion eigenvalues:
\begin{itemize}
    \item for the first type, $\{6,6,3\}$;
    \item for the second class, $\{2,2\}\times3$;
    \item for the third class, $\{3,3,3\}\times4$.
\end{itemize}
Putting them together, we have the partition function given by
\begin{equation}
    Z_g(S_3)=2\cdot 6^{2g-2}+13\cdot 3^{2g-2}+ 6\cdot 2^{2g-2}.
\end{equation}
When $g=1$, this gives 21, which is also the number of commuting triples modulo conjugation. For $g=0$, we have $Z_0(S_3)=3$, which is the number of irreps of $S_3$. This agrees with the expectation that the partition function of the 4d finite-group gauge theory on $S^2\times T^2$ is the same as the partition function of the 2d gauge theory on $T^2$ with the same gauge group $G$, which equals the number of irreps (or conjugacy classes). 

\subsection{$S_4$}

There are five conjugacy classes in $S_4$, represented by the identity $e$, transposition $a=(12)$, 3-cycle $b=(123)$, 4-cycle $c=(1234)$ and the double transposition $d=(12)(34)\sim d'=c^2=(13)(24)$.  

For commuting pairs, there turns out to be 21 conjugacy classes:
\begin{itemize}
    \item $(e,e)$ with the centralizer $S_4$ and fusion eigenvalues $\{24,24,12,8,8\}$;
    \item $(e,a)$, $(a,e)$ and $(a,a)$ with the centralizer $\bZ_2\times \bZ_2$ and fusion eigenvalues $\{4,4,4,4\}\times 3$;
    \item $(e,b)$, $(b,e)$, $(b,b)$ and $(b,b^2)$ with the centralizer $\bZ_3$ and fusion eigenvalues $\{3,3,3\}\times 4$;
    \item $(e,c)$, $(c,e)$, $(c,c)$, $(c,c^2)$, $(c^2,c)$, and $(c,c^3)$ with the centralizer $\bZ_4$ and fusion eigenvalues $\{4,4,4,4\}\times 6$;
    \item $(e,d)$, $(d,e)$ and $(d,d)$ with the centralizer D$_8$ and fusion eigenvalues $\{8,8,8,8,4\}\times 3$;
    \item $(a,d)$, $(d,a)$, $(a,ad)$ and $(d,d')$ with the centralizer $\bZ_2\times \bZ_2$ and fusion eigenvalues $\{4,4,4,4\}\times 4$.
\end{itemize}
Collecting these, we have
\begin{equation}
    Z_g(S_4)=2\cdot 24^{2g-2}+12^{2g-2}+ 14\cdot 8^{2g-2}+55\cdot 4^{2g-2}+ 12\cdot 3^{2g-2}.
\end{equation}
The partition function on $T^4$ is $Z_1(S_4)=84$, which is used for computing the dimension of sk($T^3,F_4$), while $Z_0(S_4)=5$ equals the number of irreps of $S_4$, providing a consistency check.

\subsection{$S_5$}

While it might seems intimidating to performing the analogous computation for $S_5$, which is only relevant for $E_8$, it turns out to be not too much work, as we can build upon the results for $S_4$.

There are seven conjugacy classes in $S_5$. In addition to the five classes $e$, $a=(12)$, $b=(123)$, $c=(1234)$, and $d=(12)(34)\sim c^2$, there are also the 5-cycle $f=(12345)$ and the product $g=(123)(45)$.  

For commuting pairs, there are 39 conjugacy classes, including the 21 classes of $S_4$ plus 18 new ones. Some of the 21 conjugacy classes of $S_4$ now have different centralizers:
\begin{itemize}
    \item $(e,e)$ now has centralizer $S_5$ and fusion eigenvalues $\{120,120,30,30,24,24,20\}$;
    \item $(e,a)$, $(a,e)$ and $(a,a)$ now have centralizer $\bZ_2\times S_3$ and fusion eigenvalues $\{12,12,6\}\times 6$;
    \item $(e,b)$, $(b,e)$, $(b,b)$ and $(b,b^2)$ now have centralizer $\bZ_3\times \bZ_2$ and fusion eigenvalues $\{6\}\times 24$;
\end{itemize}
while some of them remain the same:
\begin{itemize}
    \item $(e,c)$, $(c,e)$, $(c,c)$, $(c,c^2)$, $(c^2,c)$, and $(c,c^3)$ still have centralizer $\bZ_4$ and fusion eigenvalues $\{4\}\times 24$;
    \item $(e,d)$, $(d,e)$ and $(d,d)$ still have centralizer D$_8$ and fusion eigenvalues $\{8,8,8,8,4\}\times 3$;
    \item $(a,d)$, $(d,a)$, $(a,ad)$ and $(d,d')$ with the centralizer $\bZ_2\times \bZ_2$ and fusion eigenvalues $\{4\}\times 16$.
\end{itemize}
The new conjugacy classes are:
\begin{itemize}
    \item $(e,f)$, $(f,e)$, $(f,f)$, $(f,f^2)$, $(f,f^3)$ and $(f,f^4)$ with the centralizer $\bZ_5$ and fusion eigenvalues $\{5\}\times 30$;
    \item $(e,g)$, $(g,e)$, $(g,g)$, $(g,g^3)$, $(b,g)$, $(g,b)$, $(b,g^3)$, $(g^3,b)$, $(a',b)$, $(b,a')$ $(g,a')$, and $(a',g)$, where $a'=(45)$, with the centralizer $\bZ_3\times \bZ_2$ and fusion eigenvalues $\{6\}\times 72$.
\end{itemize}

Collecting these, we have
\begin{equation}
    Z_g(S_5)=2\cdot 120^{2g-2}\!+2\cdot 30^{2g-2}\!+ 2\cdot 24^{2g-2}\!+20^{2g-2}\!+12\cdot 12^{2g-2}\!+ 102\cdot 6^{2g-2}\!+12\cdot 8^{2g-2}\!+30\cdot 5^{2g-2}\!+43\cdot 4^{2g-2}.
\end{equation}
The partition function on $T^4$ is $Z_1(S_5)=206$, while $Z_0(S_4)=7$ again equals the number of conjugacy classes of $S_5$.

If one is only interested in $Z_1$, another way to obtain this answer is by noticing that the partition function of a finite-group gauge theory on $T^4$,
\begin{equation}
    Z(T^4;G)=|\Hom (\bZ^4,G)|/|G|,
\end{equation}
counts the number of commuting quadruples divided by the order of the group. For $G=S_n$, the answers fit nicely into a generating series \cite{white2013counting}
\begin{equation}
    F(x):=\sum_{n=1}^\infty Z(T^4;S_n)\cdot x^n=(1-x)^{-1} (1-x^2)^{-7} (1-x^3)^{-13} (1-x^4)^{-35} (1-x^5)^{-31}\cdots
\end{equation}
where the powers (A001001 in OEIS) are the negatives of 
\begin{equation}
    c(n)=\sum_{d|n}d^2\sigma(n/d)
\end{equation}
with $\sigma$ being the divisor sum function. Expanding this infinite product, one has
\begin{equation}
    F(x)=1+x+8x^2+21x^3+84x^4+206x^5+717x^6+\cdots
\end{equation}
This indeed gives the correct answer for $Z_1(S_0=S_1=\{e\})=1$, $Z_1(S_2=\bZ_2)=8$, $Z_1(S_3)=21$, and $Z_1(S_4)=84$, and, furthermore, gives the desired $Z_1(S_5)=206$.

We end our exploration in finite gauge theory with two simpler cases, D$_8$ and Q$_8$, which are relevant for computations in the $B$- and $D$-series.

\subsection{$\mathrm{D}_8$ and $\mathrm{Q}_8$}

The dihedral group D$_8$ is a subgroup of O(2) that acts as the symmetry of a square. It is generated by an order-4 rotation $r$ and a reflection $s$ with the relation $rsr=r^{-1}$. The eight elements of D$_8$ can be presented as $\{e,r,r^2,r^3,s,rs,r^2s,r^3s\}$, and there are 5 conjugacy classes represented by $e$, $r\sim r^3$, $r^2$, $s\sim r^2s$, and $rs\sim r^3s$. 
For commuting pairs, there turns out to be 22 conjugacy classes, which can be grouped into the following types:
\begin{itemize}
    \item $(e,e)$, $(e,r^2)$, $(r^2,e)$, and $(r^2,r^2)$ with the centralizer the entire D$_8$ and fusion eigenvalues $\{8,8,8,8,4\}\times 4$;
    \item $(e,r)$, $(r,e)$, $(r,r)$, $(r,r^2)$, $(r^2,r)$ and $(r,r^3)$  with the centralizer $\bZ_4$ and fusion eigenvalues $\{4,4,4,4\}\times 6$;
    \item $(e,s)$, $(s,e)$, $(s,s)$, $(s,r^2)$, $(r^2,s)$,  and $(s,r^2s)\sim (r^2s,s) $ with the centralizer $\bZ_2\times \bZ_2$ and fusion eigenvalues $\{4,4,4,4\}\times 6$;
    \item $(e,rs)$, $(rs,e)$, $(rs,rs)$, $(r^2,rs)$, $(rs,r^2)$, and $(rs,r^3s)$ with the centralizer $\bZ_4$ and fusion eigenvalues $\{4,4,4,4\}\times 6$;
\end{itemize}
The partition function on $\Sigma\times T^2$ is then
\begin{equation}
    Z_g(\mathrm{D}_8)=16\cdot 8^{2g-2}+ 76\cdot 4^{2g-2}.
\end{equation}
One has $Z_1=92$ and $Z_0=5$, which agrees with the five conjugacy classes of D$_8$. 

The answer for the quaternion group Q$_8$ turns out to be exactly the same, 
\begin{equation}
    Z_g(\mathrm{D}_8)=Z_g(\mathrm{Q}_8).
\end{equation}
Although we do not know a simple argument or a deeper reason for this, it is straightforward to check this explicitly.

The group Q$_8=\{\pm1,\pm I, \pm J,\pm K\}$ also has 22 conjugacy classes of commuting pairs, falling into two classes:
\begin{itemize}
    \item $(\pm1,\pm1)$ with the centralizer the entire Q$_8$ and fusion eigenvalues $\{8,8,8,8,4\}\times 4$;
    \item $(\pm1,I)$, $(I,\pm 1)$, $(I,\pm I)$, as well as 12 similar ones with $I$ replaced with $J$ and $K$, which have a $\bZ_4$ centralizer and fusion eigenvalues $\{4,4,4,4\}\times 18$.
\end{itemize}
These indeed give the same 2d TQFT as D$_8$. This equivalence should be a useful ingredient in computations for the $B$- and $D$-series, which we just scratched the surface and will leave a deeper analysis to the interested reader.

\section{\texorpdfstring{Flat $N_2$-connections on $T^3$}{Flat N_2-connections on T^3}}\label{app:NT}

In this appendix, we count components of flat
$N_2$-connections, where $N_2$ is the normalizer of a maximal torus $N(T)$ in $K=\mathrm{SU}(3)$ (or the ``magnetic version'' $N_2^\text{mag}$ inside PSU$(3)$), on $M_3=T^3$. We will count with refinement, which in the electric case is about understanding the action of $H^1(M_3,\bZ_3)\simeq \bZ_3^3$ while in the magnetic case it is given by the grading of the components by the mod-3-valued second Stiefel--Whitney class in $H^2(M_3,\pi_1(K))\simeq \bZ_3^3$.

Another convenient way of organizing the components is by relating them to the $21$ flat $S_3$-connections. 

\subsection{Structure of $N_2$}

Let $T\subset \SU(3)$ be a maximal torus, which for convenience we choose to be parametrized by diagonal unitary matrices. The Weyl group is $W = N(T)/T \simeq S_3$, and the normalizer fits into a split exact sequence
\begin{equation}
1 \to T \to N(T) \to S_3 \to 1.
\end{equation}
Thus every element of $N(T)$ has a unique decomposition
\begin{equation}
g = t \cdot w , \qquad t\in T,\; w\in S_3.
\label{gtw}
\end{equation}
Inside SU(3), $S_3$ can be represented as the standard set of permutation matrices, normalized to have determinant one. In other words, a cycle $w$ is mapped to $(-1)^{|w|}P_w$ with $|w|$ being the length of $w$ (as a word consisting of simple reflections). The conjugation action on $T$ given by $t\mapsto  w t w^{-1}$
 shuffles the diagonal entries of $t$, while $g=t\cdot w$ is in general an off-diagonal matrix with three non-vanishing entries. The group $N(T)$ is then identified with all such matrices with unit determinant.

This choice of the embedding $S_3\to N(T)$ also enables us to map flat $S_3$-connections to $N(T)$-connections. However, what is slightly more useful for classification is a map in the opposite direction.

\subsection{Classification}

We first slightly reorganize the 21 conjugacy
classes of commuting triples in $S_3$ from the previous appendix:
\begin{itemize}
\item 1 trivial class $(e,e,e)$,
\item $7=2^3-1$ classes contained in a transposition subgroup $\bZ_2=\{e,a\}$,
\item $13=(27-1)/2$ classes contained in a 3-cycle subgroup $\bZ_3=\{e,b,b^2\}$.
\end{itemize}

Similarly, a flat $N(T)$-connection on the torus is a commuting triple
\begin{equation}
(g_1,g_2,g_3)\in N(T)^3,
\end{equation}
modulo simultaneous conjugation by $N(T)$.  Writing $g_i=t_i w_i$ as in
\eqref{gtw}, one can also project $(t_1 w_1,\; t_2 w_2,\; t_3 w_3)$ to a triple $(w_1,w_2,w_3)$ in $S_3^3$. It is guaranteed to be commuting (as $T$ is abelian and normal), but this might appear ambiguous at the level of $N(T)$-conjugacy classes. However, the cycle-types of the $w_i$ are well defined, which we use to classify $N(T)$-triples. Furthermore, conjugation by $N(T)$ basically allows to set one of $t_i$ to the identity (unless all $w_i$'s are trivial) and one non-trivial $w_i$ to either $a$ or $b$. This leads to a three-fold classification.

\paragraph{The trivial class.} If $w_i=e$ for all $i$, then it is a flat $T$-connection, and they all lie in the same connected component given by $(T\times T\times T)/W$ with the diagonal Weyl group action. The center action of $H^1(T^3,\bZ_3)$ is a symmetry of this identity component. Therefore, it contributes once to the $e=0$ sector.

\paragraph{The $\bZ_2$ classes.}

Consider a triple lying in $\{e,a\}$ for a fixed transposition
$a\in S_3$.  The fixed torus $T^a$ is $1$-dimensional, along the direction of the corresponding co-weight. Therefore, we have a
\emph{continuous} family (in contrast to the O(2) case). As we can conjugate one $t_i$ to the identity, the component is two-dimensional given by $T^a\times T^a$ parametrized by two angles. Embedded into SU(3), the two elements are either $R(\theta)=$diag$\{e^{i\theta},e^{i\theta},e^{-2i\theta}\}$ or this multiplied by $a$, given by
\begin{equation}
   a\cdot R(\theta)=- \begin{pmatrix}
0 & e^{i\theta} & 0 \\
e^{i\theta} & 0 & 0 \\
0 & 0 & e^{-2i\theta}
\end{pmatrix}.
\end{equation}
Up to conjugation, there are 7 components, one for each of the $\bZ_2$-type triples in $S_3$. The action of the $\bZ_3^3$ electric symmetry preserves each component, as it just shifts $\theta_{1}$ and $\theta_2$ by a multiple of $2\pi/3$.

\paragraph{The $\bZ_3$ classes.} For triples with an element conjugate to $b$ generating $\bZ_3\subset S_3$, the $t_i$'s has to live in the $b$-fixed points in $T$, which is the center of SU(3), $\mu_3=\{1,\omega,\omega^2\}$. These central choices give $3^2=9$ distinct lifts, and no two are conjugate in
$N(T)$. The center symmetry acts transitively on the 9 points, therefore splitting it as $1+{\color{blue}8}$ with one in the $e=0$ sector and one in eight out of the 26 non-trivial electric sectors. Precisely which eight sectors are picked depends on the choice of the triple in $S_3$ (and its stabilizer in SL$(3,\bZ)$). Summing over the 13 conjugacy classes of triples restores the SL$(3,\bZ$) symmetry with the split $13\cdot (1+{\color{blue}8})=13+ 26\cdot{\color{blue}4}$, which has a four-dimensional space for each $e\neq 0$ sector.

\medskip

Putting them together, we have 
\begin{equation}
|\pi_0\cM(T^3,N_2)|=1+7+13\cdot9= 125 =21+ {\color{blue}104}.
\end{equation}

One can also work out the refinement for the magnetic version by working with the normalizer $N_2^\text{mag}$ in PSU(3). The Stiefel--Whitney class $w_{2,{(ij)}}$ measures a mod-3 obstruction for lifting a flat $N_2^\text{mag}$-connection to one in SU$(3)$. This issue is only present for the components of $\bZ_3$-type, where the triple consists of elements of the form $g=R^m \cdot b^n $ with $m,n\in\bZ_3$ with $R:=\mathrm{diag}\{1,\omega,\omega^2\}$. Such a triple in general only lift to SU(3) configurations with an 't Hooft flux, given by the commutator of the lift of  $g_i$ and $g_j$ in SU$(3)$, 
\begin{equation}
g_i g_j g_i^{-1} g_j^{-1}
= R^{m_i}b^{n_i}R^{m_j}b^{n_j}b^{-n_i}R^{-m_i}b^{-n_j}R^{-m_j}=R^{m_i-m_j} \cdot (b^{n_i} \circ R^{m_j})\cdot (b^{n_j} \circ R^{-m_i}).
\end{equation}
Here, the circle in $b^n\circ R^m$ denotes the conjugation action of $W$ on $T$, which cyclically shifts the diagonal entries of $R^m$ by $n$ positions. But such an action is equivalent to multiplying by a power of $\omega$,
\begin{equation}
    b^n\circ R^m = \omega^{-mn}\cdot R^m.
\end{equation}
Therefore, the commutator is simply $\omega^{m_in_j-m_jn_i}$ with the power being identified with the $(ij)$-component of the Stiefel--Whitney class. We then have
\begin{equation}
    w_2= (m_2 n_3 - m_3 n_2) dx^2 \cup dx^3 + (m_3 n_1 - m_1 n_3)dx^3 \cup dx^1
 + (m_1 n_2 - m_2 n_1)dx^1 \cup dx^2  = m\cup n.
\end{equation}
Here, we have used the short-hand notation $m=\sum_{i}m_idx^i$ and $n=\sum_{j}n_j dx^j$ which are $\bZ_3$-valued 1-cocycles.

Without loss of generality, we consider the 9 components in the class $(1,b,b^2)$. They are given by $(R^{m_1},b,R^{m_3}b^2)$, leading to 
\begin{equation}
    w_2=(-m_3,m_1,m_1)
\end{equation}
indeed with one trivial and 8 distinct non-trivial fluxes, $9=1+{\color{red}8}$. Summing all the 13 classes gives the same decomposition as the electric case, $13\cdot (1+{\color{red}8})=13+ 26\cdot{\color{red}4}$. Taking into the other two types, both of which have in $w_2=0$, gives the final count as
\begin{equation}
|\pi_0\cM^{m=0}(T^3,N_2^{\text{mag}})|=1+7+13=21,
\end{equation}
while for each of the 26 non-trivial $m$, we have
\begin{equation}
|\pi_0\cM^{m\neq 0}(T^3,N_2^\text{mag})|={\color{red}4}.
\end{equation}
The duality between the electric and magnetic version of $N(T)$ should be an essential ingredient that ensures the Langlands duality for the enriched skein module $\tilde \cS$ between $E_6$ and $E_6/\bZ_3$.

The 21 in the $e=m=0$ sector agrees with the number of flat $S_3$-connections, which mirrors a similar result for $N_1=\mathrm{O}(2)$. Indeed, flat $N(T)^\text{mag}$-connections with trivial $w_2$ cannot make any meaningful use of the $T$-part, and simply become $S_3$-connections up to conjugation and deformation.

Though not directly relevant for our purpose, one can ask about the similar counting problem for flat $N_n$-connections. It is similarly natural to expect that the neutral part agrees with flat $S_{n+1}$-connections. If one extrapolates from $n=1$ and $2$, naively for each non-trivial flux sector, one might expect that the number of components equals the divisor sums of $n+1$.

\section{\texorpdfstring{Counting $\bZ_N$-orbits of $\hat\fsl(N)_{k}$ weights}{Counting Z_N-orbits of sl-hat(N)_k weights}}\label{app:Weights}

In this appendix, we compute the counting function $\Gamma(G;k)$ and its refinements for $G=A_{N-1}$. We assume that $k$ is a multiple of $2N$, which is always satisfied in the cases relevant to the present work.

\subsection{Total count}

An integrable highest weight of $\widehat{\mathfrak{sl}}(N)$ at level $k$ is an $N$-tuple of affine Dynkin labels
\begin{equation}
(\mu_0,\dots,\mu_{N-1}) \in \bZ_{\ge 0}^N, \qquad \sum_{i=0}^{N-1} \mu_i = k.
\end{equation}
A generator $c$  of the center $\bZ_N$ acts on this set by the cyclic rotation
\begin{equation}
c:\quad (\mu_0,\mu_1,\dots,\mu_{N-1}) \mapsto (\mu_{N-1},\mu_0,\dots,\mu_{N-2}).
\end{equation}

Burnside's lemma allows one to rewrite 
\begin{equation}
    \Gamma(A_{N-1};k)=\frac{1}{N}\sum_{m=1}^N|\mathrm{Fix}(c^m)|,
\end{equation}
and we only need to find the number of fixed points for $c^m$.

\subsubsection*{Fixed points of $c^m$}

Let $d = \gcd(m,N)$. The rotation $c^m$ has $d$ cycles, each of length $N/d$. A weight fixed by $c^m$ is constant on each cycle and can be labeled by the $d$-tuple $(x_1,\dots,x_d)$ with
\begin{equation}
x_1+\cdots+x_d = \frac{k\cdot d}{N}.
\end{equation}
Then the number of fixed points is
\begin{equation}
\bigl|\mathrm{Fix}(c^m)\bigr| = \binom{\frac{k\cdot d}{N} + d - 1}{d-1}.
\end{equation}

Grouping the sum over $m$ according to $d$ gives
\begin{equation}\label{GammaA}
\Gamma(A_{N-1};k)=\frac{1}{N}\sum_{m=1}^N \binom{\frac{k\cdot d}{N} + d - 1}{d-1}=\frac{1}{N}\sum_{d|N}\varphi\left(\frac{N}{d}\right)\binom{\frac{k\cdot d}{N} + d - 1}{d-1},
\end{equation}
where $\varphi(n)$ denotes Euler's totient function that counts positive integers up to $n$ that are co-prime with $n$.

One can quickly check that when $N=2$ and $k=4g-4$, we have 
\begin{equation}
    \Gamma(A_1;4g-4)=\frac{1}{2}\left(1+\binom{4g-4 + 1}{1}\right)=2g-1.
\end{equation}
And for $N=3$ and $k=6g-6$, we have
\begin{equation}
    \Gamma(A_2;6g-6)=\frac{1}{3}\left(2+\binom{6g-6 + 2}{2}\right)=6g^2-9g+4.
\end{equation}
For general $N$, the asymptotics for large $g$ is controlled by the $d=N$ term, leading to 
\begin{equation}
    \Gamma(A_{N-1};(2g-2)\cdot N)\sim \frac{N^{N-1}}{N!}\cdot  (2g)^{N-1} + \ldots
\end{equation}
We thus see that this polynomial has fractional coefficients for all $N>2$.

\subsection{Refinement}

We refine the count of weights by the central character of $\bZ_N$ on the weight space, 
\begin{equation}
   \mu=(\mu_0,\mu_1,\dots,\mu_{N-1}) \mapsto r(\mu)=\sum_{m=0}^{N-1}m\cdot\mu_m \bmod N.
\end{equation}
This is preserved by the action of $c$ as long as $N|k$.

To count the number of weights carrying character $r \in \bZ_N$, consider the generating function
\begin{equation}
    F(q,t)=\sum_{\mu\in \bZ_{\ge 0}^N}q^{\sum_i \mu_i}t^{\sum_j\mu_j\cdot j}=\prod_{j=0}^{N-1}\frac{1}{1-qt^j},
\end{equation}
where $t$ is taken to be an $N$-th root of unity. Then setting $t=1$ and taking the coefficient of the $q^k$ term gives the total number of weights,
\begin{equation}
    \left[F(q,t=1)\right]_{q^k}=\binom{k+N - 1}{N-1}.
\end{equation}
 We can project out the number of weights with central character $r$ using a discrete Fourier transform,
\begin{equation}
    a_r = \frac{1}{N} \sum_{n=0}^{N-1} e^{-2\pi i r n/N} \cdot\left[\prod_{j=0}^{N-1}\frac{1}{1-q\cdot e^{2\pi nj/N}}\right]_{q^k}.
\end{equation}
For a given $n$, take $\ell = \gcd(n,N)$, and one has
\begin{equation}
    \prod_{j=0}^{N-1}\left(1-q\cdot e^{2\pi nj/N}\right)=(1-q^{N/\ell})^\ell.
\end{equation}
Taking the coefficient of the $q^k$ term of its inverse gives
\begin{equation}\label{arsum}
    a_r=\frac{1}{N} \sum_{n=0}^{N-1} e^{-2\pi i r n/N} \cdot \binom{\frac{k\cdot \ell}{N} + \ell - 1}{\ell-1}.
\end{equation}
This is quite similar to the expression \eqref{GammaA} for $\Gamma$ except for the phase. 

Now, we now should ``combine'' the two to count the number of orbits $\Gamma^r(A_{N-1};k)$ for each central character $r$. This is achieved by replacing each binomial in \eqref{GammaA} by a sum like \eqref{arsum}.

This follows from a key observation that, for a fixed point of $c^m$, the central character of the $d$-tuple $x=(x_0,\ldots,x_{d-1})$ is
\begin{equation}\label{factorization}
    r(\mu)\equiv s_d(x)\cdot \frac{N}{d} \pmod N
\end{equation}
when $2N|k$. Here, we have defined 
\begin{equation}
s_d(x):=\sum_{j=0}^{j=d-1} j\cdot x_j\in\bZ_d
\end{equation} which gives back $r$ when $d=N$. The factorization \eqref{factorization} means that for such a fixed point, $r$ lives in a $\bZ_d$ subgroup of $\bZ_N$. Then $x$ can be viewed as a weight for $A_{d-1}$ at level $\frac{k\cdot d}{N}$, and we know from \eqref{arsum} how to count the classes with different central character $s_d$,
\begin{equation}
    b_{s,d}=\frac{1}{d} \sum_{n=0}^{d-1} e^{-2\pi i s n/d} \cdot \binom{\frac{k\cdot \ell}{N} + \ell - 1}{\ell-1}, \quad \ell=\gcd(n,d).
\end{equation}
Notice that $b$ also depends on $k/N$, which we have suppressed to avoid clutter. We have the $r$-th sector receiving a contribution from $b_{s,d}$ when $r$ is in the $\bZ_d$ subgroup (i.e.~$N|rd$). In other words, $s={rd/N}$ needs to be integer. We then have
\begin{equation}\label{Gammar}
    \Gamma^r(A_{N-1};k)=\frac{1}{N}\sum_{\substack{d|N\\N|rd}}\varphi\left(\frac{N}{d}\right)b_{rd/N,d},
\end{equation}
where the first sum is over divisors $d$ of $N$ subject to the condition that $N|rd$. 
The expression for $b_{s,d}$ can be slightly simplified by reorganizing it as a sum over divisors of $d$,
\begin{equation}
    b_{s,d}=\frac{1}{d} \sum_{n|d}c_{d/n}(s)  \cdot \binom{\frac{k\cdot \ell}{N} + \ell - 1}{\ell-1}, \quad \ell=\gcd(n,d),
\end{equation}
where 
\begin{equation}
    c_a(s) =\sum_{\substack{1\le m\le a\\(m,a)=1}}e^{-2\pi i ms/a}
\end{equation}
is ``Ramanujan's sum'' given by including a phase factor in the sum for $\varphi$. 

For a general $N$, this seems to be a rather complicated expression with double summations (although they are both quite ``small''). For several special cases, it simplifies significantly. One such case is when $r$ is co-prime with $N$, then $d=N$ and we only have a single sum over divisors of $N$. Furthermore, when $N=p$ is a prime, both sums collapse to at most two terms, and we have
\begin{equation}
    \Gamma^{r\neq0}(A_{p-1};k)=\frac{1}{p}b_{r,p}=\frac{1}{p^2} \left(c_{p}(r)+c_{1}(r)\binom{k + p - 1}{p-1}\right)=\frac{1}{p}b_{r,p}=\frac{\binom{k + p - 1}{p-1}-1}{p^2},
\end{equation}
which are the same for all $r$ as long as $r\neq 0$. On the other hand,
\begin{equation}
    \Gamma^{0}(A_{p-1};k)=\frac{1}{p}\left((p-1)b_{0,1}+b_{0,p}\right)=1 + \frac{1}{p^2}
\left(
\binom{k+p-1}{p-1} - 1
\right),
\end{equation}
which is one more than $\Gamma^r$ for other $r$.

In the above, we have used $c_1(r)=1$ and 
\begin{equation}
    c_p(r)= \begin{cases}p-1, & \text{if $r=0$,}\\
    -1, & \text{otherwise.}\end{cases}
\end{equation}

We now list the results for $N=2$ and $3$, taking the level to be $k=(2g-2)N$. For $N=2$, we have 
\begin{equation}
    \Gamma^{r}(A_{1};4g-4)= \begin{cases}g, & \text{$r=0$,}\\
    g-1, & \text{$r=1$.}\end{cases}
\end{equation}
For $N=3$, we have 
\begin{equation}
    \Gamma^{r}(A_{2};6g-6)= \begin{cases}(2g-1)(g-1)+1, & \text{$r=0$,}\\
    (2g-1)(g-1), & \text{$r=1,2$.}\end{cases}
\end{equation}
These reproduce results we have found previously.

Now we use the formula to compute the count in the $N=4$ case used in the main text. A direct computation gives
\begin{equation}
    \Gamma^0(A_3;k)=\frac{1}{4}(\varphi(4)b_{0,1}+\varphi(2)b_{0,2}+\varphi(1) b_{0,4})=\frac{k^3+6k^2+20k+96}{96}.
\end{equation}
The denominator is $96=4^2\times 3!$ as expected, and it evaluates to integers when $8|k$. For $r=2$, we have
\begin{equation}
     \Gamma^2(A_3;k)=\frac{1}{4}(b_{1,2}+b_{2,4})=\Gamma^0(A_3;k)-1.
\end{equation}
The counts are the same for $r=1$ and 3, given by
\begin{equation}
     \Gamma^1(A_3;k)=\Gamma^3(A_3;k)=\frac{1}{16}\cdot b_{1,4}=\frac{k^3+6k^2+8k}{96}=\Gamma^2(A_3;k)-\frac{k}{8}.
\end{equation}

\bigskip

\paragraph{Conflict of Interest.} The author declares that they have no relevant conflict of interest.

\paragraph{Data Availability.} Data-sharing is not applicable to this article, as no datasets were generated or analyzed during the current study.

\providecommand{\doihref}[2]{\href{#1}{#2}}
\providecommand{\arxivfont}{\tt}

\bibliographystyle{utphys} 
\bibliography{refs}

\end{document}